\documentclass{article}

\usepackage{arxiv}

\usepackage[utf8]{inputenc} 
\usepackage[T1]{fontenc}    
\usepackage{hyperref}       
\usepackage{url}            
\usepackage{booktabs}       
\usepackage{amsfonts}       
\usepackage{nicefrac}       
\usepackage{microtype}      
\usepackage{lipsum}		
\usepackage{graphicx}
\usepackage{natbib}
\usepackage{doi}

\usepackage{graphicx}
\usepackage{multicol}
\usepackage{placeins}
\usepackage{xcolor}
\usepackage{amsmath,amssymb}

\usepackage{mathptmx}
\usepackage{verbatim}
\usepackage{multirow}
\usepackage{cuted}

\DeclareMathAlphabet{\pazocal}{OMS}{zplm}{m}{n}
\DeclareUnicodeCharacter{394}{$\Delta$}
\newcommand{\rs}{r_\mathrm{s}}
\newcommand{\dm}{\mathrm{\Delta}M}
\newcommand{\dmm}{\mathrm{\Delta}M/M}
\newcommand{\drs}{\mathrm{\Delta}\rs}
\newcommand{\cl}{\pazocal{L}}
\newcommand{\ce}{\mathcal{E}}

\newcommand{\diff}{\mathrm{d}}
\newcommand{\dd}{\mathrm{d}}

\usepackage{placeins}

\newcommand\actaa{\ref@jnl{AcA}}
\newcommand\aapr{\ref@jnl{A\&A~Rv}}

\DeclareMathAlphabet{\pazocal}{OMS}{zplm}{m}{n}

\title{Geodesic Model of HF QPOs Tested for Black Holes in Spacetimes Reflecting the Effect of Surrounding Dark Matter}

\author{Zden\v{e}k Stuchl\'{\i}k \\
	Research Centre for Theoretical Physics and Astrophysics\\
	Institute of Physics, Silesian University in Opava\\
	Bezru\v{c}ovo n\'am.~13, 746\,01, Opava, CZ \\
	\texttt{zdenek.stuchlik@physics.slu.cz} \\
	\And
	Jaroslav Vrba \\
	Research Centre for Theoretical Physics and Astrophysics\\
	Institute of Physics, Silesian University in Opava\\
	Bezru\v{c}ovo n\'am.~13, 746\,01, Opava, CZ \\
	\texttt{jaroslav.vrba@physics.slu.cz} \\
}

\renewcommand{\shorttitle}{Geodesic Model of HF QPOs in Spacetimes with  Dark Matter}

\hypersetup{
pdftitle={Geodesic Model of HF QPOs Tested for Black Holes in Spacetimes Reflecting the Effect of Surrounding Dark Matter},
pdfsubject={},
pdfauthor={Zden\v{e}k Stuchl\'{\i}k, Jaroslav Vrba},
pdfkeywords={Supermassive black holes, Dark matter, Quasiperiodic Oscillations},
}

\begin{document}
\maketitle

\begin{abstract}
Using the simple but robust model of a shell of dark matter (DM) around a Schwarzschild black hole (BH), represented by the mass ratio of the shell and BH $\dmm$, the shell extension $\drs$ and its inner radius $\rs$ , we study the influence of DM on the spacetime structure and geodesic motion, and provide a classification of the BH+DM shell spacetimes according to the properties of the stable circular geodesics governing Keplerian disks. We focus our attention on the epicyclic motion around circular geodesics that can be related to observational phenomena in X-ray radiation from Keplerian accretion disks, assumed to be influenced by the DM shell only gravitationally. We give the frequencies of the orbital and epicyclic motions and discuss their properties in terms of the parameters governing the DM shell. Using the frequencies in relevant variants of the standard geodesic model of high-frequency quasiperiodic oscillations (HF QPOs), we test the role of DM by fitting the HF QPO data from some microquasars and active galactic nuclei with supermassive BHs where no variant of the geodesic model applied in the standard vacuum BH background is able to explain the data. We thus provide a robust review of the applicability of the geodesic model of HF QPOs, and also provide limits on the amount of DM around a BH. We demonstrate that the geodesic model could be well applied to most observations of active galactic nuclei, with strong restrictions on the amount of invisible matter around BHs.
\end{abstract}


\section{Introduction}
The cosmological tests related to properties of the relict radiation fluctuations \citep{Ada-Dur-Gue:2013:ASTRA:,Ade-Pla-Col:2014:ASTRA:} along with measurements of the accelerated expansion of the universe \citep{Rie-etal:2004:APJ:}, and with additional studies of the structure of the Universe \citep{Bah-Ost-Per:1999:Sci:,Cal-Kam:2009:Nat:}, or of galaxies and galaxy clusters \citep{Bosma:1981:AstronJ:,Bar-etal:2015:MNRAS:,Sar-etal:2014:ApJL:}, give clear indications on the relevance of dark energy and dark matter (DM) for the expansion of the universe and the behavior of galaxies and their clusters \citep{Linde:1990:PPIC:,Kra-Tur:1995:GRG:,Per-Sal:1996:MNRAS:,Stu-Hle-Nov:2016:PRD:,Salucci:2019:AAPR:}. Dark energy can be well represented by the relict cosmological constant that implies significant effects even on the level of large galaxies in relation to central supermassive black holes (BHs) \citep{Stu-Sla-Hle:2000:ASTRA:,Rez-Zan-Fon:2003:ASTRA:,Stu:2005:MPLA:,Stu-Sla-Kov:2009:CLAQG:,Stu-Sche:2011:JCAP:,Stu-Hle-Nov:2016:PRD:}, especially in connection to the so-called \textit{static radius} giving a natural limit on gravitationally bound systems in the expanding universe \citep{Stu:1983:BAC:,Stu-Hle:1999:PRD:,Stu-Char-Sche:2018:EPJC:,Stu-etal:2020:Uni:}. \footnote{An alternative related to the quintessential dark energy is represented by the Kiselev BHs \citep{Kis:2003:CLAQG:} and their rotational generalization \citep{Tos-Stu-Ahm:2017:EPJP:,Abd-Tos-Stu:2017:IJMPD:}.}

The influence of the cosmological constant (dark energy) is most relevant at the edges of large galaxies, but the influence of the DM could be in principle also relevant also in the immediate vicinity of the central supermassive BHs of large galaxies. Of course, the
effect of DM could also be, in some circumstances, significant
near the stellar-mass BHs in binary systems (microquasars).
Therefore, it is of crucial interest to estimate the possible role of
DM in close vicinity of the event horizon of BHs. Note that
metrics combining BH vacuum metrics with the influence of
standard models of DM halos could be expected irrelevant for
the phenomena observed in close vicinity to the central BH, as
they are tuned by explaining phenomena (orbital velocity
related to rotational curves) in galaxy regions distant from the
center. In fact, we have to use a model enabling us to reflect
trapping of any kind of DM by the strong gravity of the
central BH.

Using a very simple and robust model of DM located in the
vicinity of a BH, Konoplya \citep{Konoplya:2019:PLB:} presented a rough estimate
of the possible role of DM on the BH shadow. In this model of
DM distributed around a Schwarzschild BH, which was implied for rough studies of the role of the DM halos \citep{Leu-Liu-Sue:1997:PRL:,Kon-Zhi:2011:RevModPhys:,Car-Pan:2017:NatAstr:,Kon-Zhi-Stu:2019:PRD:,Bar-Car-Pan:2017:PHYSR4:}, the DM shell is represented by its total mass $\dm$ related to the mass of the central BH $M$, while distribution of its mass is represented by the extension of the DM region $\Delta \rs$ and location of its inner edge $\rs$. A more complex model of a BH surrounded by a DM halo taking into account the phenomenological model of DM halos \citep{Nav-Fre-Whi:1997:ApJ:}, can be found in \cite{Jus-Jam-etal:2019:PRD:}. Recently a new \textit{fluid-hairy} (or f-hairy) exact BH solution with DM halo was introduced in \cite{Car-etal:2021:arxiv:}, with the influence of the assumed DM around BHs reflecting the Hernquist model of DM connected to the central part of galaxies. 

An efficient test field for the astrophysical relevance of DM around supermassive BHs seems to be the geodesic model of high-frequency quasiperiodic oscillations (HF QPOs), which explains well data coming from microquasars \citep{Tor-etal:2011:ASTRA:} or neutron stars \citep{Tor-etal:2010:ApJ:,Tor-etal:2012:ApJ:,Stu-etal:2015:ActaA:}, but totally fails in the case of data from active galactic nuclei where supermassive BHs are assumed \citep{Smi-Tan-Wag:2021:ApJ:}. The geodesic model has a variety of variants that are all based on the frequencies of the orbital epicyclic test particle motion in the BH (or some of its mimickers) gravitational field, and on combinations of these frequencies. The model thus includes the possibility of oscillatory motion of radiating sources, hot spots on Keplerian disks, or structures (e.g. toroids) oscillating with frequencies equal to those of the epicyclic motion \citep{Stu-Kot-Tor:2013:ASTRA:}. The idea of the geodesic model goes back to the relativistic precession (RP) model of hot spots of Keplerian disks introduced by Stella and his collaborators \citep{Ste-Vie-Mor:1999:ApJ:}. The idea of relevance of resonant effects between the oscillatory modes was introduced by Abramowicz and Kluzniak and their collaborators \citep{Klu-Abr:2001:ACTAASTR:,Tor-Abr-etal:2005:AA:}. A detailed overview of the geodesic model can be found in \cite{Stu-Kot-Tor:2013:ASTRA:}; on the importance of the resonant phenomena in various variants of the geodesic model \cite{Tor-Abr-etal:2005:AA:}. Modifications of the geodesic model to the orbital and epicyclic frequencies corrected due to interactions of slightly charged matter of the disk with external electromagnetic fields around the BHs were introduced in \cite{Kol-Stu-Tur:2015:CLAQG:} and \cite{Stu-Kol:2016:EPJC:,Tur-Stu-Kol:2016:PRD:} and applied in \cite{Kol-Tur-Stu:2017:EPJC:,Pan-Kol-Stu:2019:EPJC:,Stu-etal:2020:Uni:}. An alternative string oscillation model \citep{Stu-Kol:2012:JCAP:} was also considered for an explanation of the HF QPOs observed in microquasars \citep{Stu-Kol:2014:PRD:,Stu-Kol:2016:ApJ:}.

In fact, the twin HF QPOs observed in microquasars serve as a very efficient tool for restricting the free parameters of BH solutions constructed in the framework of alternative theories of gravity, regular BHs constructed in general relativity combined with nonlinear electrodynamic theories, around magnetized BHs, or generally for BH mimickers. There is a large variety of such studies of which we mention several cases: regular BHs related to modified gravity \citep{Ray-etal:2021:Gal:}, regular BHs related to nonlinear electrodynamics \citep{Ray-Abd-Han:2021:PRD:,Ray-etal:2022:PDU:,Ray-Ahm-Bok:2022:IJMPD:}, noncommutative BHs \citep{Ray-Bok-Ahm:2022:CQG:}, braneworld BHs \citep{Kot-Stu-Tor:2008:CQG:,Stu-Kot:2009:GRG}, BHs in Gauss-Bonnet gravity \citep{San-etal:2020:PDU:,Ray-Bar-Abd:2022:IJMDP:} variety of BH mimickers \citep{Xin-etal:2021:EPJC:,All-Sha:2021:JCAP:,Stu-Vrb:2021:Universe:,Stu-Vrb:2021:EPJP:}; magnetized BHs \citep{San-etal:2022:arx:,Tur-Stu-Kol:2016:PRD:,Stu-etal:2020:Uni:}.

Using the resonance epicyclic variant of the geodesic model,
we demonstrate a possible positive role of DM treated in the
framework of the f-hairy BH solution \citep{Car-etal:2021:arxiv:} to explain the HF QPO data related to the supermassive BHs in active galactic nuclei \citep{Stu-Vrb:2021:JCAP:}. Note that in a
similar positive role in fitting the QPO data from active galactic
nuclei we demonstrated using the assumption of central
attractor represented by wormholes—namely, in \cite{Stu-Vrb:2021:Universe:}--for a very simple model of reflection symmetric wormholes \citep{Sim-Vis:2019:JCAP:}, and in \cite{Stu-Vrb:2021:EPJP:} for more realistic Einstein-Dirac-Maxwell wormhole model \citep{Bla-sal-Kno-Rad:2021:PhRvL:,Chur-etal:2021:JCAP:,Kon-Zhi:2022:prl:}, which did not require an exotic form of matter guaranteeing the stability of the wormhole. Of course, an explanation of HF QPOs observed in active galactic nuclei based on the DM models would be of higher relevance
from the point of view of astrophysics.

We believe that it would be very useful to realize a detailed
study of the HF QPO geodesic model, applying its widely
discussed variants in a representative model of BH spacetime
modified by the presence of DM. We expect that the simple but
robust model of the DM shell modifying the spherically
symmetric Schwarzschild BH spacetime \citep{Kon-Zhi:2011:RevModPhys:,Car-Pan:2017:NatAstr:,Konoplya:2019:PLB:} represents the most convenient model for our purposes, despite its roughness, as it contains robust estimates of the properties of the DM around BHs -- namely on the amount of DM as related to the mass of the central BH, its distribution as related to the BH, and its influence on the spacetime geometry near the BH event horizon.

There are two advantages of this rough model of the DM influence on the BH spacetime \citep{Car-Pan:2017:NatAstr:,Konoplya:2019:PLB:}. The first one is based on the fact that the model is governed by independent fundamental parameters characterizing the shell: the mass ratio $\dm/M$, inner edge position $\rs$ and extension $\drs$. The second one is that it can be related to any form of DM--that binding the whole galaxy, but also its variants of local character, as non-shining stars (brown dwarfs), stellar-mass BHs, primordial BHs, non/radiating dust, etc, all being captured by the central supermassive BH gravity. In other words, if based on DM influencing the structure of galaxies, the models the reflecting the influence of DM on vacuum BH spacetimes are focused only on one case of DM, or only on fixed models of DM influence on some regions of the galaxy, usually its outer parts, while the rough model considered here enables inclusion of all phenomena that could represent DM in close vicinity to the BH where the HF QPOs are observed. 

The most critical point of the BH+DM shell model is selection of the inner edge of the shell. Generally, it is assumed that $\rs\geq 2M$, and in the previous study of the influence on the BH shadow \citep{Konoplya:2019:PLB:}, the $\rs=2M$ was selected to provide observable effect of DM on the BH shadow. Of course, the choice $\rs = 2M$, corresponding to the BH event horizon, is a physically unrealistic choice for stationary and stable configurations. For this reason, we shall concentrate our attention in this paper on physically realistic possibilities where the inner edge is restricted by existence of circular geodesics in the BH spacetime. We could consider the three important radii of the circular geodesics in the vacuum Schwarzschild spacetime -- namely, by the radius of the photon circular orbit ($\rs=3M$), bound circular orbit ($\rs=4M$) and stable circular orbit ($\rs=6M$) of the Schwarzschild BH background. From the point of view of astrophysics, the most realistic is naturally the choice of $\rs=4M$, as it limits all matter energetically bound to the central BH - this choice will be applied in our study, and especially in the tests of the geodesic model for the observational data.

In the present paper, we study how the DM shell region influences the circular motion of test particles representing a Keplerian accretion disk orbiting in the spacetime governed by the configuration of the BH and the shell of surrounding DM (BH+DM shell model), influencing the Keplerian disk only due to its gravity. The crucial point of our study is how the epicyclic oscillations of the orbital motion of the Keplerian disk matter, modified by the DM shell, could influence selected variants of the geodesic model of HF QPOs, related to the X-ray radiation oscillations observed in microquasars, or around supermassive BHs in active galactic nuclei.

First, we place the reality conditions on the BH+DM shell
spacetimes, which guarantee the stationarity of the shell and
exclude the dynamical possibilities of the spacetime structures
located above the event horizon of the central BH. Then we
determine the circular geodesic motion—frequencies of the
orbital motion, its radial and vertical epicyclic oscillations, and discuss their properties in dependence on the parameters of the DM shell. We use these frequencies in the geodesic model of HF QPOs \citep{Stu-Kot-Tor:2013:ASTRA:,Stu-etal:2020:Uni:}: (1) in its selected
astrophysically relevant variants, especially in the epicyclic
resonance (ER) variant, where the epicyclic frequencies are
identified with the observed upper and lower frequencies of the
twin peak HF QPOs, (2) in its its sub-variants and the RP
variant, where the orbital frequency enters into play, and (3) in
its sub-variants to test applicability of geodesic model in
explaining the HF QPOs in the two recently discussed cases.
First, we put limits on the parameters characterizing the DM
shell implied by the reality conditions, and by the requirement
of existence of important circular geodesics of the spacetime;
this enables to find regions allowing for the existence of stable
circular geodesics necessary for the epicyclic motion related to
the twin HF QPOs observed in the microquasars, where the
geodesic models work well for vacuum BH spacetimes \citep{Tor-etal:2011:ASTRA:,Stu-Kol:2016:ASTRA:}. Second, and most importantly, we test the possibility of explaining the QPOs observed around supermassive BHs in active galactic nuclei, where a total failure of the variants of the geodesic model related to isolated (vacuum) BHs was demonstrated \citep{Smi-Tan-Wag:2021:ApJ:}. For all the sources presented in \cite{Smi-Tan-Wag:2021:ApJ:} we give the range of the parameters of the BH+DM shell spacetimes enabling explanation of the HF QPOs observed in these sources by the selected variants of the geodesic model. Our results could thus considered as a very rough but robust limit on the allowed range of the amount of DM around supermassive BHs in the considered sources, if applied to an explanation of the observed HF QPOs. On the other hand, in the case of microquasars the strong restrictions on the allowed maximum mass of the DM  around the central BH, which does not destroy the satisfactory results of the geodesic model for vacuum BH spacetimes, are implied. 

Throughout the paper we use space-like signature \mbox{$(-,+,+,+)$}, and a geometric system of units in which $G = c = 1$; we restore them when we need to compare our results with observational data. Greek indices run from $0-3$, Latin indices from $1-3$.

\section{Schwarzschild BH surrounded by DM shell}

The simplest, very rough, but robust model of a spherically symmetric spacetime representing a BH surrounded by a shell of DM was introduced in \cite{Kon-Zhi:2011:RevModPhys:} and \cite{Car-Pan:2017:NatAstr:} and can be expressed in the Schwarzchild-like form using the mass function $m(r)$ \citep{Konoplya:2019:PLB:}:  

\begin{eqnarray}\label{e:met}
	\mathrm{d}s^2=&-&\left(1-\frac{2m(r)}{r} \right)\mathrm{d}t^2+\left(1-\frac{2m(r)}{r} \right)^{-1}\mathrm{d}r^2\nonumber\\
	&+&r^2\big(\mathrm{d}\theta^2 + \sin^2\theta \mathrm{d}\varphi^2\big),
\end{eqnarray}
where 
\begin{equation}
	m(r)= 
	\begin{cases}
		M, \quad r \le \rs\\
		M + \dm\left( 3-2\frac{r-\rs}{\drs} \right)\left(\frac{r-\rs}{\drs} \right)^2,\quad \rs\leq r \leq \rs + \drs ,\\
		M + \dm,\quad r > \rs + \drs. 
	\end{cases}
	\label{e:mr}
\end{equation}
\begin{figure*} [h]
	\includegraphics[width=\linewidth]{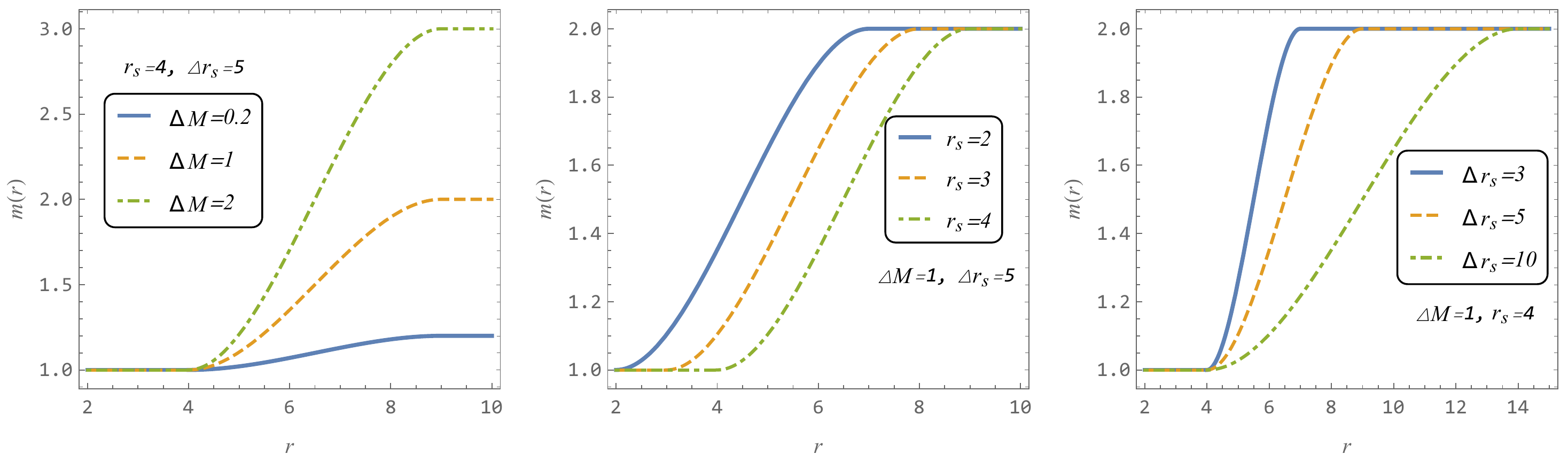}
	\caption{Mass function $m(r)$ (\ref{e:mr}) illustrated for various parameters $\dm$, $\rs$ and $\drs$. We put $M=1$ for simplicity and in the following figures, if not stated otherwise.}
	\label{f:f1}
\end{figure*}
$M$ denotes the BH mass parameter, while $\dm$ denotes the mass of the DM shell that is distributed in the region of extension $\drs$ with the inner edge located at the radius $\rs \geq 2M$. Some illustrations of the DM cloud mass profile are presented in Figure\ref{f:f1}.
\subsection{Reality Condition on the BH+DM shell spacetime}

The considered spacetime describes a central Schwarzschild BH with an exterior modified by the presence of invisible matter concentrated in a shell with a given extension and fixed inner radius. Its geometry is governed by the lapse function
\begin{eqnarray}
   f(r)=1-\frac{2m(r)}{r}
\end{eqnarray}
with $m(r)$ given by Equation (\ref{e:mr}). In the following (if not stated otherwise), we assume for simplicity $M=1$, i.e., mass and radius coordinates are expressed in units of BH mass $M$.

Of course, in order to describe a stationary (static) geometry at the exterior of the central Schwarzschild BH, the lapse function has to fulfill the condition $f(r)>0$ at $r>2$. However, for any $\rs>2$ and fixed $\dm>0$, there exists a dynamic region of the spacetime characterized by $f(r)<0$ at some part of the region $r>2$, if $\drs$ is low enough. Such geometries are, of course, forbidden as they give physically unrealistic spacetimes.

\begin{figure} 
  \centering
	\includegraphics[width=0.34\linewidth]{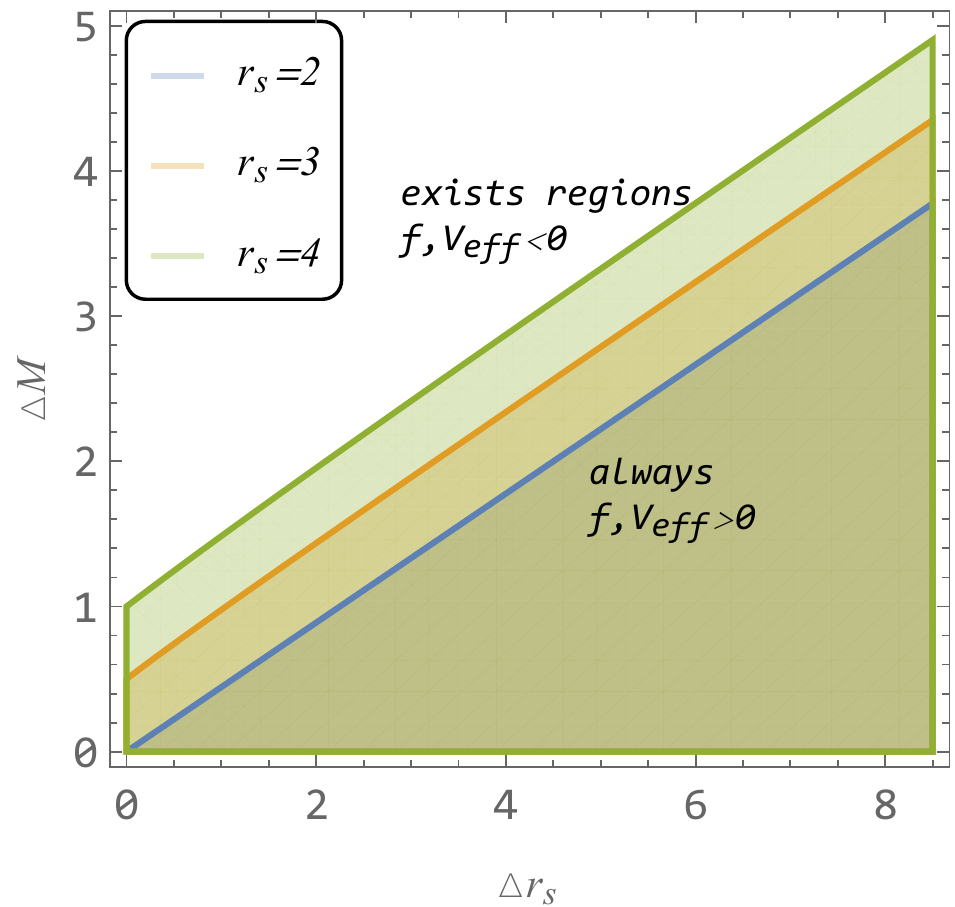}
	\caption{Solution of Eq. (\ref{e:real}) for $\rs=2-4$ showing in $\drs - \dm$ parameter space the acceptable combinations of parameters to have realistic spacetimes with a positive lapse function at $r>2$.}
	\label{f:real}
\end{figure}

The boundary of the realistic and forbidden BH+DM shell spacetimes is in the parameter space $\rs,\drs$ and $\dm/M$ given by the condition $f(r_\mathrm{min})=0$, where $r_\mathrm{min}$ is the location of the minimum of the lapse function given by the condition $\dd f(r)/\dd r=0$. The minimum $f_{min}$ has to fulfill the cubic equation taking the form 
\begin{eqnarray}
  &&4 f_{min}^3 \drs^3+f_{min}^2 \left(9 \dm \drs^2-108 \dm \drs \rs-108 \dm \rs^2-12 \drs^3\right)+ \nonumber \\
  &&f_{min} \left(-216 \dm^2 \rs-18 \dm \drs^2+216 \dm \drs \rs-216 \dm \drs+216 \dm \rs^2-432 \dm \rs+12 \drs^3\right)+ \nonumber\\
   &&216 \dm^2 \rs-432 \dm^2+9 \dm \drs^2-108 \dm \drs \rs+216 \dm \drs-108 \dm \rs^2+432 \dm \rs-432 \dm-4 \drs^3=0.
   \label{e:real}
\end{eqnarray}
The resulting relation for the boundary of the forbidden BH +DM shell spacetime is from the numerical solution of the equations and is presented in Figure \ref{f:real}

The regions of well-defined geometries are represented in Figure \ref{f:real} for fixed typical values of $\rs$. Notice that the allowed regions of the parameter space, giving the stationary spacetimes external to the central BH, increase with increasing value of the parameter $\rs$, which is crucial for our investigation. Our solutions will be relevant only in the allowed regions of the parameters, i.e., under the $\dm(\drs)$ lines given for fixed considered values of $\rs$. Note that the stationarity of the whole configuration (the geometry + the matter shell) is forbidden for solutions with $\rs\in \{2,3\}$ as stationary states of moving matter are not allowed there.

The influence of the DM shell on the physical properties of the BH+DM shell spacetime strongly depends on the parameters as shown in the case of the BH shadow \citep{Konoplya:2019:PLB:}, where it was demonstrated that the modifications of the BH shadow shape due to the DM shell will only be significant if $\rs=2$, and for relatively large shell masses ($\dm \geq 10$). Here, we focus our attention on the case of the inner edge of the shell at $r_{s}=4$, corresponding to the position of the marginally bound circular geodesic of the Schwarzschild spacetime represented from the point of view of astrophysical phenomena; this is the most natural choice, as it reflects in an appropriate  way the possibility of successive gravitational binding of incoming matter by the strong gravitational field of the BH. 

We thus focus our study on the influence of DM on test particle epicyclic motion related to the Keplerian disk orbiting in the common gravitational field of the central BH and the DM shell with a shell inner radius at $\rs=4$. The Keplerian disk is orbiting in a central plane of the spherical BH-shell system. We consider not only $\dm>1$, as in \cite{Konoplya:2019:PLB:}, but also values of the mass of the shell $\dm \leq 1$ in order to cover whole the range of the possible of DM influence on the epicyclic oscillatory motion. 
	
\FloatBarrier

\section{Test particle motion}

The motion of test particles with nonzero or zero rest mass is governed by the geodesics of the spacetime. Here, we concentrate our attention on circular geodesics. The geodesics of a spacetime are governed by the geodesic equation for the four-momentum (wavevector) $p^{\mu}$ ($k^{\mu}$) of the massive (massless) particle. The geodesic equation reads as
\begin{equation}
\frac{\diff^2 x^\mu}{\diff \tau^2} + \Gamma^\mu_{\rho\sigma}\,\frac{\diff x^\rho}{\diff \tau} \frac{\diff x^\sigma}{\diff \tau}=0
\end{equation}
and the accompanied norm condition for the geodesics of the test particle having rest mass $m$ or photon with $m=0$ reads as
\begin{equation}
u^\mu u_\mu = g_{\mu\nu}\, u^\mu u^\nu = -\epsilon
\end{equation}
where the parameter $\epsilon = m^2$ is for massive particles and $\epsilon = 0$ is for photons and other massless particles. 

Because of the spherical symmetry of the spacetime, the geodesic motion is fixed to central planes; the equatorial plane $\theta = \pi/2 = const$ can be selected, if only one particle (or flat Keplerian disk) is under consideration. Due to the additional spacetime symmetries, two constants of the motion exist along with the fixed central plane -- namely, stationarity of the spacetime implies conservation of the covariant energy and axial symmetry implies conservation of the axial angular momentum 
\begin{equation}
  E = -p_t \qquad L = p_{\phi}.
  \label{e:conserv}
\end{equation}

\subsection{The photon motion and circular null geodesics}

\begin{figure*}[ht] 
	\includegraphics[width=\linewidth]{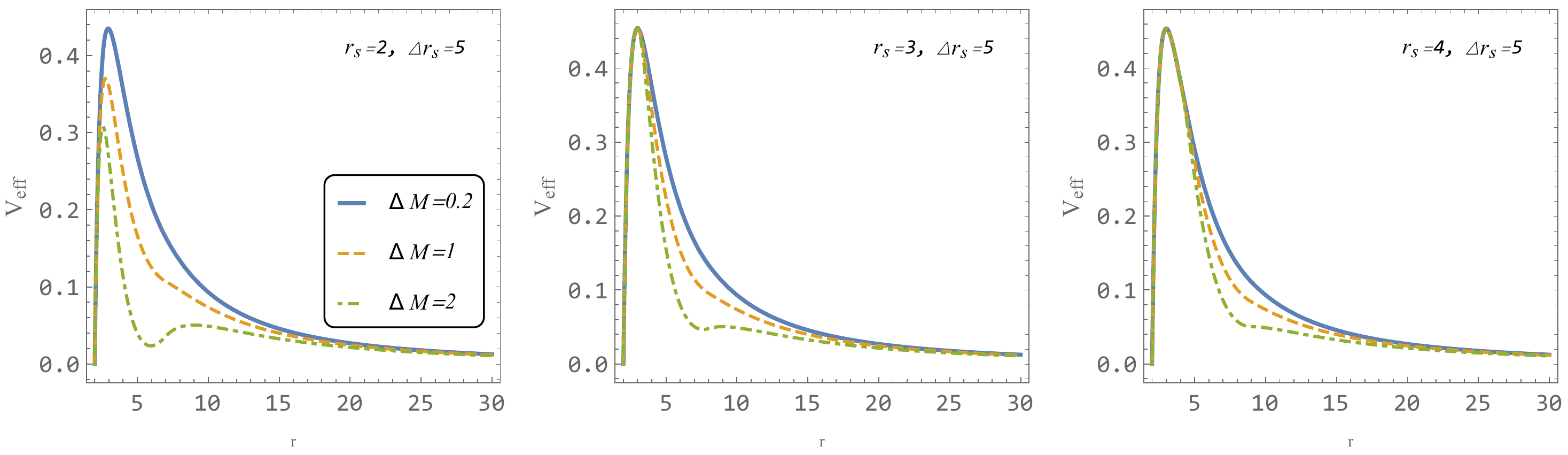}
	\caption{Effective potentials of massless test particles for various parameters, $\dm$, $\rs$, and $\drs$. Here, the range of the DM is fixed to $\drs=5$ and $\rs$ takes all the considered values $2-4$.}
	\label{f:f2}
\end{figure*}

\begin{figure*} 
	\includegraphics[width=\linewidth]{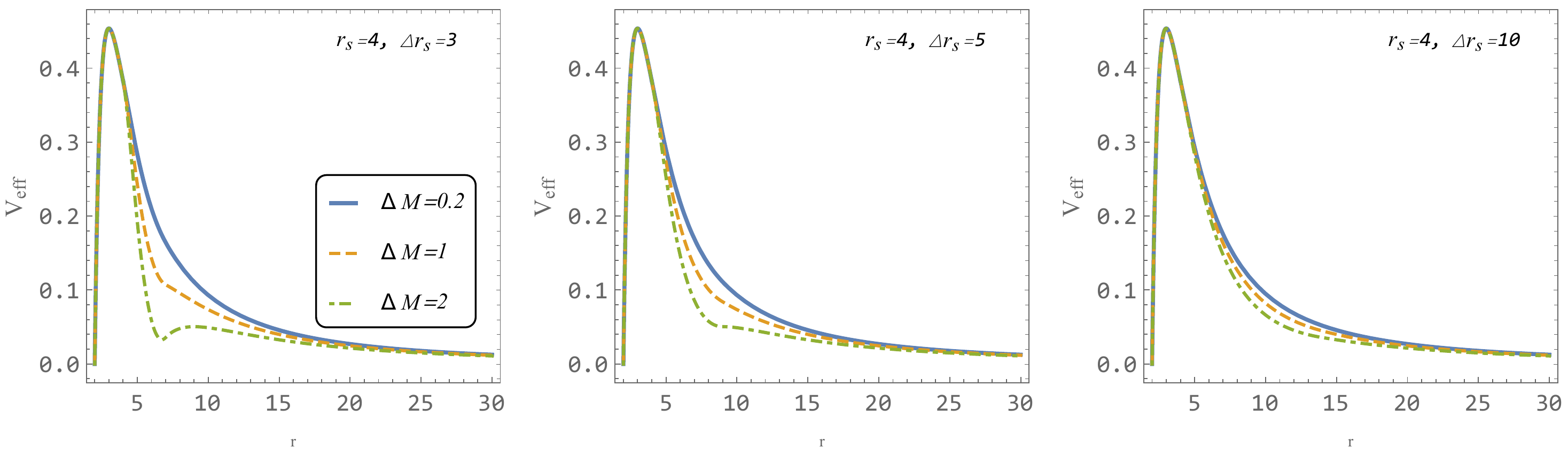}
	\caption{Effective potentials of massless test particles for various parameters, $\dm$, $\rs$, and $\drs$. Here $\rs=4$ while the range takes values $\drs=3,10$ and $20$.}
	\label{f:f3}
\end{figure*}

The motion of massless particles is independent of their energy, and for this reason it, can be fully determined by a single motion constant (see, e.g. \cite{Mis-Tho-Whe:1973:Gravitation:,Stu-Char-Sche:2018:EPJC:})-here we use the inverse impact parameter defined by the relation 
\begin{eqnarray}\label{e:const1}
b = \frac{E}{L} . 
\end{eqnarray}
The effective potential governing the motion of the massless particles can be then related to the inverse impact parameter instead of the energy and can be given in the form  
\begin{equation}
  V_\mathrm{eff}=\frac{f(r)}{2}\frac{1}{g_{\theta\theta}(r)}=\frac{\left[r-2m(r)\right] }{2 r^3}
\end{equation}
where the lapse function 
\begin{eqnarray}\label{f:lapse}
  f(r) = 1 - \frac{2m(r)}{r} 
\end{eqnarray}
contains the mass function $m(r)$ defined by Eq.(\ref{e:met}). \footnote{Clearly, the effective potential is well-defined only in the stationary parts of the BH+DM shell spacetime.} We provide some examples of the behavior of the photon effective potential in Figures \ref{f:f2} and \ref{f:f3} -- its behavior indicates an unexpected property, namely that even three circular null geodesics (photon orbits) could occur above the BH horizon. In order to keep the possibility of comparing our results to those of \cite{Konoplya:2019:PLB:}, we study the case of $\rs=2$ and $\rs=3$ along with the relevant one of $\rs=4$. 
The circular null geodesics are defined by the condition 

\begin{equation}
  \diff V_\mathrm{eff}/\diff r=-\frac{r \left[ m'(r)+1\right]- 3m(r)}{r^4} = 0 , 
\end{equation}
where 
\begin{eqnarray}
	m'(r)= dm(r)/dr = \frac{6 \dm (r-\rs) (\drs-r+\rs)}{\drs^3} \qquad\text{if } \rs<r<\rs+\drs, 
\end{eqnarray}
while $m'(r)=0$ if $2\leq r \leq \rs$ or $r>\rs+\drs$.

If the condition 
\begin{equation}
  \dm \geq \dm_\mathrm{ph(e)}(\rs,\drs) \equiv \frac{\rs-3+\drs}{3}
\end{equation}
is satisfied, a photon circular orbit at $r_\mathrm{ph(e)}=3(1+\dm)$ arises above the DM shell and two additional photon circular orbits exist under the outer edge of the DM shell. For $\rs\geq3$, the inner photon circular orbit is located at $r_\mathrm{ph(i)}=3$, while the mediate photon circular orbit is at $r_\mathrm{ph(m)}$ located inside the DM shell being implicitly determined by the function 
%
\begin{equation}
  \dm_\mathrm{ph}(r;\rs,\drs)=\frac{\drs^3 (r-3)}{3 (r-\rs) \left(\drs r-3 \drs \rs+2 r \rs-2 \rs^2\right)}.
  \label{e:dmph}
\end{equation}
For $\rs<3$, both $r_\mathrm{ph(i)}$ and $r_\mathrm{ph(m)}$ are governed by the function Equation (\ref{e:dmph}). The inner and outer circular null geodesics are unstable against radial perturbations, and the mediate orbit at $r_\mathrm{ph(m)}$ is stable against radial perturbations. For $M<\dm_\mathrm{ph(e)}$, only the unstable inner photon circular orbit survives. 

\begin{figure*} 
	\includegraphics[width=\linewidth]{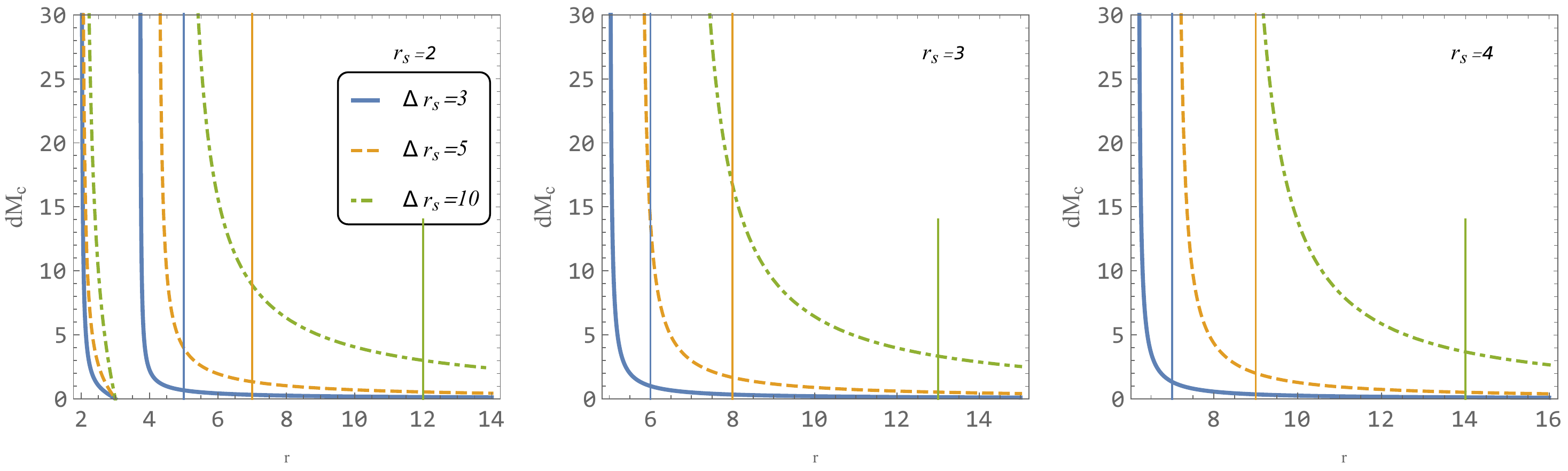}
	\caption{Radial profile of parameter $\dm$ representing the circular photon orbit for various parameters $\rs$ and $\drs$, where the vertical lines indicate the end of the DM halo $\rs+\drs$.}
	\label{f:f8A}
\end{figure*}

\begin{figure*} 
	\includegraphics[width=\linewidth]{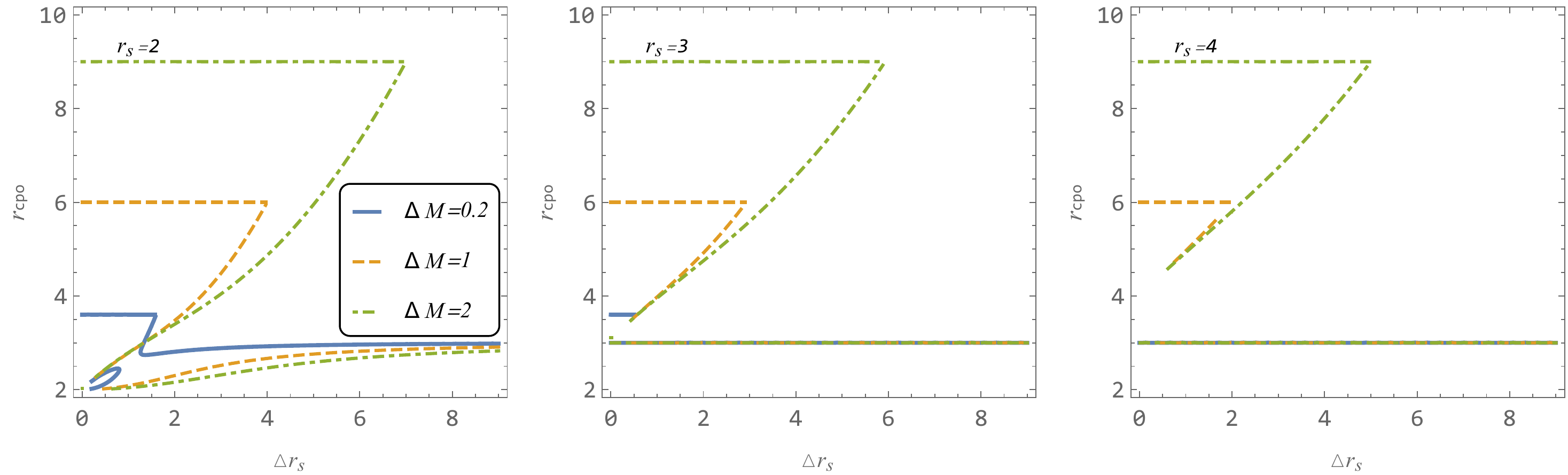}
	\caption{Radii of the dependence of the circular photon orbits in dependence on parameter $\drs$ for fixed values of the parameters $\dm$ and $\rs$.}
	\label{f:f4}
\end{figure*}

The zero-point of the characteristic function $\dm_\mathrm{ph}(r;\rs,\drs)$ is located as expected at $r=3$; it has two divergent points located at $r_{d1}=\rs$ and 
\begin{equation}
  r_{d2}=\rs\frac{3\drs+2\rs}{\drs+2\rs}. 
\end{equation}
Of course, in the case of $\rs\geq3$ only the second divergent point of the characteristic function is relevant. The behavior of the characteristic function of photon circular motion $\dm_\mathrm{ph}(r;\rs,\drs)$ is illustrated in Figure \ref{f:f8A}. We must recall that the photon circular orbits governed by the function $\dm_\mathrm{ph}(r;\rs,\drs)$ are relevant only in the interval $\{\rs,\rs+\drs\}$. We can see that three circular photon orbits exist if $\dm$ is sufficiently high where the critical point is governed by \\$\dm_\mathrm{ph}(r;\rs,\drs)$. The behavior of the location of the photon circular orbits is demonstrated for selected values of the spacetime parameters in Figure \ref{f:f4}. The parameters of
the impact of the circular photon orbits are determined by the
effective potential taken at the extremal point. 

\FloatBarrier

\FloatBarrier

\begin{figure*}[h]
  \centering
	\includegraphics[width=0.66\linewidth]{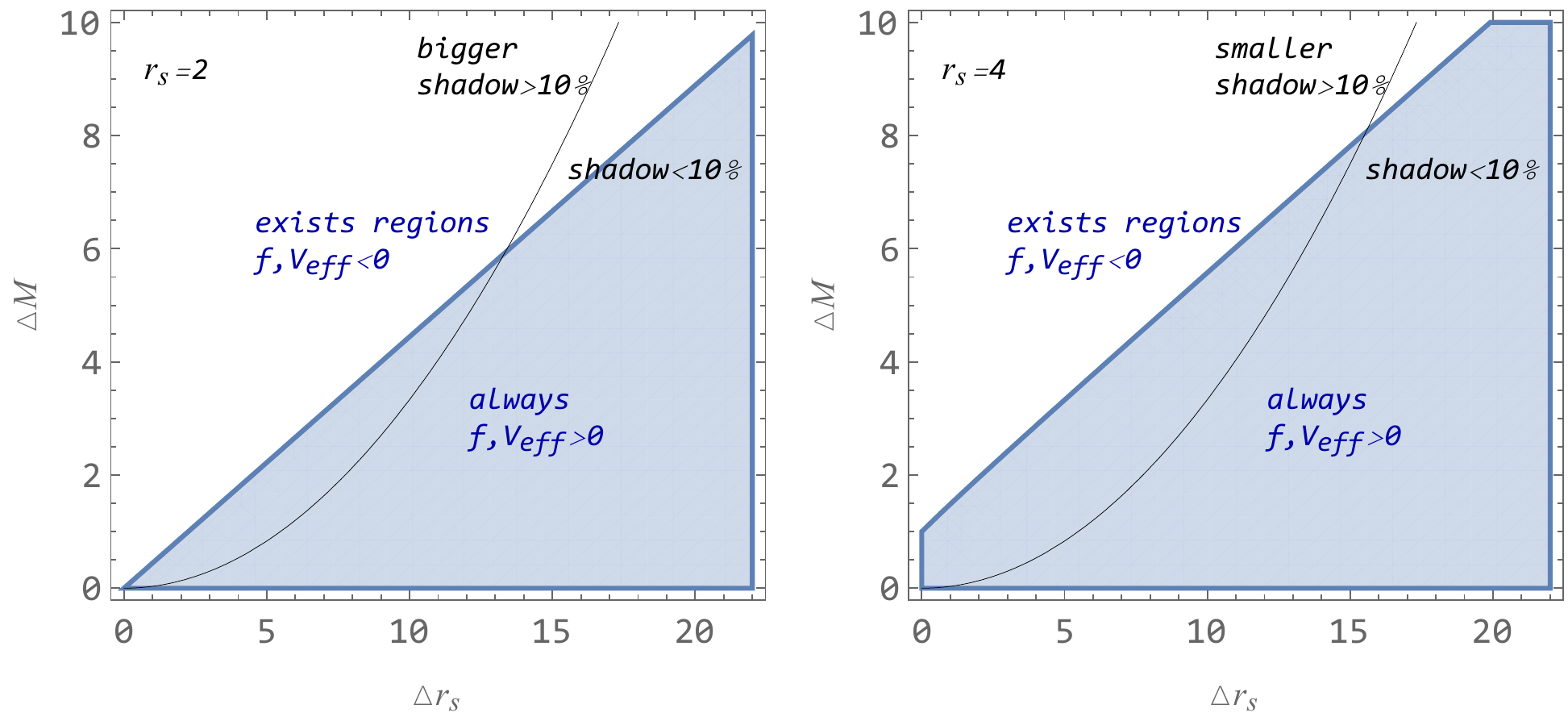}
	\caption{Plot of the Konoplya relation for the change in shadow due to the DM shell with the reality condition on the BH+DM shell spacetime.}
	\label{f:shad}
\end{figure*}

The circular null geodesics play a crucial role in establishing the BH shadow. In the case of the special metric investigated here, the influence of the DM shell on the BH shadow has been discussed in \cite{Konoplya:2019:PLB:}; therefore, we will not repeat the discussion. We only recall the main results of the study; namely, that for the relevant influence we must use the shell with  $\rs=2$, and the relevant influence of the DM shell of mass $\dm$ on the size of the BH shadow starts if an effective radius of the shell is reached that is given by the condition \citep{Konoplya:2019:PLB:} 
\begin{equation}
  \drs=\sqrt{3\dm}. 
		\label{e:shadow}
\end{equation}
These estimates have to be confronted with the reality conditions that imply a limit on the minimal extension of the DM shell $\Delta r_\mathrm{s crit}(\dm)$ as given in Figure \ref{f:shad}. We can thus see that the \textit{Konoplya curve} leaves the region for realistic spacetimes with a dependence on both $\drs$ and $\rs$. For $\rs=2$ considered in \cite{Konoplya:2019:PLB:}, the critical point is located at $\dm\dot=5.92$, $\drs\dot=13.33$.

\subsection{Motion of Massive Particles and Its Circular Geodesics}

In the case of the motion of a massive particle with rest mass $m_0$, it is convenient to introduce the specific energy and the specific axial angular momentum by the relations  
\begin{eqnarray}\label{e:const}
\mathcal{E} = \frac{E}{m_0}, \qquad
\cl = \frac{L}{m_0}.
\end{eqnarray}

\subsubsection{Effective potential}
Considering the motion of a test particles in the equatorial plane ($\theta=\pi/2$), the characteristics of the radial motion can be governed in the standard way \citep{Mis-Tho-Whe:1973:Gravitation:} by the effective potential related to the specific energy taken for the massive test particles of the form (recall assumed condition $M=1$)
\begin{equation}
  V_\mathrm{eff}(r;\rs,\drs,\dm,\cl)=f(r)\left[\frac{\cl^2}{g_{\theta\theta}(r)}+1\right]=\frac{\left[r-2m(r)\right] \left(\cl^2+ r^2\right)}{r^3}.
  \label{e:veff}
\end{equation}
%
In the following we focus attention only on the physically natural choice of the inner edge $\rs=4$, corresponding to the location of the marginally bound circular geodesic of the central Schwarzschild BH; the other parameters $\dm$ and $\drs$ are assumed in the region implying the static shell configurations outside the BH (see Figure \ref{f:real}).

The behavior of the effective potential (Equation \ref{e:veff}) is fully governed by the interplay of the parameters $\dm,  \drs$ and the specific angular momentum $\cl$. In fact, in the dependence on the relation of $\dm$ and $\drs$, several fundamentally different regimes of the behavior of the effective potential and Keplerian accretion are possible. This enables separation of the considered physically relevant BH + DM shell spacetimes into six classes demonstrating physically different behavior of circular geodesics and especially stable circular geodesics governing the behavior of Keplerian disks. Assuming a fixed value of $\drs$, there exist critical values of $\dm_\mathrm{(crit)}$ separating the BH+DM shell spacetimes into the six classes.

In order to obtain the required spacetime classification, we have to discuss two basic new phenomena that can occur in the considered BH + DM shell spacetimes. 

First, we discuss the possible existence of stable and unstable equilibrium positions of test particles (matter) -- such particles are at rest as related to the distant observers (and the central BHs) due to the balance of the gravitational influence of the BH and the DM shell. \footnote{We can assume an increasing concentration of matter descending to the stable equilibrium position inside the shell -- this process can represent consequent increase of shell mass on long timescales.} The equilibrium positions correspond to the local extrema of the effective potential with $\cl=0$. Thus, we must find the spacetimes allowing for the existence of the local extrema of the effective potential 
\begin{equation}
  V_\mathrm{eff}(r,\rs,\drs,\dm,\cl=0)=\frac{r-2m(r)}{r^3}.
\end{equation}
In order to find the case of limiting spacetimes, we must simultaneously solve the conditions of local extrema and the inflex point. Using the conditions $\dd V_\mathrm{eff}(\cl=0)/ \dd r=0$ and $\dd^2 V_\mathrm{eff}(\cl=0)/ \dd r^2=0$, we find for $\rs=4$ the critical value of the shell mass in its dependence on its extension $\drs$ in the form
\begin{equation}
  \dm_\mathrm{crit(\cl=0)}(\rs=4,\drs)=\frac{4\drs}{24+\drs} . 
\end{equation}
Then, for $\dm<\dm_\mathrm{crit(\cl=0)}$ there are no extrema of the effective potential (no equilibrium position in the spacetime), while for $\dm>\dm_\mathrm{crit(\cl=0)}$ there are two equilibrium positions; the inner one being unstable and the outer one being stable against radial perturbations. 
 
Second, in addition to the possible existence of two local extrema of the general effective potential in the vacuum Schwarzschild spacetimes, and the existence of its inflexion point for $\cl=\cl_\mathrm{ms}$, corresponding to the marginally stable circular orbit, in the BH+DM shell spacetime there can even exist four local extrema of the effective potential, demonstrating a possible complex relation to the marginally stable circular geodesics. This also implies a nonstandard structure of the Keplerian disks due to the possible existence of two separated families of the stable circular orbits. 

For values of $\cl$ large enough, the effective potential  $V_\mathrm{eff}(r)$ can demonstrate the standard shape with maximum corresponding to an unstable circular orbit with $\ce \gg 1$ and a minimum with $\ce<1$.
However, for $\cl=\cl_\mathrm{inf(i)}$ an inflexion point arises on the descending part of $V_\mathrm{eff}(r)$, and for $\cl<\cl_\mathrm{inf}$ additional maximum (external) and minimum (internal) of $V_\mathrm{eff}(r)$ arise. With the further decrease of $\cl$, an inflexion point arises for $\cl=\cl_\mathrm{inf(e)}$ where the external maximum coincides with the external minimum, thus giving the marginally stable circular orbit of the external family of stable circular orbits representing the standard Keplerian disk of the BH+DM shell configuration. However, the inner stable circular orbit survives at $\cl<\cl_\mathrm{inf(e)}$ down to $\cl=0$, or down to an inner inflexion point at $\cl=\cl_\mathrm{inf(i)}$. The sequence of the inner stable circular orbits represents a nonstandard Keplerian disk—the sequence is localized near the outer edge of the shell and finishes at the inner inflexion point, or in the stable equilibrium position thus giving a final state of matter from the nonstandard Keplerian disk. The creation of such a nonstandard Keplerian disk should be caused due to matter infalling from the inner edge of the standard outer Keplerian disk, if the effective potential corresponding to $\cl=\cl_\mathrm{inf(e)}$ generates a barrier forbidding a fall into the central BH. 

In order to close the classification of the BH+DM shell spacetimes, we have to introduce critical values of the mass parameter related to the possible existence of the marginally stable, marginally bound, or circular photon orbit at the exterior of the shell. In the case of marginally stable circular geodesics, we thus have to look for the condition satisfying the relation $r_\mathrm{ms(e)}=6(1+\dm)\geq \rs +\drs$. This condition reads (for $\rs=4$)
\begin{equation}
  \dm\geq\dm_\mathrm{ms(e)}(\rs=4,\drs)\equiv \frac{\drs-2}{6}.
\end{equation}
The condition for the location of the marginally bound circular geodesic located above the shell reads as
\begin{equation}
  \dm\geq\dm_\mathrm{mb(e)}(\rs=4,\drs)\equiv \frac{\drs}{4},
\end{equation}
while in the case of the unstable photon circular geodesic the condition reads as
\begin{equation}
  \dm\geq\dm_\mathrm{ph(e)}(\rs=4,\drs)\equiv \frac{1+\drs}{3} . 
\end{equation}

\begin{figure*} 
  \centering
	\includegraphics[width=0.7\linewidth]{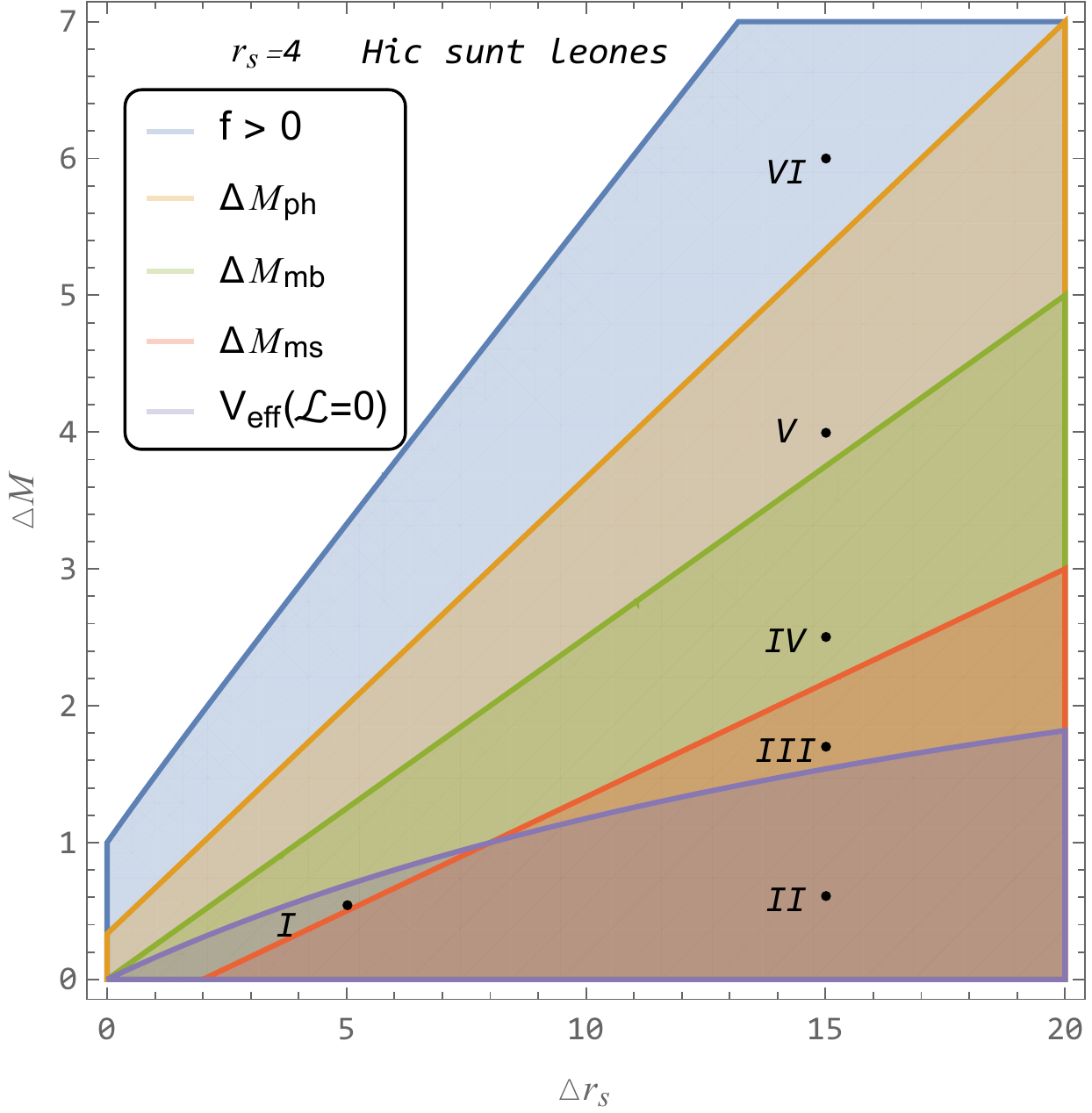}
	\caption{Classification of the BH+DM shell spacetimes based on the properties of circular geodesics (Keplerian disks) is represented for fixed parameter $\rs=4$. The behavior of the effective potential is for the given six classes is demonstrated in Figure \ref{f:veffcl}}
	\label{f:class}
\end{figure*}

\begin{figure*} 
	\includegraphics[width=\linewidth]{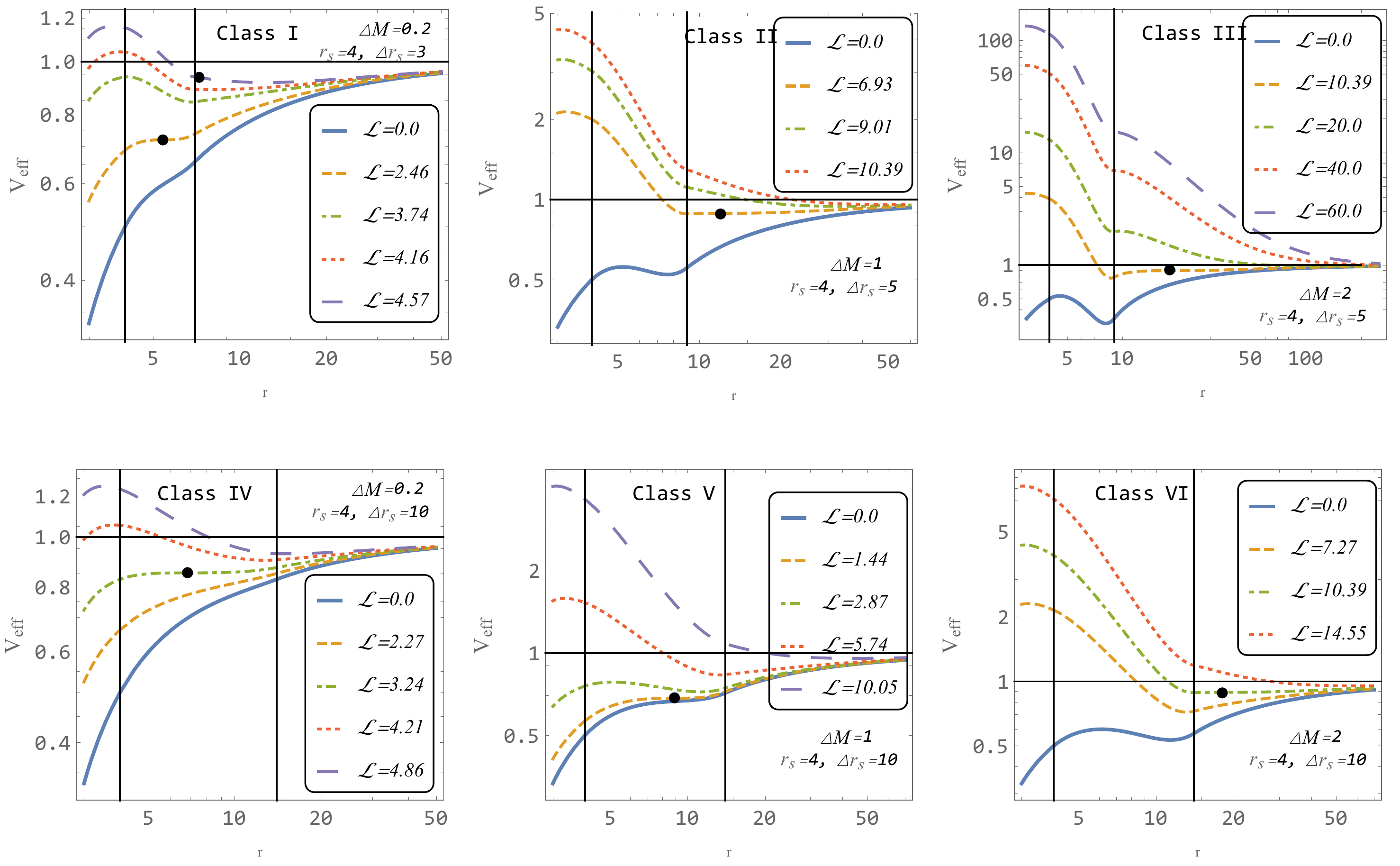}
	\caption{Effective potentials demonstrating properties of the circular geodesics for each of the six introduced classes of the BH+DM shell spacetimes as distributed in Figure \ref{f:class}.}
	\label{f:veffcl}
\end{figure*}

Using the above presented relations, we can separate the parameter space of the BH+DM shell spacetime into six classes as demonstrated in Figure \ref{f:class}. The behavior of the effective potential in these six classes is determined in Figure \ref{f:veffcl} for a representative selection of the spacetime parameters $\dm$, $\drs$ and $\rs=4$. Note that in the spacetimes of the class VI, with $\dm>\dm_\mathrm{ph(e)}(\rs=4,\drs)$, the minimum of the effective potential (stable circular orbits) exists for $\cl\rightarrow \infty$ and in this limit it corresponds to the stable circular photon orbit.


Properties of the circular geodesics, their stability, and related properties of Keplerian disks can be for the six classes of the BH+DM shell spacetimes can be given in the following summarized way.

\textbf{Class I:} Two Keplerian disks (sequences of stable circular orbits) exist. The outer one finishes at an outer marginally stable orbit $r_\mathrm{ms(e)}>\rs+\drs$ at an inflexion point (the infalling matter could be trapped if a maximum of the effective potential related to the corresponding $\cl$ has energy larger than those corresponding to the inflexion point), the inner disk finishes at the inner marginally stable orbit $r_\mathrm{ms(i)}\in (\rs,\rs+\drs)$ and matter can fall into the BH.

\textbf{Class II:} There is one Keplerian disk with the inner edge at the marginally stable circular geodesics located inside the shell. The matter falls from the disk into the BH.

\textbf{Class III:} There is one Keplerian disk finishing at the stable equilibrium radius where the matter of the accretion disk concentrates. No matter falls into the BH.

\textbf{Class IV:} Two Keplerian disks exist. The external Keplerian disk finishes at an inflex point of the effective potential whose maximum can cause trapping of the matter inside the DM shell possibly causing the possibly creation of the inner Keplerian disk that finishes in the stable equilibrium radius where the accreting matter is concentrated. No matter falls into the BH.

\textbf{Class V:} Two Keplerian disks exist with the same properties as in class IV spacetimes. In addition, a marginally bound circular orbit can exist outside the shell.

\textbf{Class VI:} Two Keplerian disks exist with properties related to accretion of the same characteristics as in classes IV and V. However, the inner Keplerian disk can exist for values of specific energy and specific angular momentum unlimited from above (approaching the stable circular photon orbit).

Note that in relation to natural conditions for accretion from large distances (as related to observed QPOs), the last three classes can be considered equivalent.

\FloatBarrier

\subsubsection{Circular orbits}
The circular geodesics are determined by the extrema of the effective potential, i.e., by the condition
\begin{equation}
	\frac{\diff V_{\mathrm{eff}}}{\diff r}=0,
\end{equation}
where
\begin{equation}
  \diff V_\mathrm{eff}/\diff r=\frac{r \left[\left(\cl^2+ r^2\right) m'(r)+\cl^2\right]-m(r) \left(r^2+3 \cl^2\right)}{r^4}. 
\end{equation}
This implies the radial profile of the particle-specific angular momentum $\cl$ in the form
\begin{equation}
  \cl_\mathrm{c}=\frac{r\sqrt{m(r)-r\, m'(r)}}{\sqrt{r+r\, m'(r)-3 m(r)}}.
  \label{e:lc}
\end{equation}
The radial profile of the specific energy then takes the form 
\begin{equation}
  \ce_\mathrm{c}=\frac{r - 2m(r)}{\sqrt{r[{r+r\, m'(r)-3 m(r)]}}}.
  \label{e:Ec}
\end{equation}
The angular (Keplerian) frequency of the orbital motion as related to distant static observers $\Omega_\mathrm{K}=\dd\phi/\dd t$ is given by the relation 
\begin{eqnarray}
	\Omega_\mathrm{K} = \frac{-g_{tt}\cl_c}{r^2\mathcal{E}_c}
\end{eqnarray}
Clearly, the radial profiles of the specific angular momentum and the specific energy diverge at radii corresponding to the circular photon orbits. Note that the specific angular momentum $\cl_{c}$ can be taken with both $+/-$ signs due to two possible (equivalent in the spherically symmetric spacetimes) orientations of the circular motion, while the specific energy $\ce_{c}$ has to be taken only with the positive sign, if we consider the particles in positive root (future-directed) states (for details see \cite{Mis-Tho-Whe:1973:Gravitation:,Bic-Stu-Bal:1989:BAC:}). 

\begin{figure*}[h]
	\includegraphics[width=\linewidth]{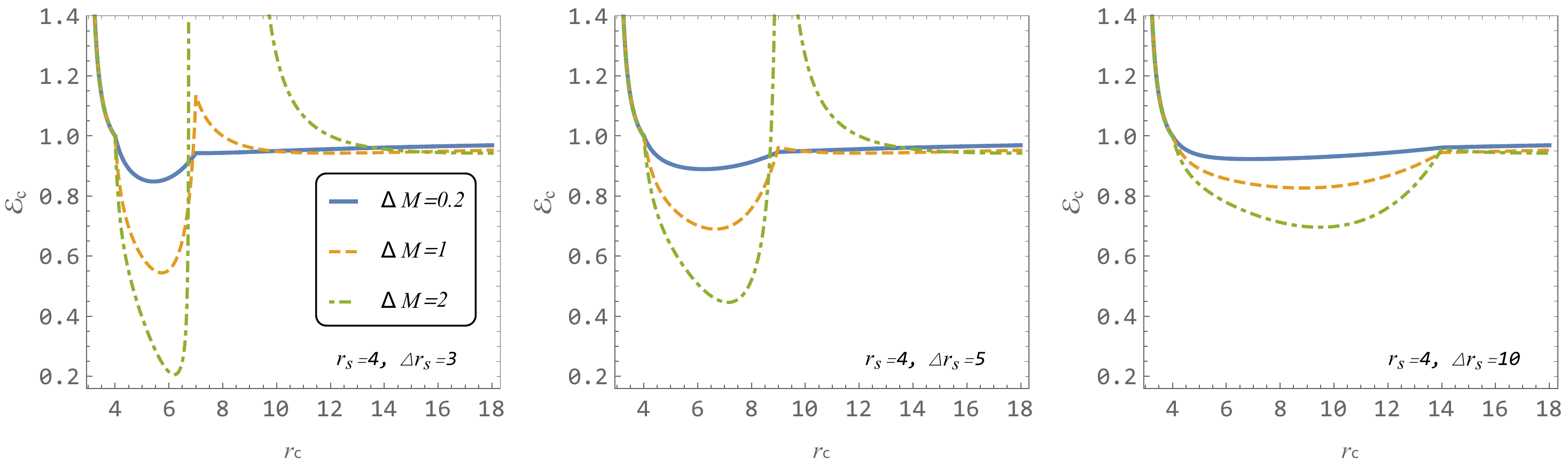}
	\caption{Radial profile of the specific energy of the circular orbits for fixed $\rs=4$ and various typical values of the parameters $\dm$ and $\drs$.}
	\label{f:f8}
\end{figure*}

\begin{figure*}[h]
	\includegraphics[width=\linewidth]{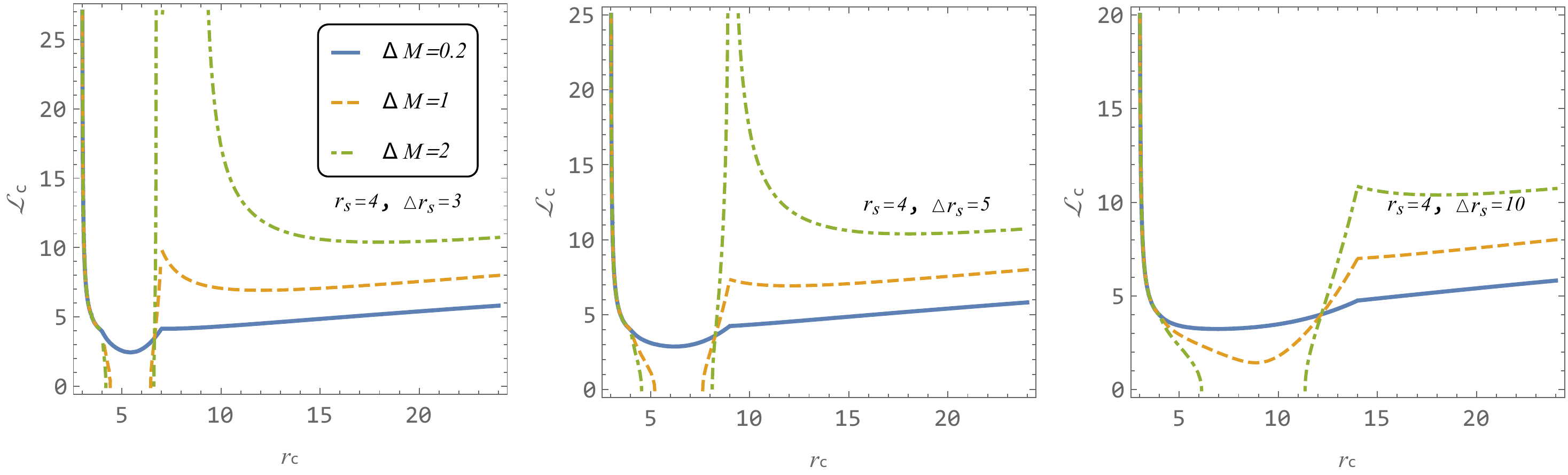}
	\caption{Radial profile of the specific angular momentum of the circular orbits for typical values of the parameters $\dm$, $\drs$ and fixed $\rs=4$.}
	\label{f:f9}
\end{figure*}

\begin{figure*}[h]
	\includegraphics[width=\linewidth]{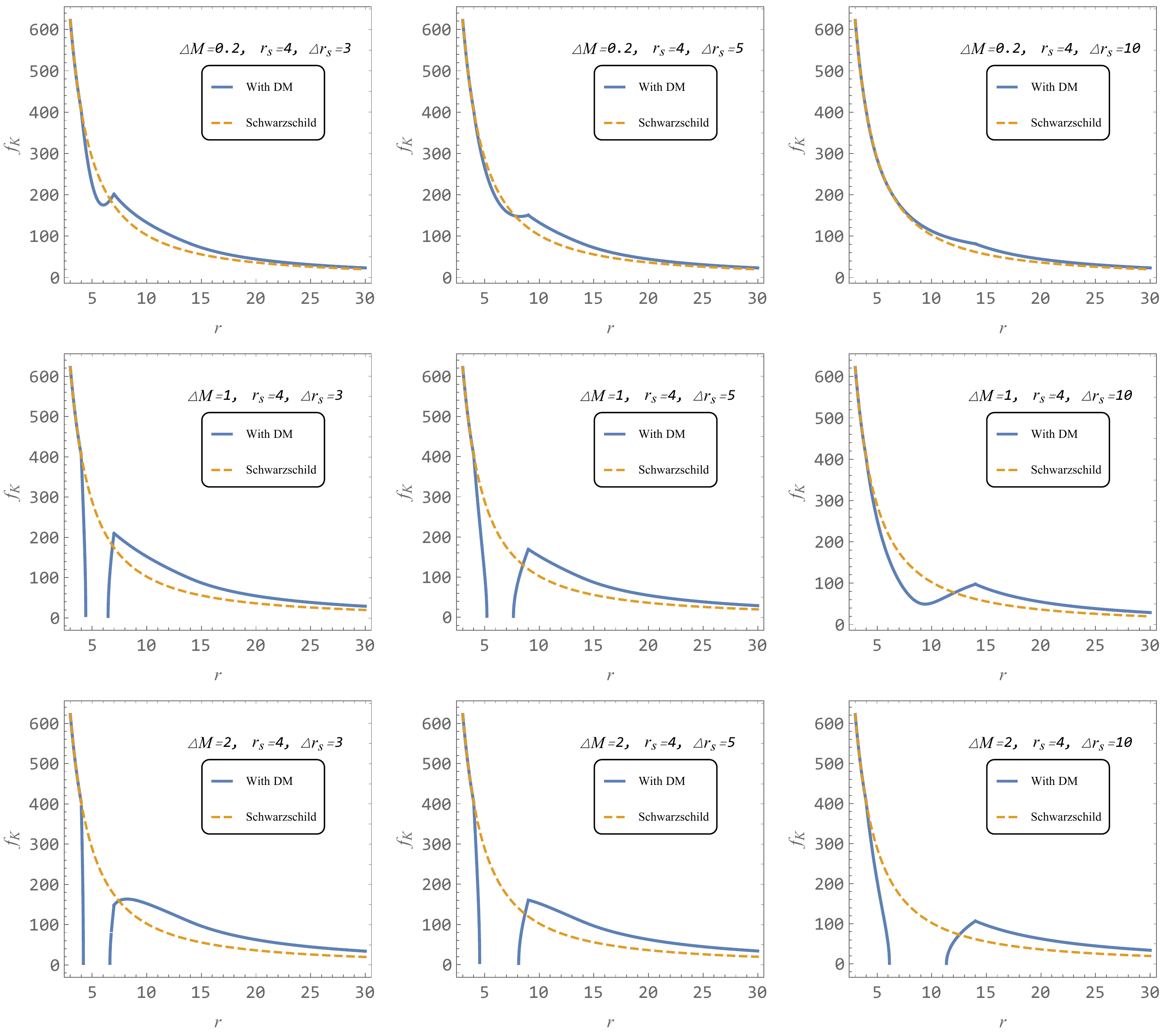}
	\caption{Radial profile of the Keplerian frequency for typical values of the parameters $\dm$, $\drs$ and fixed $\rs=4$. The frequency is given in SI units (hertz) under the assumption of BH mass $M=10M_{\odot}$.}
	\label{f:f9B}
\end{figure*}

The radial profiles of the specific energy, the specific angular momentum, and the Keplerian angular frequency, of the circular geodesics are for fixed $\rs=4$ and  typical values of $\drs$ and $\dm$ are illustrated in Figures \ref{f:f8} - \ref{f:f9B}. Notice increasing binding (decrease of $\ce_c$) of the circular orbits inside the shell with increasing compactness ($\dm/\drs$) of the shell, and vanishing of $\cl$ and $\Omega_K$ at the equilibrium positions. 
We have to stress that the increasing parts of the $cl_\mathrm{c}$ and $ce_\mathrm{c}$ radial profiles correspond to the stable circular geodesics, while the decreasing parts correspond to the unstable circular geodesics. 

Finally, we determine the critical radii of the circular geodesics, namely, for the marginally bound and marginally stable orbits; the limiting photon circular geodesics were discussed above.
\FloatBarrier

\subsection{Marginally bound orbits}

It is instructive to introduce also the notion of the marginally bound orbits that are in the asymptotically flat spacetimes, as is also our case, defined by the simple condition $\ce_c = 1$ governing the possibility to marginally reach infinitely distant regions of the spacetime. However, in the some classes of the BH+DM shell spacetimes the situation is more complex in comparison with the standard vacuum Schwarzschild BH spacetimes, as the condition $\ce_c = 1$ can give incomplete and even false information because there are bound circular orbits with specific energy $\ce_\mathrm{c}>1$ that can occur above the inner inflexion point of the effective potential, or there could exist a stable circular orbit with $\ce_\mathrm{c}=1$, which are not marginally bound. 

We can summarize the results for the case of the BH+DM shell spacetimes with $\rs=4$ in the following way: there is a inner marginally bound orbit at $r_\mathrm{mb(i)}=\rs=4$. There is a outer marginally bound orbit above the shell at $r_\mathrm{mb(0)}=4(1+\dm)$ if $\drs<4\dm$. Inside the shell, at $4<r<4+\drs$, the marginally bound orbit is located at a radius implicitly given by an appropriately defined mass function, as in the case of the photon circular orbits. The standard condition for the existence of the marginally bound obits can be expressed as follows:
\begin{equation}
	\mathcal{E}_c= 
	\begin{cases}
		\frac{r-2}{ \sqrt{r(r-3)}}=1, \quad r \le \rs\\
		\frac{-6 \dm \drs (r-\rs)^2+4 \dm (r-\rs)^3+\drs^3 (r-2)}{\drs^{3/2} \sqrt{r \left(-3 \dm \drs (r-3 \rs) (r-\rs)-6 \dm \rs (r-\rs)^2+\drs^3 (r-3)\right)}}=1,\quad \rs\leq r \leq \rs + \drs ,\\
		\frac{r-2-2 \dm}{\sqrt{r(r-3 \dm-3)} }=1,\quad r > \rs + \drs. 
	\end{cases}
	\label{e:m1}
\end{equation}
The critical case of the $\ce=1$ inside the DM shell at $4<r<4+\drs$ implies the definition of the mass function that implicitly governs the position of the orbits with $\ce=1$ located inside the shell. The mass function can be expressed in the following way:
\begin{equation}
	\dm_\mathrm{mb}= 
	\begin{cases}
		- ,\quad r \leq \rs\\
		\frac{\drs^3}{8 (r-\rs)^3 (3 \drs-2 r+2 \rs)^2}\left( h_1 + r\sqrt{h_2} \right),\quad \rs\leq r \leq \rs + \drs ,\\
		\frac{r^{3/2}-r-3 \sqrt{r}+2}{3 \sqrt{r}-2},\quad r > \rs + \drs, 
	\end{cases}
	\label{e:mb}
\end{equation}
where
\begin{eqnarray}
   h_1=&&3 \drs \left(3 r^2-r (\rs+8)+8 \rs\right)-2 (r-\rs) \left(4 r^2-r (\rs+8)+8 \rs\right),\\ \nonumber
   h_2=&&9 \drs^2 \left(9 r^2-2 r (3 \rs+16)+\rs (\rs+32)\right)-12 \drs (r-\rs) \left(12 r^2-r (7 \rs+40)+\rs (\rs+40)\right)+\\ \nonumber 
   &&4 (r-\rs)^2 \left(16 r^2-8 r (\rs+6)+\rs (\rs+48)\right) . 
\end{eqnarray}

\begin{figure} 
  \centering
	\includegraphics[width=0.35\linewidth]{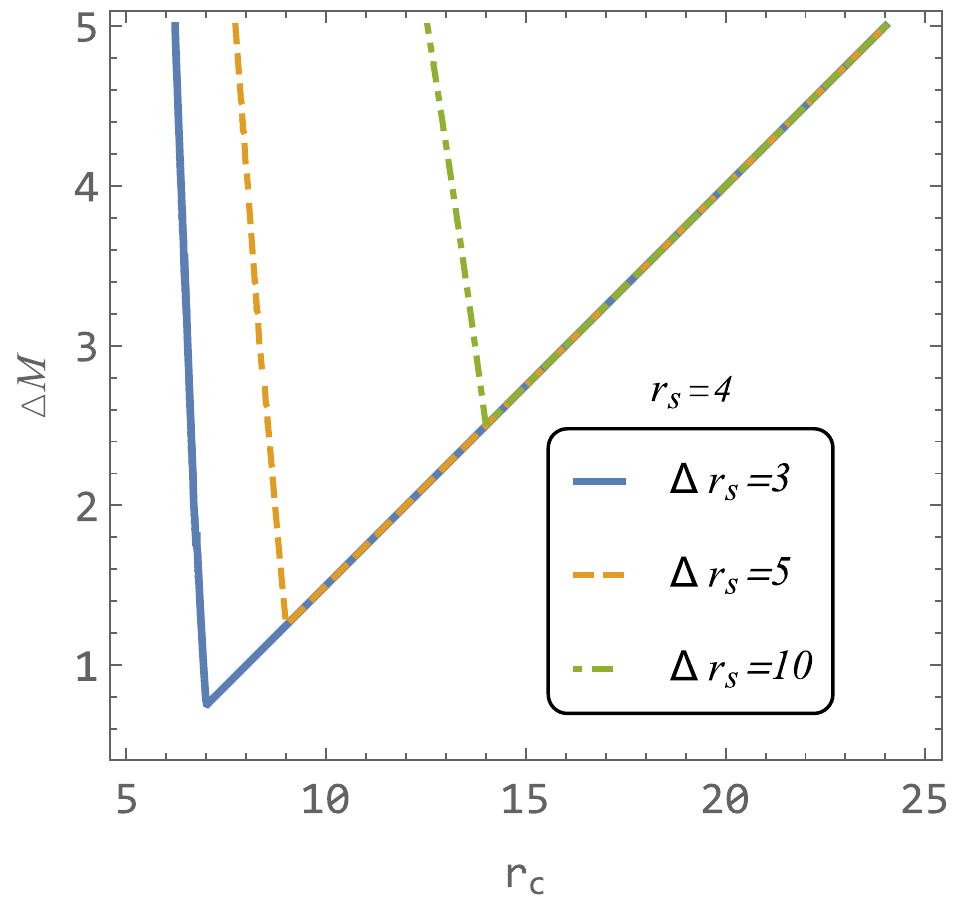}
	\caption{Function of $\dm_\mathrm{mb}(r;\drs,\rs)$ for $\drs=\{3,5,10\}$ and fixed $\rs=4$. Note that only the increasing (right) part of the mass function is relevant for the location of the marginally bound orbits. There is an additional section of the marginally bound orbits related to the outer maximum of the effective potential being limited by the inflexion point of the effective potential.}
	\label{f:mb}
\end{figure}

Clearly, the solution $\dm=0$ at $r=4$ corresponds to the vacuum Schwarzschild spacetime. The general behavior of the mass function related to the marginally bound orbits is strongly dependent on the influence of the DM shell on the spacetime structure and the related effective potential of the geodesic motion. The radii of the marginally bound circular orbits are given in Figure \ref{f:mb}.
We have to stress that the decreasing (left) part of the mass function $\dm_\mathrm{mb}(r;\drs,\rs)$ corresponds to the stable circular orbits with $\ce=1$, thus being irrelevant, while the increasing (right) part is the relevant one, as it corresponds to the unstable circular orbits with $\ce=1$. The other family of marginally bound orbits is related to the existence of the outer inflexion point of the effective potential being identical to the marginally stable circular orbits. However, the marginally bound orbits of this kind must fulfill the condition $\ce>1$. 

\subsection{Marginally stable circular orbits} 

Detection of the twin HF QPOs is possible only in the spacetime regions where stable circular geodesics are allowed. Therefore, we have to find the marginal radii limiting the existence of stable circular geodesics that are governed by the inflexion points of the effective potential, or by the stable equilibrium positions represented by the minimum of $V_\mathrm{eff}(\cl=0)$, if it exists. We have to keep in mind that there are six classes of the BH+DM shell spacetimes - in some of them, two regions of stable circular orbits (Keplerian disks) can exist, in some of them only one such region exists, while their inner edge can be determined by both inflexion radii or stable equilibrium radii. The outer region of stable orbits contains the static radii (equilibrium points) where particles static relative to distant observers (with $\cl=0$) can exist. The situation can be summarized in the following way.

\textit{Class I:} Inner radius of the outer Keplerian disk at $r_\mathrm{ms(0)}=6(1+\dm)$, and inner radius of the inner disk located inside the shell at $r_\mathrm{ms(i)}$ determined below.

\textit{Class II:} Inner radius of the Keplerian disk located inside the shell at $r_\mathrm{ms(i)}$

\textit{Class III:} Inner radius of the Keplerian disk located at the stable equilibrium position $r_\mathrm{e(o)}$ given above.

\textit{Class IV-VI:} The outer Keplerian disk has the inner edge at $r_\mathrm{ms(i)}$ inside the shell, while the inner disk has the inner edge at $r_\mathrm{e(o)}$.

The marginally stable orbits (MSO) located inside the shell are determined by the additional (to $\dd V_\mathrm{eff}/\dd r=0$ condition 
\begin{equation}
\diff^2 V_\mathrm{eff}(r_\mathrm{c})/\diff r^2 = 0,
\end{equation}
where
\begin{eqnarray}
   \frac{\diff^2 V_\mathrm{eff}}{\diff r^2} = \frac{r\, h(r)+2 m(r) \left(6 \cl^2+ r^2\right)}{r^5},
\end{eqnarray}
and
\begin{equation}
  h(r) = \left[r\ m''(r) \left(\cl^2+ r^2\right)-2 m'(r)\left(r^2+3 \cl^2\right)-3 \cl^2\right],
\end{equation}
with 
\begin{eqnarray}
	m''(r)=\frac{6 \Delta M (\Delta \rs -2 r+2 \rs)}{\Delta \rs^3} \qquad\text{if } \rs<r<\rs+\Delta \rs.
\end{eqnarray}

The stable equilibrium point (minimum of the effective potential $V_\mathrm{eff}(r,\cl=0)$) can be considered as the innermost stable circular orbit, with vanishing orbital velocity or frequency, representing the inner edge of the non-standard Keplerian disk; of course, this point is not marginally stable.

\begin{figure*} 
    \centering
	\includegraphics[width=0.34\linewidth]{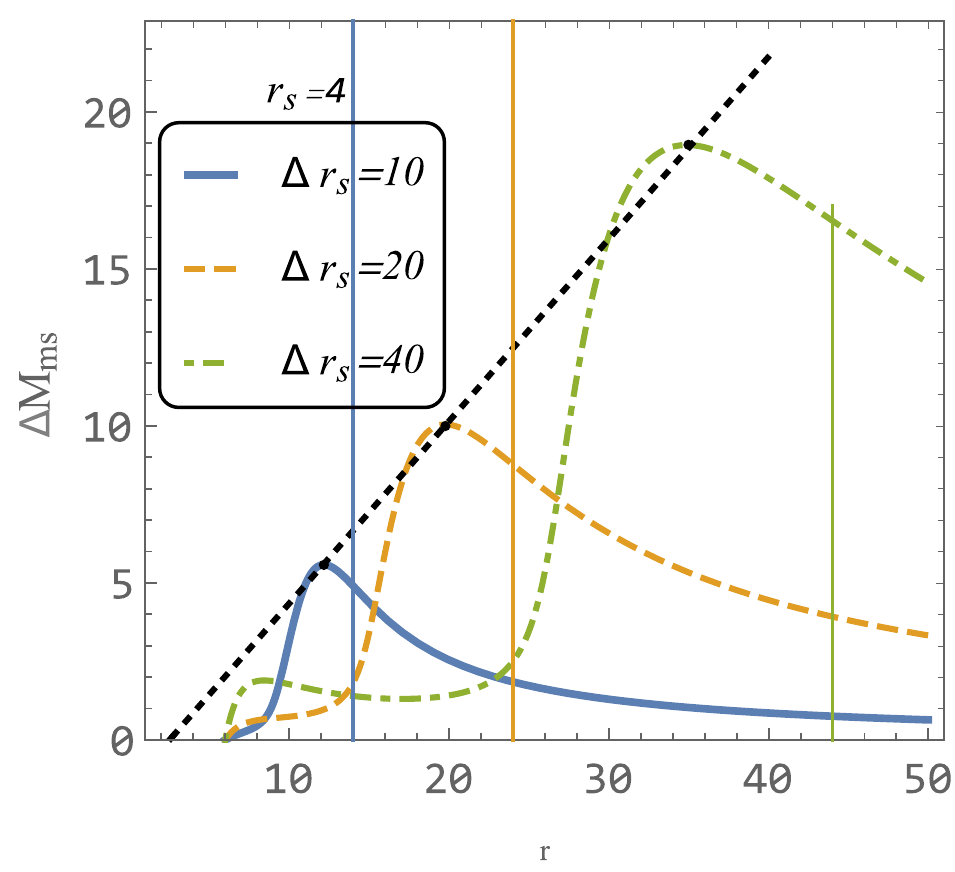}
	\caption{Radial profile of the mass function $\dm_\mathrm{MS0}(r;\rs,\drs)$ giving the MSO orbits for $\rs=4$ and typical values of the parameter $\drs$.}
	\label{f:f9A}
\end{figure*}

\begin{figure} 
  \centering
	\includegraphics[width=0.35\linewidth]{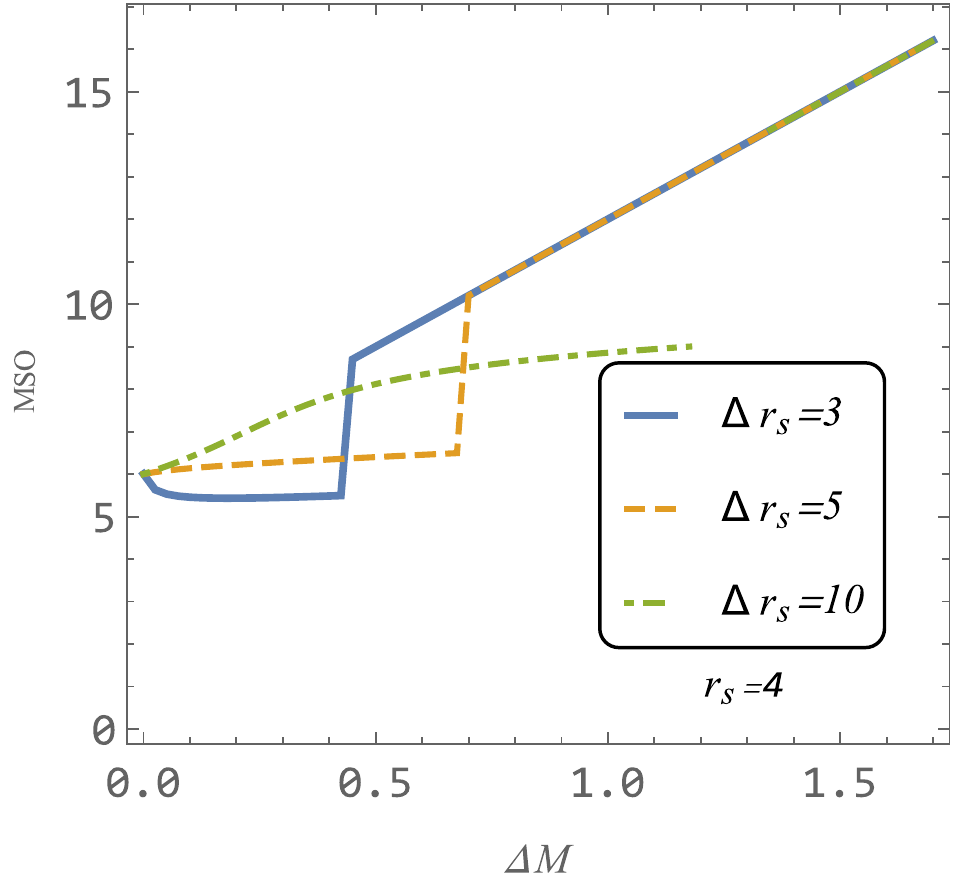}
	\caption{Position of the MSO given in the dependence on the mass parameter $\dm$ for fixed $\rs=4$ and various values of the extension parameter $\drs$.}
	\label{f:f10}
\end{figure}

\begin{figure} 
  \centering
	\includegraphics[width=0.35\linewidth]{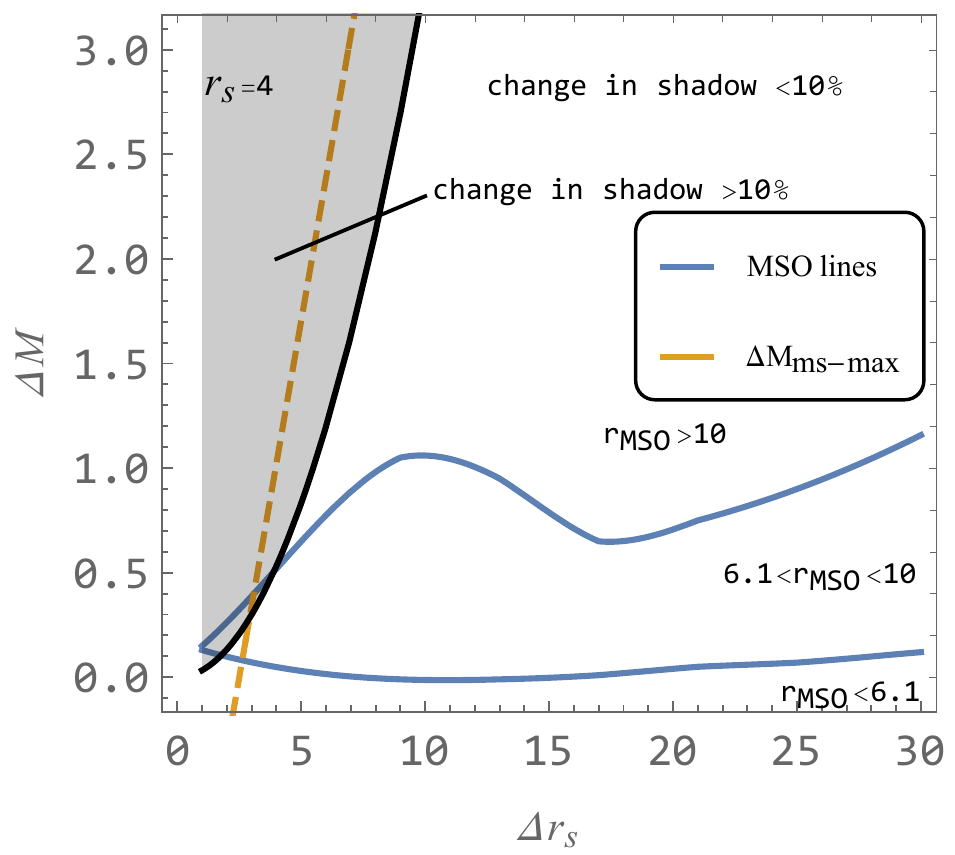}
	\caption{The $\dm$ - $\drs$ space divided (by two blue curves) into three areas according to the MSO location, for the fixed parameter $\rs = 4$. MSO is at a distance greater than $10\,M$ for parameters in the upper region. MSO is below $6.1\,M$ in the bottom area. The region allowing for the existence of marginally stable orbits is located under the orange line. The black curve represents the Konoplya curve separating the relevance of the influence on the BH shadow.}
	\label{f:f11}
\end{figure}

In the region related to the DM shell, $\rs<r<\rs+\drs$, position of the innermost stable circular orbit can be, as in the case of photon circular orbits and marginally bound circular orbits, conveniently expressed implicitly in terms of a characteristic function related to the parameter $\dm$. The characteristic mass function can be now expressed in the form  
\begin{eqnarray}
&&\dm_\mathrm{MS0}(r;\rs,\drs)=\\
&&\frac{2 \drs^3 (r-6)}{r\sqrt{k_\mathrm{s}}+3 \drs \left[3 (r-4) r^2-(r-12) \rs^2-4 \rs r\right]-16 r^4+6 r^3 (3 \rs+10)-72 r^2 \rs-2 r \rs^2 (\rs+6)+24 \rs^3}\nonumber
\end{eqnarray}
with 
\begin{eqnarray}
&&k_\mathrm{s}=9 \drs^2 \left(9 r^4-80 r^3-6 r^2 (\rs-8) (\rs+4)+48 r (\rs-4) \rs+(\rs-4)^2 \rs^2\right)+\\ \nonumber
&&12 \drs \{-24 r^5+9 r^4 (3 \rs+22)+8 r^3 \left[(\rs-34) \rs-54\right]-12 r^2 \rs \left(\rs^2+\rs-70\right)+48 r \rs^2 (2 \rs-9)\\ \nonumber
&&\qquad+(\rs-6) (\rs-4) \rs^3\} +\\ \nonumber
&&4 (r-\rs)^2 \left(64 r^4-16 r^3 (\rs+30)+3 r^2 (\rs (76-5 \rs)+300)+2 r (\rs-6) \rs (\rs+66)+(\rs-6)^2 \rs^2\right).
\end{eqnarray}
We have to stress that only the BH+DM shell spacetimes with well-defined function $\dm_\mathrm{MS0}(r;\rs,\drs)$ are appropriate for our consideration, as only in those spacetimes the stable circular geodesics allowing for existence of HF QPOs are possible. 
Clearly, the zero-point of this characteristic function is located at $r=6$ corresponding to the MSO of the Schwarzschild geometry. However, even this simplified expression is too complex to realize a qualitative discussion of its properties. We thus represent its behavior graphically only. In all the considered cases, the function $\dm_\mathrm{MS0}(r;\rs,\drs)$ demonstrates a maximum giving limit on the existence of the relevant spacetimes. We give the limiting maximal values of the mass function $\dm_\mathrm{MS0}(r;\rs,\drs)$ in terms of the extension parameter of the DM shell. 
The behavior of the characteristic mass function \\$\dm_\mathrm{MS0}(r;\rs=4,\drs)$ is illustrated for the typical cases in Figure \ref{f:f9A}. We stress again that the validity of the characteristic function is relevant only in the region $\rs<r<\rs+\drs$. The position of the MSO is given in the dependence on the mass parameter $\dm$ for typical values of $\drs$ and fixed $\rs=4$ in Figure \ref{f:f10}. The radial profiles are relevant in the region corresponding to the influence of the DM shell; for the fixed $\rs=4$, the maximum of the function $\dm_\mathrm{MSO}(r;\rs,\drs)$ increases nearly linearly with increasing $\drs$, as shown in Figure \ref{f:f9A}, thus giving the dependence of the mass of the shell allowing for the existence of stable circular geodesics on the other two parameters of the shell. Using the maxima of the curves $\dm_\mathrm{MS0}(r;\rs=4,\drs)$ in Figure \ref{f:f9A}, we can separate the $\dm - \drs$ parameter space in the region where the inflexion points of the effective potential and the MSO orbits are allowed, as shown in Figure \ref{f:f11}. 

\FloatBarrier

It is necessary to emphasize that with an appropriate choice of the parameters, stable and unstable circular orbits are pushed to a large distance from the central object. This phenomenon occurs when there is a large accumulation of matter under a given orbit, e.g., MSO located in the Schwarzschild spacetime at $r = 6M$, is shifted to $r = 6 (M + \dm)$ in the BH+DM shell spacetime. However, this phenomenon is not supported yet by observations of astrophysical objects, and the nearest orbits are often measured around $6M$ (of course, exceptional cases are not observationally fully excluded also due to the possible role of electromagnetic interaction \citep{Stu-etal:2020:Uni:}). Figure \ref{f:f11} shows in the $\dm - \drs$ parameter space the regions of these parameters that allow the MSO position sufficiently close to the central object, in contrast to regions where the MSO is significantly shifted to large values. 
For completeness, we also include at this place also the distribution of the regions allowing a substantial influence on the size of the BH shadow discussed in \cite{Konoplya:2019:PLB:}. The restriction on the parameters $\drs$ and $\dm$ allowing for significant modifications of the BH shadow, introduced in \cite{Konoplya:2019:PLB:} and given by Eq. \ref{e:shadow} enables us to make a comparison with the shifting of the MSO in such spacetimes. The resulting comparison of the shadow limits and the restrictions on the MSO position is shown in Figure \ref{f:f11}. We can see that the regions of the significant shadow modifications are entering the region of a substantial shift of the MSO positions. Therefore, we also have to seriously consider spacetimes with such choice of the parameters seriously also in our study of the HF QPOs. \footnote{We are convinced ourselves that the profiles related to various choices of the inner edge of the DM shell, $\rs=2,3,4$, imply similar radial profiles of the mass function $\dm_\mathrm{MSO}(r;\rs,\drs)$ constructed for typical values of the extension parameter $\drs$.} 
\FloatBarrier


\section{Epicyclic orbital motion and its frequencies}
The HF QPOs observed in the microquasars (see \cite{McCli-Rem:2006:BHbinaries:}, or in active galactic nuclei, see \cite{Smi-Tan-Wag:2021:ApJ:} for an overview of the observed cases), where the central object is usually assumed to be a stellar-mass BH in the first case, and a supermassive BH in the second, give one of the most efficient ways to test gravity in the strong-field limit, even in its extreme form being concentrated on the general relativity models, or on alternative gravity theories \citep{Bambi:2018:ADP:}. Of course, strong magnetic fields could also play an extremely important role, as demonstrated recently in \cite{Stu-Kol:2016:EPJC:,Kol-Tur-Stu:2017:EPJC:,Stu-etal:2020:Uni:,Tur-etal:2020:ApJ:,Tur-Zaj-etal:2020:ApJ:}. On the other hand, the possible role of some alternative objects to BHs is also intensively studied, e.g., for naked singularities and super-spinars \citep{Stu-Sche:2012:CLAQG:,Stu-Sche:2013:CLAQG:}, or wormholes \citep{Stu-Vrb:2021:Universe:,Stu-Vrb:2021:JCAP:}. Here, we focus attention on the possible role of DM concentrated around an accreting BH--assuming a vacuum Schwarzschild BH spacetime modified by a simple but robust model of the DM shell influencing the accreting matter only gravitationally. 

We present a detailed investigation of the standard geodesic model in its important variants, which were studied in the case of microquasars as being relevant for explanation of the observed HF QPOs \citep{Stu-Kot-Tor:2013:ASTRA:,Stu-Kol:2016:ASTRA:}. We have chosen these variants of the geodesic model to reflect in the simplest way the role of the DM shell in the behavior of the epicyclic motion and the fitting of observational data from microquasars and active galactic nuclei. We are then able to give rough limits on the amount of DM concentrated around the assumed central BH in active galactic nuclei, or put restrictions on the amount of DM around the BHs in binary systems of microquasars. 

If a particle moving along a stable circular orbit at $r_c$ in the equatorial plane $\theta=\pi/2$ is slightly perturbed, it starts epicyclic oscillations. The displacement for small perturbation in radial and latitudinal directions can be defined as $r=r_\mathrm{c}+\delta r$ and $\theta=\pi/2+\delta\theta$. The equations governing the epicyclic motion in the radial and latitudinal directions around the stable position at $r_c$ are equivalent to the equation of harmonic oscillator and for the particular directions can be presented in the form 
\begin{eqnarray}
\delta\ddot{ r}+\bar{\omega}_r^2 \delta r = 0, \ \ \ 
\delta\ddot{ \theta}+\bar{\omega}_\theta^2 \delta\theta = 0.
\end{eqnarray}

The frequencies of the epicyclic oscillatory motion can be determined from the effective potential; here, we use the standard Hamiltonian formalism applied in \cite{Kol-Stu-Tur:2015:CLAQG:,Stu-Kol:2016:EPJC:,Kol-Tur-Stu:2017:EPJC:}, and an alternative perturbation approach can be found in \cite{Ali-Gal:1981:GRG:,Tur-Stu-Kol:2016:PRD:}. For completeness, we also present here also the frequency of the orbital motion, i.e., the azimuthal (Keplerian) frequency $\omega_\phi = \omega_K$. We first give the radial, latitudinal, and azimuthal frequencies of the epicyclic and related orbital motion, $\bar{\omega_r}$, $\bar{\omega_\theta}$, $\bar{\omega_{\phi}}$, as measured by the local observer at $r_c$. 

The Hamiltonian is generally defined as
\begin{eqnarray}\label{e:ham}
H=\frac{1}{2}g^{\alpha\beta}p_\alpha p_\beta+\frac{m^2}{2} 
\end{eqnarray}
and it is convenient to split the Hamiltonian (Equation \ref{e:ham}) into its dynamic and potential parts
\begin{eqnarray}
H=H_{\textrm{dyn}}+H_{\textrm{pot}},
\end{eqnarray}
where
\begin{eqnarray}
H_{\textrm{dyn}}&=& \frac{1}{2}\Big(g^{rr}p_r^2 +g^{\theta\theta}p_\theta^2\Big),\\
H_{\textrm{pot}}&=& \frac{1}{2}\Big(g^{tt}\mathcal{E}^2+g^{\phi\phi}\cl^2+1\Big).\label{e:hpot}
\end{eqnarray}
The potential part of the Hamiltonian serves as the effective potential giving thus the radial and latitudinal epicyclic frequencies $\bar{\omega}_r$ and $\bar{\omega}_\theta$. The orbital (azimuthal) frequency $\bar{\omega}_\phi$ can be obtained directly from Eq. (\ref{e:conserv}) and radial profiles of the specific energy and specific angular momentum of the circular orbit. The resulting formulas for the orbital frequency and both the epicyclic frequencies read as
\begin{eqnarray}
\label{e:bom}
\bar{\omega}_\phi &=& \frac{\cl_\mathrm{c}}{g_{\theta\theta}}, \\ \nonumber 
\bar{\omega}_r^2 &=& \frac{1}{g_{rr}}\frac{\partial^2 H_{\textrm{pot}}}{\partial r^2},\\ \nonumber
\bar{\omega}_\theta^2 &=& \frac{1}{g_{\theta\theta}}\frac{\partial^2 H_{\textrm{pot}}}{\partial \theta^2}.
\end{eqnarray}
Using Equations (\ref{e:met}), (\ref{e:lc}), (\ref{e:const}), and (\ref{e:hpot}) in (\ref{e:bom}), we obtain the explicit form of the locally measured frequencies:
\begin{eqnarray}
\label{e:bomegas}
\bar{\omega}_\theta &=& \bar{\omega}_\phi = \frac{ \sqrt{m(r)-r\, m'(r)}}{r \sqrt{r\, m'(r)+r-3 m(r)}},\\ \nonumber
\bar{\omega}^2_r &=& \frac{r\, m(r) \left[2 r m''(r)+6 m'(r)+1\right] - r^2 \left[r m''(r) - 3 m'(r)\right]+6 m(r)^2}{r^3 \left[r m'(r)-3 m(r)+r\right]}.
\end{eqnarray}

However, the epicyclic motion has to be related to distant observers who really observe and measure the frequencies of the motion; therefore, the form (Equation \ref{e:bomegas}) related to the local observer has to be transformed to those corresponding to the frequencies measured by the static observers at infinity, expressed in the standard (SI) units (hertz). The desired frequencies are obtained from (Equation \ref{e:bomegas}) by the standard scaling with the redshift factor between the orbiting particle and observers at infinity (or large distance), giving the transformation in the following form: 
\begin{eqnarray}\label{e:trfreq}
f_{i}=\frac{1}{2\pi}\frac{c^3}{G\,M}\frac{\bar{\omega_{i}}}{-g^{tt}\mathcal{E}},
\end{eqnarray}
where $i=phi, r,$ and $\theta$. For $\dm = 0$, we arrive at the know results corresponding to the epicyclic motion in the Schwarzschild spacetime \citep{Kat-Fuk-Min:1998:BOOK:,Tor-Stu:2005:ASTRA:}. The resulting frequencies (Equation \ref{e:trfreq}) related to the distant observers can be applied in fitting the observational data of the HF QPOs. The geodesic models were introduced in the framework of general relativity by Stella and Vietri in their relativistic precession model \citep{Ste-Vie-Mor:1999:ApJ:}, the idea of possible role of resonant phenomena was introduced in \cite{Klu-Abr:2001:ACTAASTR:}.
\begin{figure*} 
	\includegraphics[width=\linewidth]{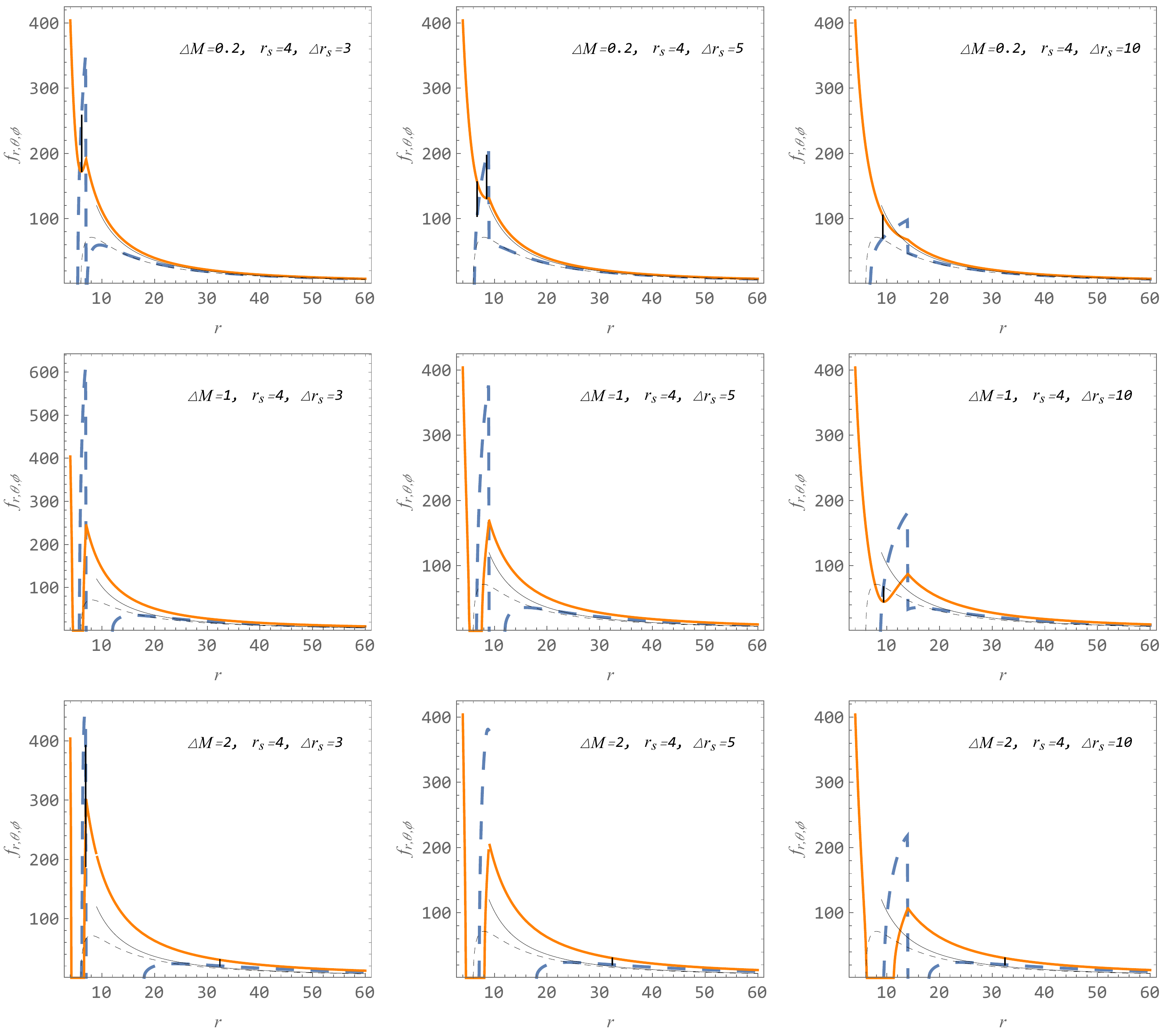}
	\caption{Radial profiles of the radial epicyclic frequency $f_r$ (blue dashed) and the orbital frequency $f_\phi$ that is equal to the latitudinal epicyclic frequency $f_\theta$ (orange) of test particles in the BH+DM shell spacetimes are given for $\rs=4$, and various parameters $\drs$ and $\dm$ in units of $M$ for $M=10\, M_\odot$. Black is epicyclic frequencies $\omega_r$ (dashed) and $\omega_\phi = \omega_\theta$ (solid) in the vacuum Schwarzschild spacetime. Vertical black lines indicate positions of the frequency ratio $3:2$.}
	\label{f:f13}
\end{figure*}
\FloatBarrier

We illustrate behavior of the resulting radial profiles of the orbital and epicyclic frequencies in the BH+DM shell spacetimes in Figure \ref{f:f13} for $\rs=4$ and typical selection of the parameters $\dm$ and $\drs$. Due to the spherical symmetry of the BH+DM shell spacetimes, there is $f_\theta = f_\mathrm{K}$, nevertheless we can see a complex variety of possible character of the radial profiles of the two considered frequencies, being strongly influenced by the properties of the DM shell, even in the considered fixed case of $\rs=4$. The standard behavior given by the profiles representing the vacuum Schwarzschild BH spacetime is enriched by the cases where two regions of stable circular geodesic orbits are possible; note that the stable circular geodesics are in the regions where the radial epicyclic frequency is well-defined. Moreover, there are also cases, where even $f_\theta=f_r$, or the latitudinal frequency $f_\theta = f_\mathrm{K}$ radial profile, demonstrates a local maximum and minimum $f_\theta = 0$ at the equilibrium position, thus representing cases not allowed in the vacuum Schwarzschild spacetimes \citep{Tor-Stu:2005:ASTRA:}. Therefore, we can find several radii, where the frequency ratio of the latitudinal and radial epicyclic frequencies equals $3:2$, i.e., the crucial value for the observed twin HF QPOs in both microquasars and active galactic nuclei \citep{Tor-Abr-etal:2005:AA:,Stu-Kot-Tor:2013:ASTRA:}. Moreover, for the frequency ratio of the epicyclic frequencies we can have now both possibilities allowed -- $f_{\theta}/f_r = 3/2,2/3$ similarly to the case of the epicyclic oscillations around slightly charged matter orbiting a magnetized BH \citep{Stu-etal:2020:Uni:}. 

We thus see that the presence of a DM shell could substantially extend the possibility of the geodesic model to fit the observational twin HF QPO data, as in the pure Schwarzschild spacetime the frequency ratio $3:2$ is allowed only in one radius. We now realize the rough data fitting of the observed mean frequencies (not considering the measurement errors) for all the relevant variants of the geodesic HF QPO model that can be used both for hot spots and for oscillating slender tori \citep{Stu-Kol:2016:ASTRA:} -- the case of tori could be conveniently considered as the observation of the SgrA* and M87 innermost regions demonstrate very extended hot spots that could easily be identified with radiating slender tori generated in a Keplerian accretion disk. We focus attention on the case of the BH+DM shell spacetime with the inner edge located at $\rs=4$ representing the most relevant possibility from the point of view of astrophysical phenomena as discussed in the introduction.

\section{Variants of the Geodesic Model of HF QPOs Applied to Data Observed in Microquasars and around Supermassive BHs}

In observational astrophysics, the HF QPOs are of very high importance as they enable well-founded predictions of parameters of the BHs (or their alternatives) in the accretion systems, i.e., in microquasars representing binary systems containing a stellar-mass BH, or in active galactic nuclei where supermassive BHs are expected. Frequencies of the HF QPOs are observed in hundreds of hertz in microquasars, and by up to 10 orders lower around the supermassive BHs. The inverse-mass scaling of the observed frequencies that is typical for the relations governing the epicyclic frequencies of the orbital motion \citep{Rem-McCli:2005:ARAA:}, makes the models of HF QPOs related to the orbital motion and its epicyclic oscillations, so-called geodesic model of HF QPOs, to be very promising and attractive, especially in connection with the fact that the HF QPOs are often observed in the rational ratio \citep{McCli-Rem:2006:BHbinaries:}. Namely, the observed ratio of $3:2$ \citep{Tor-etal:2011:ASTRA:}, indicating the presence of resonant phenomena \citep{Klu-Abr:2001:ACTAASTR:} makes them very relevant due to the effect of parametric resonance where the frequency ratio $3:2$ corresponds to the strongest resonant phenomena \citep{Tor-Abr-etal:2005:AA:}. A detailed description of the geodesic model can be found in \cite{Stu-Kot-Tor:2013:ASTRA:}, and for
the inclusion of an electromagnetic interaction, \citep{Kol-Tur-Stu:2017:EPJC:,Stu-etal:2020:Uni:,Tur-Stu-etal:2020:ApJ:}. In these models, both the upper $f_\mathrm{u}$ and lower $f_\mathrm{l}$ observed frequencies are considered as a combination of the orbital and epicyclic frequencies of the orbital motion (note that these frequencies are are relevant for oscillations of the slender tori \citep{Rez-Yos-Zan:2003:MNRAS:}). 

\subsection{Variants of the Geodesic Model of HF QPOs}

We first present in Table \ref{t:tab3} the list of the variants of the geodesic model studied in our paper. The list corresponds to the variants considered in the extensive study of the fitting the data of HF QPOs in the microquasar GRO J1655-40 in \cite{Stu-Kol:2016:ASTRA:}. The considered combinations of the epicyclic and orbital frequencies are originally given as related to the Kerr spacetimes, where all the radial profiles of these frequencies are reflected by different formulas. However, in the spherically symmetric background, as is the BH+DM shell spacetime considered in the present paper, the radial profiles of the epicyclic vertical frequency and the azimuthal (Keplerian) or orbital frequency coincide, causing a significant reduction of the considered variants of the geodesic model. The following identities are thus relevant in Table \ref{t:tab3}: RP $\equiv$ RP1 $\equiv$ RP2, ER1 $\equiv$ RP, and ER3 $\equiv$ TD. This significantly reduced (from $11$ to $7$) the number of variants of the geodesic models we have to test for the BH+DH shell spacetimes in fitting the data.
\begin{table*}[ht]
	\begin{center}
		\begin{tabular}{| l l l |}
			\hline
			model & $\nu_U$ & $\nu_L$  \\
			\hline \hline
			RP & $\nu_{\rm \phi}$  & $\nu_{\rm \phi}-\nu_r $  \\
			RP1 & $\nu_{\rm \theta}$ & $\nu_{\rm \phi} - \nu_r $  \\
			RP2 & $\nu_{\rm \phi}$  & $\nu_\theta -\nu_r $  \\
			\hline 
			ER  & $\nu_\theta$  & $\nu_r $ \\
			ER1 & $\nu_{\rm \theta}$ & $\nu_{\rm \theta}-\nu_r $ \\
			ER2 & $\nu_{\rm \theta}-\nu_{r}$ & $\nu_r $  \\
			ER3 & $\nu_{\rm \theta}+\nu_{r}$ & $\nu_{\rm \theta} $ \\
			ER4 & $\nu_{\rm \theta}+\nu_{r}$ & $\nu_{\rm \theta}-\nu_{r}$  \\
			ER5 & $\nu_r $  & $\nu_{\rm \theta}-\nu_{r}$ \\
			\hline
			TD & $\nu_{\rm \phi} + \nu_r$  & $\nu_{\rm \phi}$  \\
			WD & $2\nu_{\rm \phi} - \nu_r$ & 2($\nu_{\rm \phi}-\nu_r) $ \\
			\hline
		\end{tabular}
		\caption{
			The variants of geodesic models of twin HF QPOs as introduced and discussed in \cite{Stu-Kol:2016:ASTRA:}.
			\label{t:tab3}
		} 
	\end{center}
\end{table*}

The first introduced version of the geodesic model is the RP model proposed in \cite{Ste-Vie-Mor:1999:ApJ:} where $f_\mathrm{u} = f_{\phi}=f_\mathrm{K}$ and $f_\mathrm{l} = f_\mathrm{K} - f_\mathrm{r}$. High relevance is attributed also to the simple epicyclic resonance model discussed in \cite{Tor-Abr-etal:2005:AA:} where $f_\mathrm{u} = f_{\theta}$ and $f_\mathrm{l} = f_\mathrm{r}$. 

\subsection{Testing the Geodesic Model Variants for the Frequency
Ratio $3:2$} 

A wide variety of different frequency ratios of twin HF QPOs observed in microquasars and active galactic nuclei is possible, along with more complex frequency patterns, as discussed in detail in \cite{Stu-Kot-Tor:2013:ASTRA:}. However, the most common (almost exclusively, with the exception, e.g., of the GRS 1915 microquasar \citep{Tor-Abr-etal:2005:AA:} or SgrA* \citep{Asch:2004:AAP:,Stu-etal:2005:PRD:}) frequency ratio in observed twin HF QPOs in both microquasars and active galactic nuclei is the $3:2$ ratio of the upper and lower observed frequency. Therefore, we concentrate here on the test of the relevance of the geodesic model in this special case. 

We give the radial profiles of the frequency ratio of the considered variants of the geodesic model, in the dependence on the parameters of the considered BH+DM shell spacetime, and look for the possible existence of the $3:2$ values of the frequency ratio in the profiles. This simple method simultaneously allows a direct selection of acceptable variants of the geodesic model, and the estimate of acceptable regions of the parameter space of the considered spacetime. The results are presented in the Appendix by Figures \ref{f:fmod1}-\ref{f:fmod4}.

Now we can use the selected geodesic model variants along with the related regions of the spacetime parameters implied by the test in the direct fitting to the observational data. From the obtained results we thus consider in the following the epicyclic resonance and its variants, and the RP where all the variants degenerated due to the spherical symmetry of the spacetime. 

\FloatBarrier
\subsection{Fitting to the Observational Data and Related Rough
Limits on the DM}

We demonstrate, in a way that is both illustrative and relevant from the point of view of astrophysical phenomena, the role of the spacetime parameters governing the DM shell in the fitting to observational data related to observed $3:2$ twin HF QPOs in microquasars, and around supermassive BH in active galactic nuclei, as discussed recently in \cite{Smi-Tan-Wag:2021:ApJ:}, where it is demonstrated that all variants of the geodesic model of twin HF QPOs are not able to fit data in the case of the supermassive BHs. All the considered sources with supermassive BHs in active galactic nuclei studied in \cite{Smi-Tan-Wag:2021:ApJ:} are summarized in Table \ref{t:tab1}. In the microquasars case, we consider sources GRG 1915-105, XTE 1530-564 and GRO1655-40. 

Note that in the BH+DM shell spacetimes, the above-defined identification of the upper and lower frequencies also allows the equivalent ratio 2:3 caused by the special characteristics of the radial profiles of the epicyclic frequencies that occur in the spacetimes with a special selection of the parameters of the DM shell. We are thus testing if the DM shell could play a positive role in fitting the data for the supermassive BHs, and to what extent it is not destroying the satisfactory fits obtained by the geodesic model in the case of microquasars, or if it could mimic the role of the BH spin in microquasars, thus giving clear (although very rough) limits on the parameters of the assumed DM shell for each of the sources considered in our study. 

We have to stress that our goal here is a description of the trends and the potential possibility of the DM shell modifications of the BH geometry to establish the ability of the whole variety of relevant variants of the geodesic model in fitting the twin HF QPO data observed around supermassive BHs in active galactic nuclei. For this reason, we realize a rough search
for the parameters characterizing the BH+DM shell spacetime that enable the fitting related to the mean values of the observed frequencies. Because of the approximate characterization of the BH+DM shell metric, and lack of the BH rotation, we are not considering the errors in the frequency measurements, and give only rough estimates of the metric parameters enabling the fitting leading to a basic orientation in the possibilities of the role of DM; we plan a more detailed study related to some selected sources under consideration of the rotational variant of the BH+DM shell spacetimes. 

\begin{figure*} 
	\includegraphics[width=\linewidth]{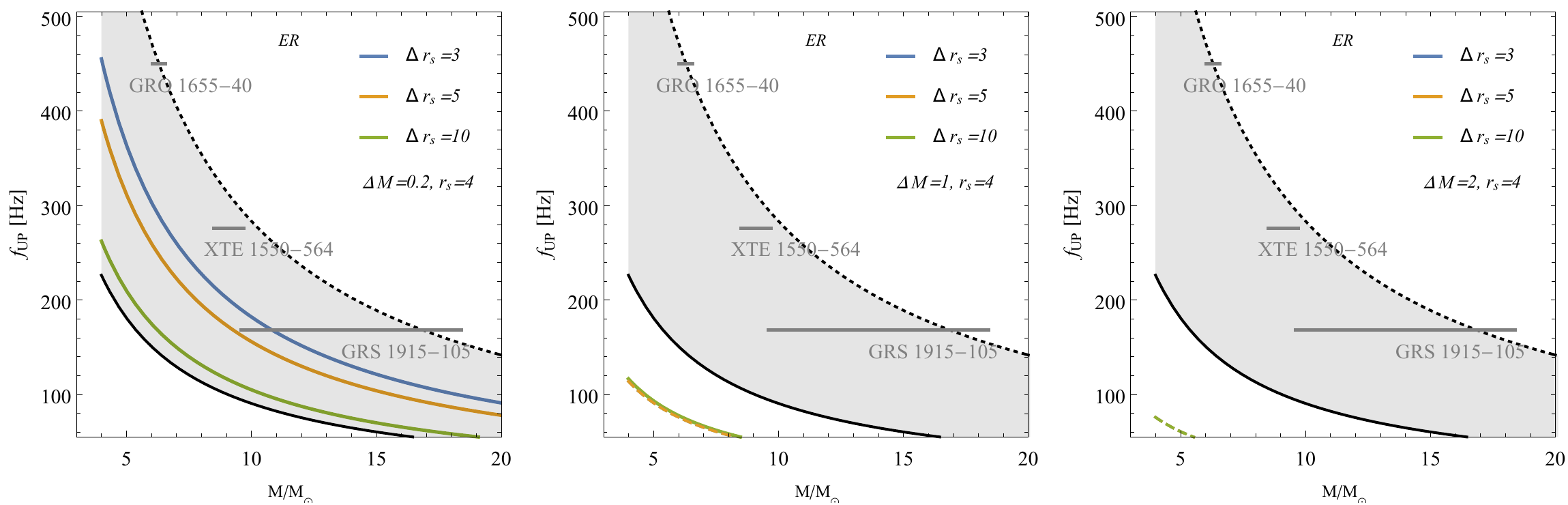}
	\includegraphics[width=\linewidth]{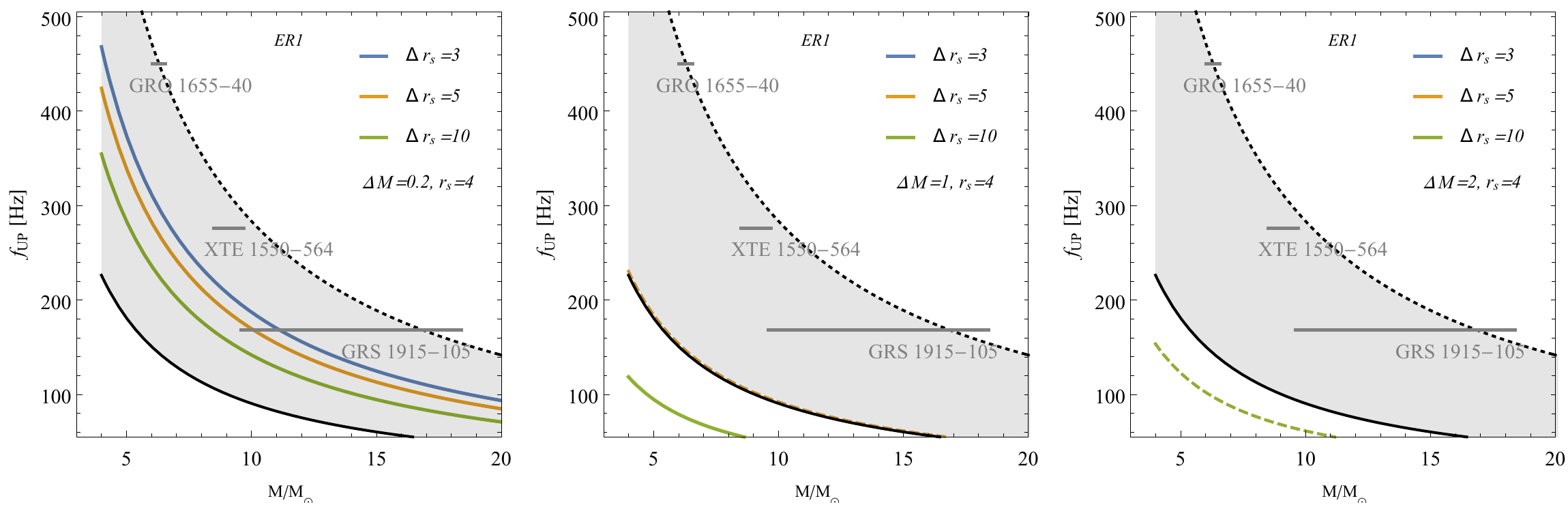}
	\includegraphics[width=\linewidth]{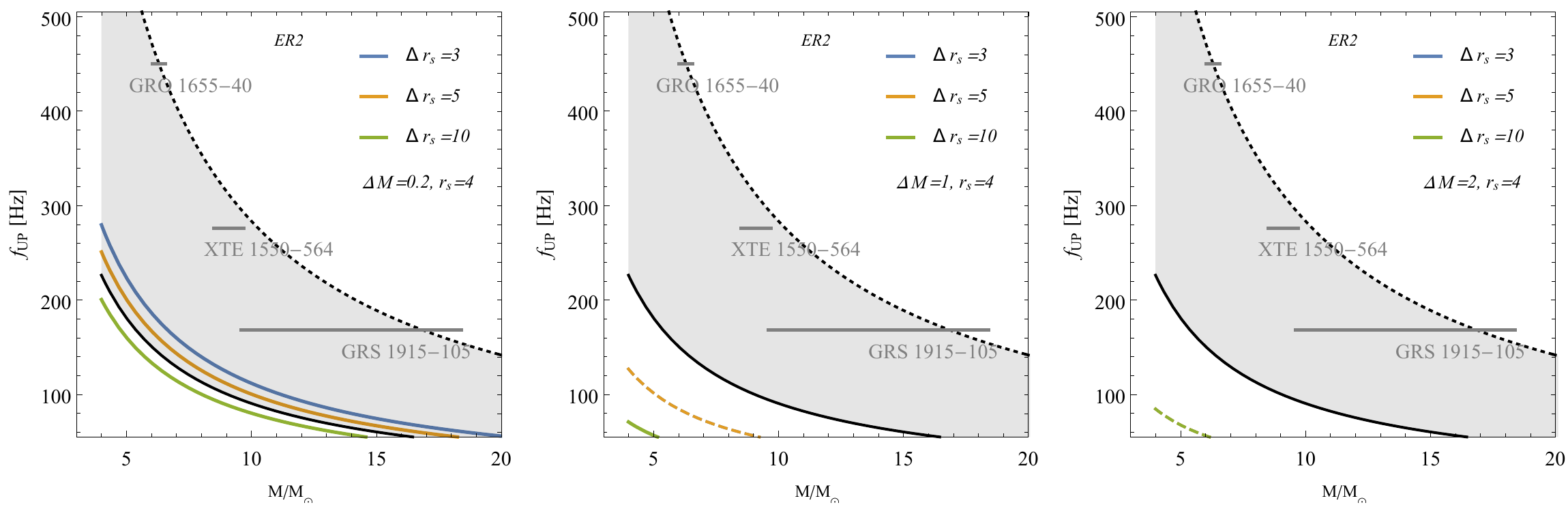}
	\caption{Dependence of the upper frequency on the mass of the central object compared to three selected observed microquasars as given by the ER, ER1, and ER2 variants of the geodesic model. The solid curves describe $f_\theta:f_r=3:2$ for various parameters $\dm$, $\drs$, and $\rs=4$. Notice that the DM shell of mass $\dm=1$ is fully excluded. The mass has to be lower in order to give acceptable fits; however, in the case of $\dm=0.2$, the fits are acceptable and they could mimic the role of the BHs spin demonstrated by the shaded region of the diagram.}
	\label{f:f14}
\end{figure*}

\begin{figure*} 
	\includegraphics[width=\linewidth]{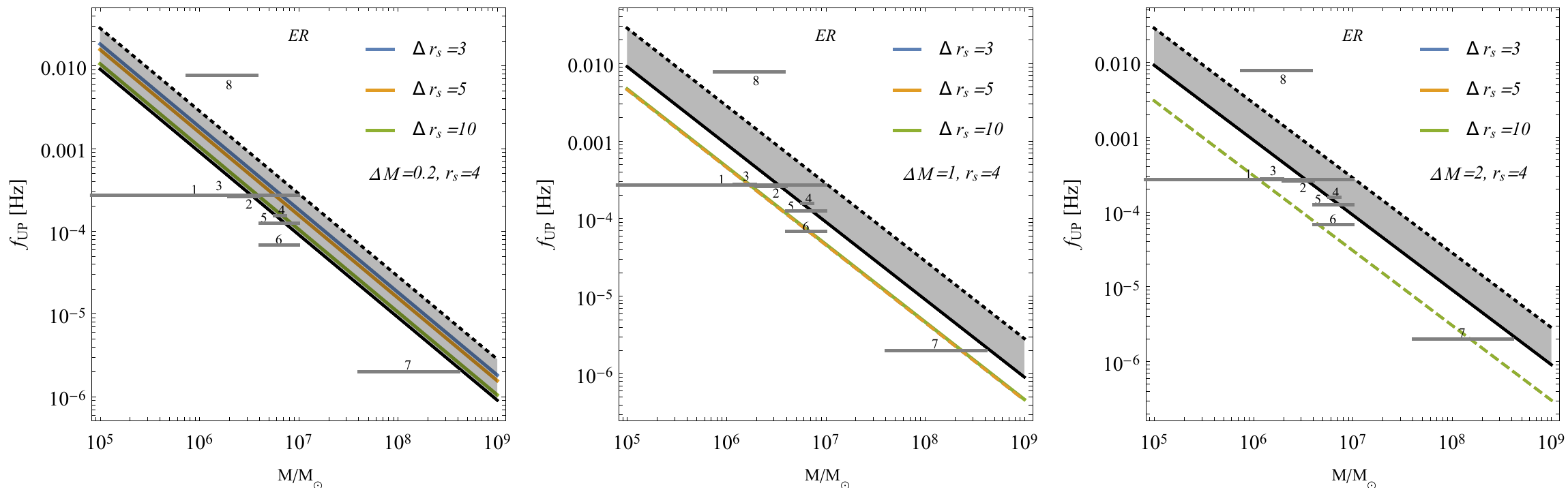}
	\includegraphics[width=\linewidth]{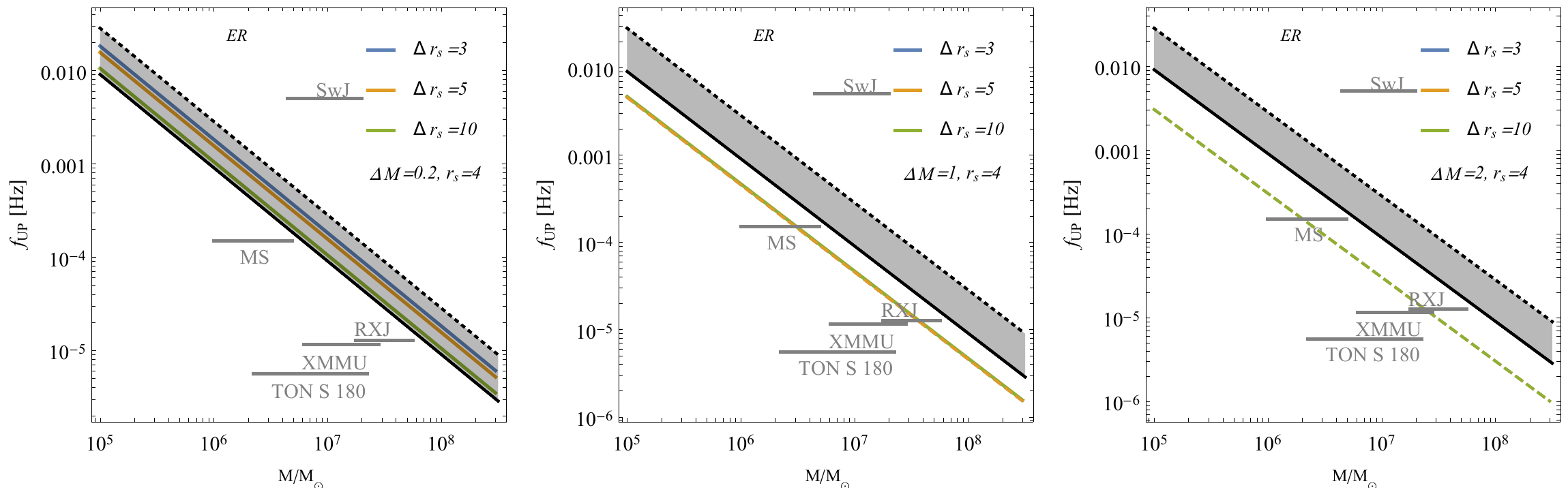}
	\caption{Dependence of the upper frequency on the mass of the central BH in the active galactic nuclei studied in \cite{Smi-Tan-Wag:2021:ApJ:}, given for the ER variant of the geodesic model. The solid curves describe $f_\theta:f_r=3:2$ appearing inside the DM shell, and the dashed curves describe $f_\theta:f_r=3:2$  appearing outside the DM shell for various parameters $\dm$, $\drs$ and $\rs=4$.}
	\label{f:f16}
\end{figure*}

In order to fit the data to the model and obtain restrictions on the spacetime parameters from the restriction on the BH mass obtained by methods different from the frequency measurements, we apply the method introduced in \cite{Stu-Kot-Tor:2013:ASTRA:} for all the considered variants of the geodesic model, taking into account the selected three microquasars, and all the sources discussed in \cite{Smi-Tan-Wag:2021:ApJ:}. The results of the fitting procedure are presented in Fig \ref{f:f14} for microquasars, and in \mbox{Fig \ref{f:f16}} for the case of supermassive BHs assumed in active galactic nuclei. 
 
We have tested all the variants of the geodesic model, as presented in the table 1, for the fitting to the observational data -- we present typical relevant examples of the fitting procedure for properly selected values of the parameters $\dm$, $\drs$ and fixed $\rs=4$, demonstrating the fitting schemes for all the seven considered variants of the geodesic model. 

\begin{figure*} 
	\includegraphics[width=\linewidth]{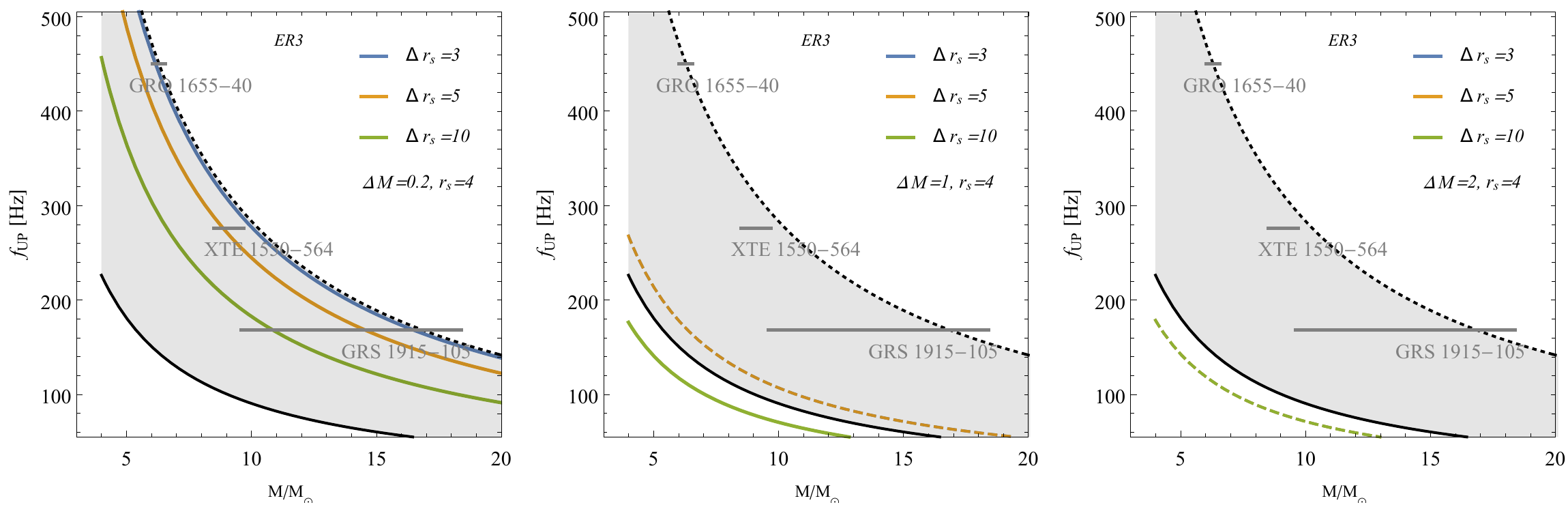}
	\includegraphics[width=\linewidth]{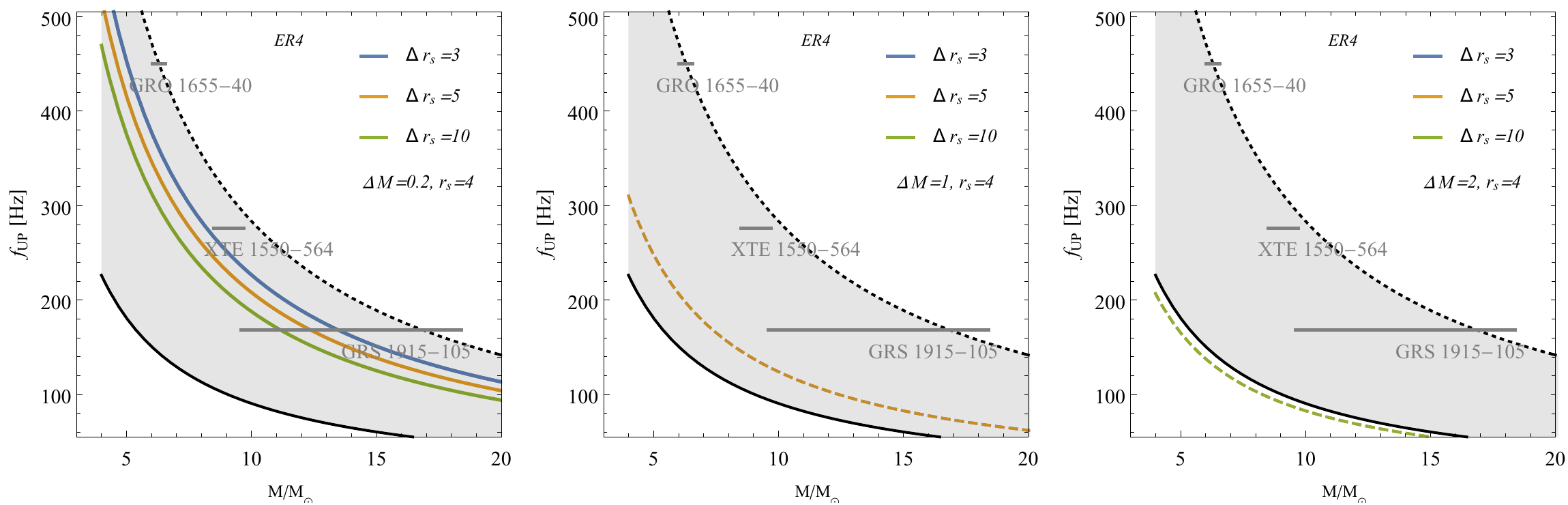}
	\includegraphics[width=\linewidth]{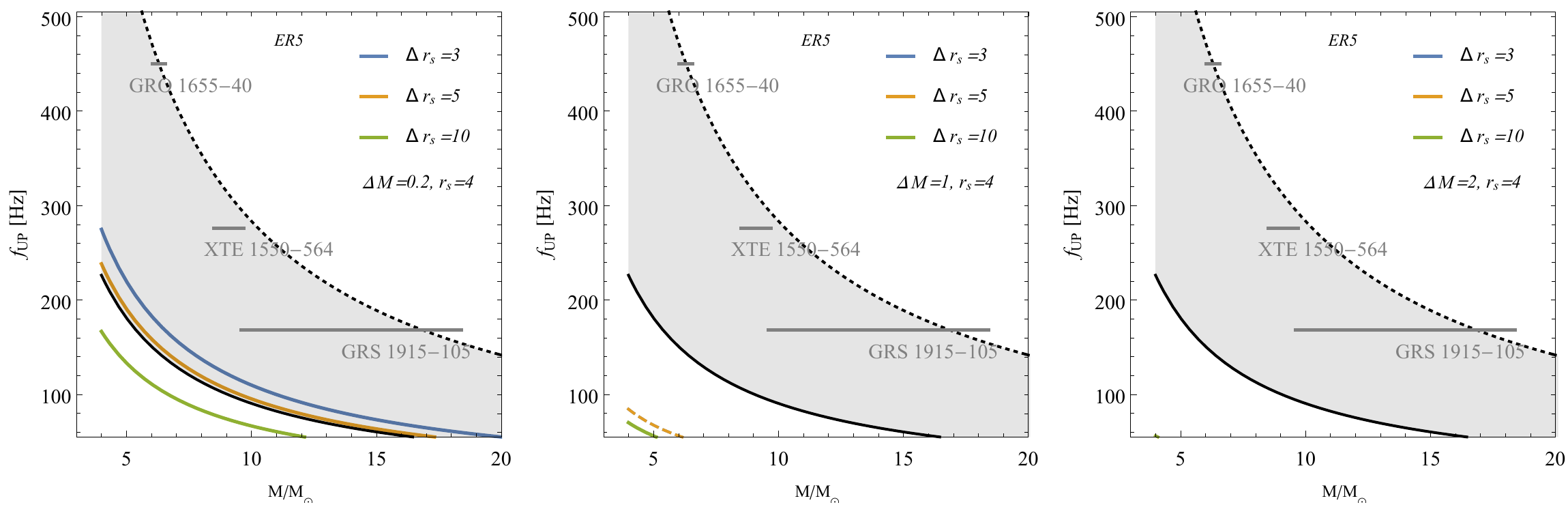}
	\caption{Dependence of the upper frequency on the mass of the central object compared to three selected observed microquasars as given by the ER3, ER4, and ER5 variants of the geodesic model. The solid curves describe $f_\theta:f_r=3:2$ for various parameters $\dm$, $\drs$, and $\rs=4$. Notice that the DM shell of mass $\dm=1$ is fully excluded. The mass has to be lower in order to give acceptable fits; however, in the case of $\dm=0.2$, the fits are acceptable and they could mimic the role of the BHs spin demonstrated by the shaded region of the diagram.}
	\label{f:f14b}
\end{figure*}

\begin{figure*} 
	\includegraphics[width=\linewidth]{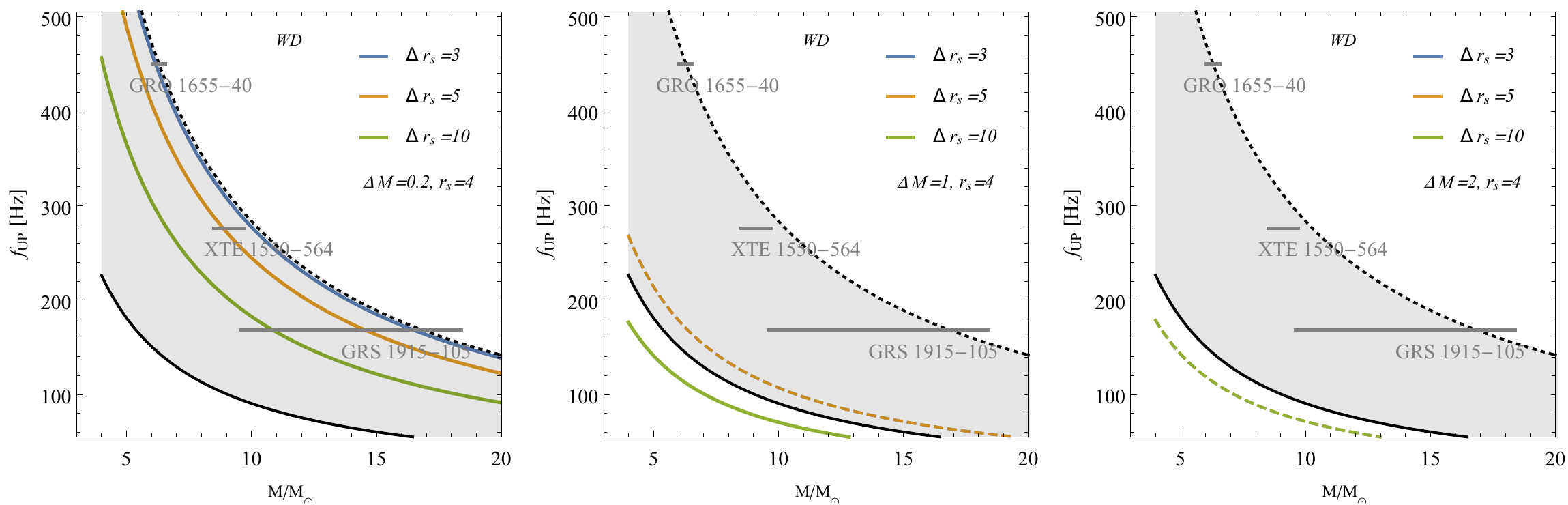}
	\caption{Dependence of the upper frequency on the mass of the central object compared to three selected observed microquasars as given by the WD variant of the geodesic model. The solid curves describe $f_\theta:f_r=3:2$ for various parameters $\dm$, $\drs$ and $\rs=4$. Notice that the DM shell of mass $\dm=1$ is fully excluded. The mass has to be lower in order to give acceptable fits; however, in the case of $\dm=0.2$ the fits are acceptable and they could mimic the role of the BHs spin demonstrated by the shaded region of the diagram.}
	\label{f:f14c}
\end{figure*}

The results of the fittings demonstrated in Figures \ref{f:f14}, \ref{f:f14b}, \ref{f:f14c} for the seven variants of the geodesic model applied in the case of microquasars demonstrate clearly that introducing the DM shell generally decreases the possibility of matching the observed
data. As for  $\dm=1$ and $\dm=2$, the fits are completely beyond the required data, but they also give an important additional information, as in the case of $\dm=0.2$ they could well mimic the influence of the BH spin, being compact enough, i.e., having the extension $\drs$ low enough. The role of the spin is reflected by the shadowed area of the figures, where the lower (upper) bound corresponds to dimensionless spin $a=0$ ($a=1$). Notice the exceptional cases of ER3, ER4 and WD with $\dm=1$ and $\drs=5$ where the fitting lines enter the region of the spin mimicking, but they still are not able to reach the possibility of fitting the data related to the three considered microquasars. 

\begin{figure*} 
	\includegraphics[width=\linewidth]{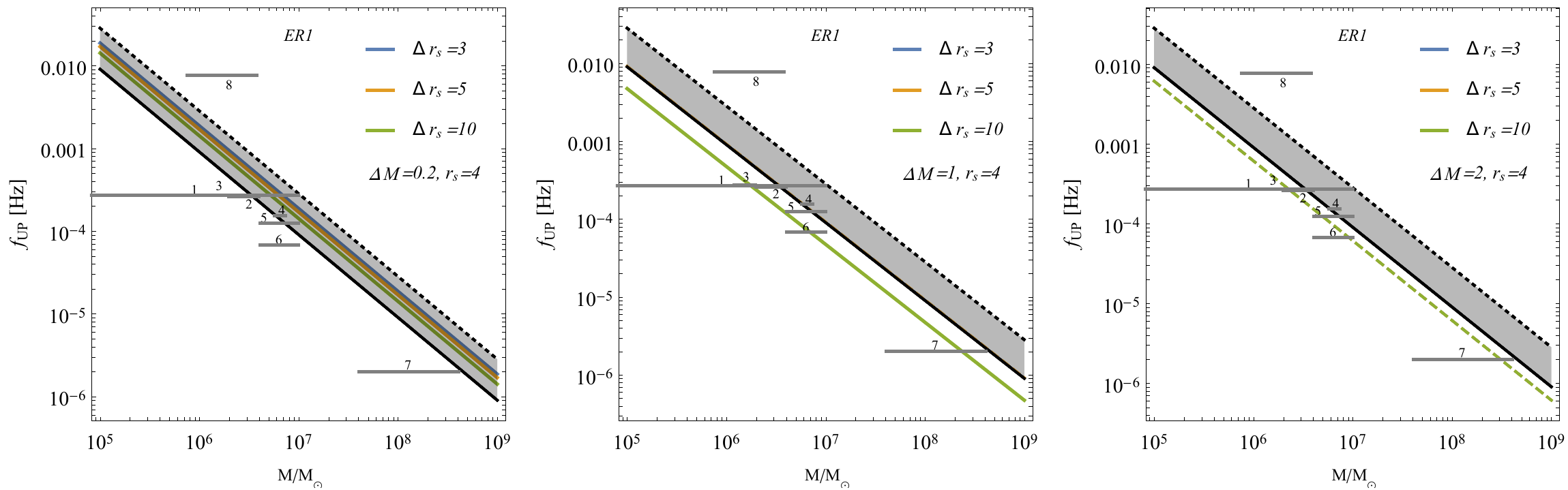}
	\includegraphics[width=\linewidth]{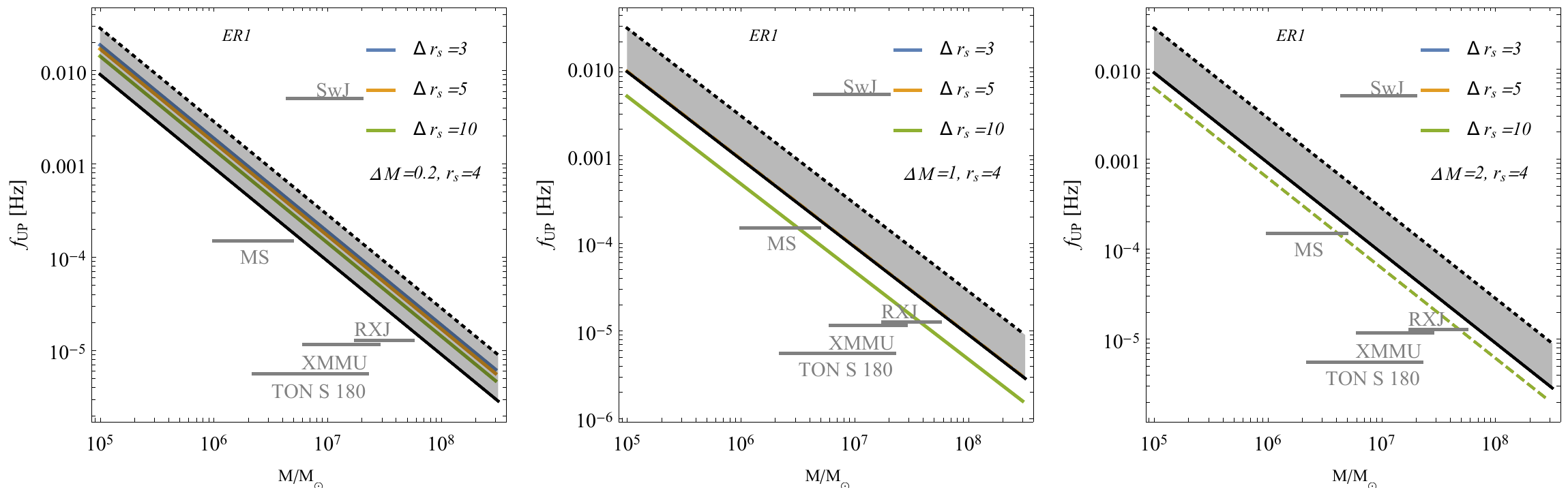}
	\caption{Dependence of the upper frequency on the mass of the central BH in the active galactic nuclei studied in \cite{Smi-Tan-Wag:2021:ApJ:}, given for the ER1=RP variant of the geodesic model. The solid curves describe $f_\theta:f_r=3:2$ appearing inside the DM shell, and the dashed curves describe $f_\theta:f_r=3:2$  appearing outside the DM shell for various parameters $\dm$, $\drs$ and $\rs=4$.}
	\label{f:f16a}
\end{figure*}

\begin{figure*} 
	\includegraphics[width=\linewidth]{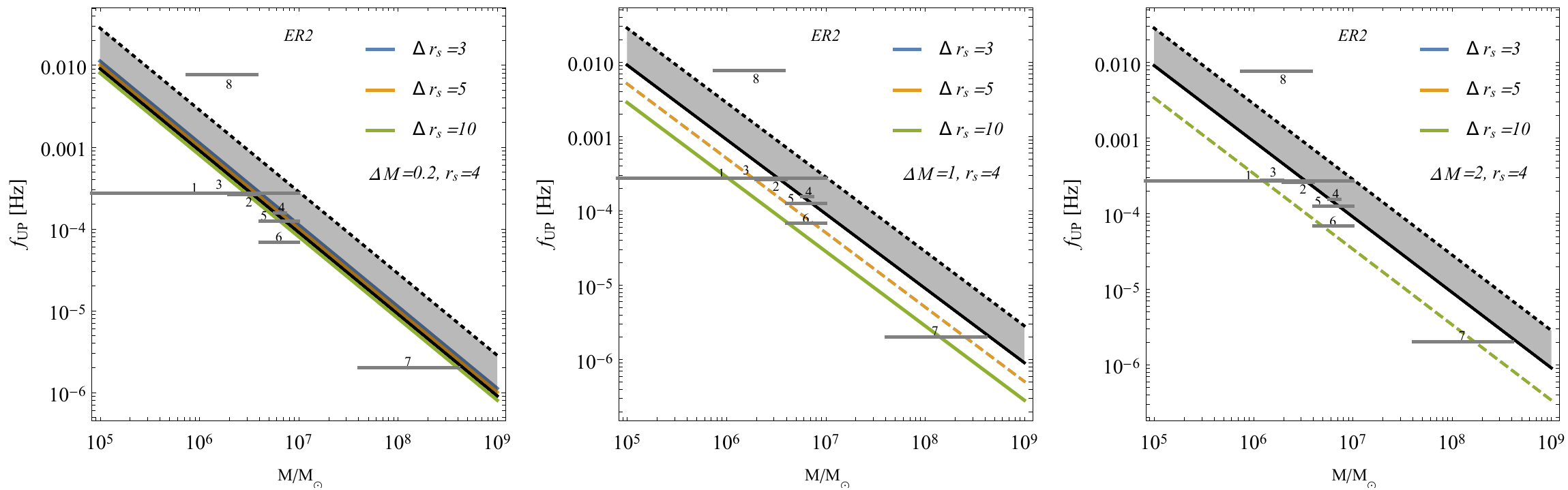}
	\includegraphics[width=\linewidth]{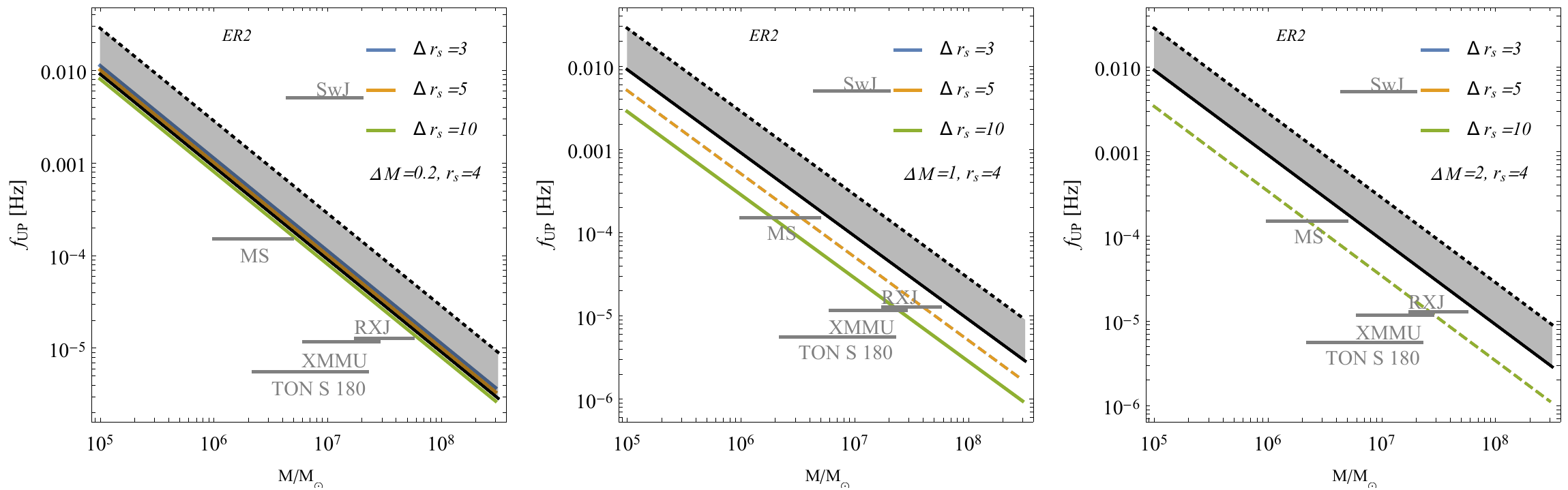}
	\caption{Dependence of the upper frequency on the mass of the central BH in the active galactic nuclei studied in \cite{Smi-Tan-Wag:2021:ApJ:}, given for the ER2 variant of the geodesic model. The solid curves describe $f_\theta:f_r=3:2$ appearing inside the DM shell, and the dashed curves describe $f_\theta:f_r=3:2$  appearing outside the DM shell for various parameters $\dm$, $\drs$ and $\rs=4$.}
	\label{f:f16b}
\end{figure*}

\begin{figure*} 
	\includegraphics[width=\linewidth]{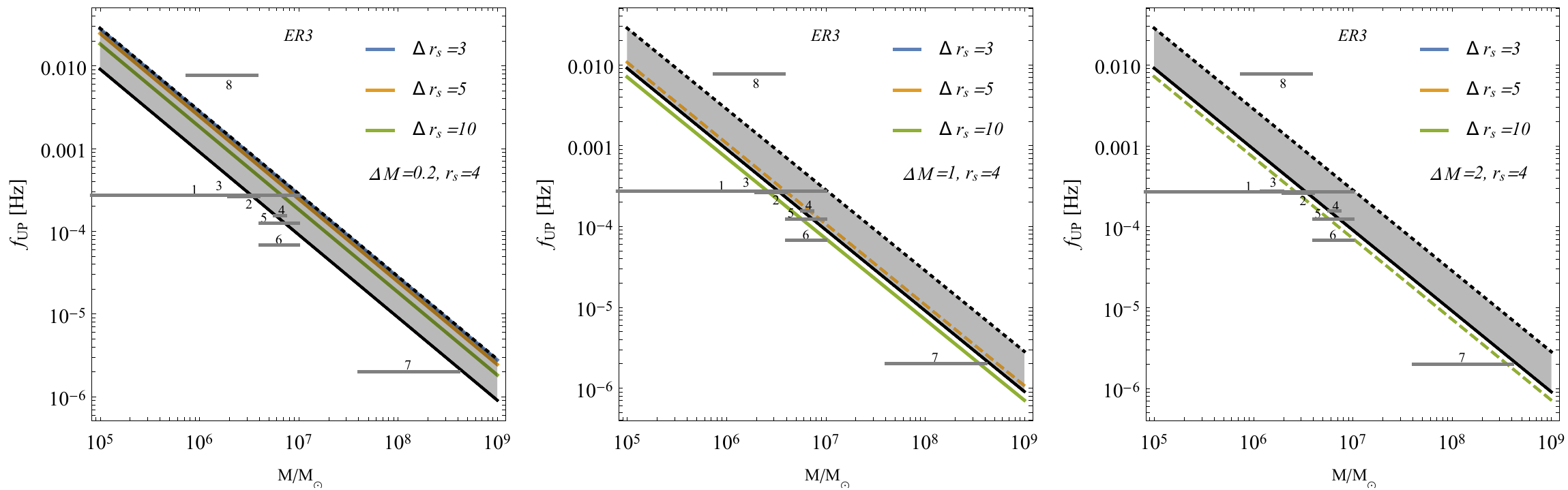}
	\includegraphics[width=\linewidth]{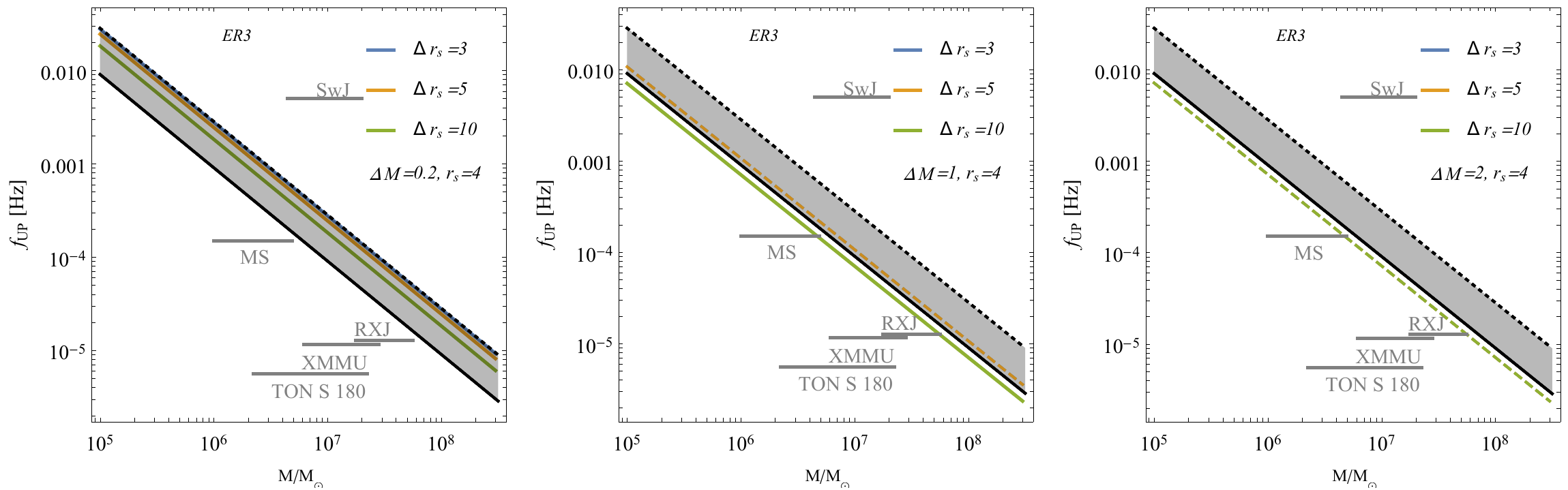}
	\caption{Dependence of the upper frequency on the mass of the central BH in the active galactic nuclei studied in \cite{Smi-Tan-Wag:2021:ApJ:}, given for the ER3=TD variant of the geodesic model. The solid curves describe $f_\theta:f_r=3:2$ appearing inside the DM shell, and the dashed curves describe $f_\theta:f_r=3:2$  appearing outside the DM shell for various parameters $\dm$, $\drs$ and $\rs=4$.}
	\label{f:f16c}
\end{figure*}

\begin{figure*} 
	\includegraphics[width=\linewidth]{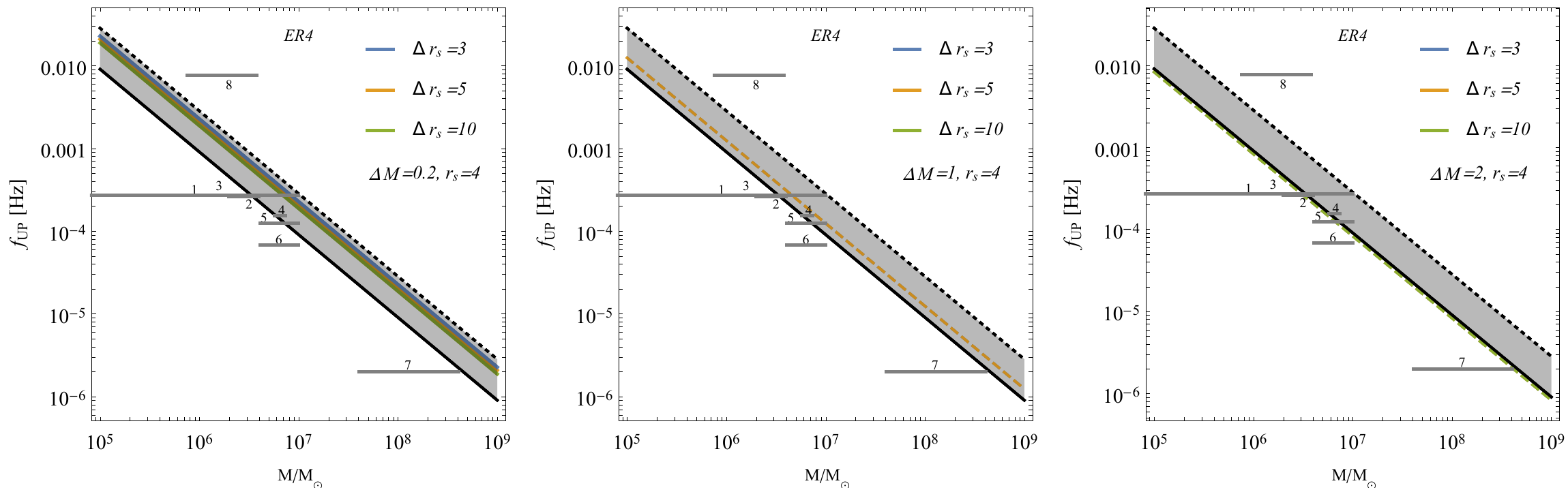}
	\includegraphics[width=\linewidth]{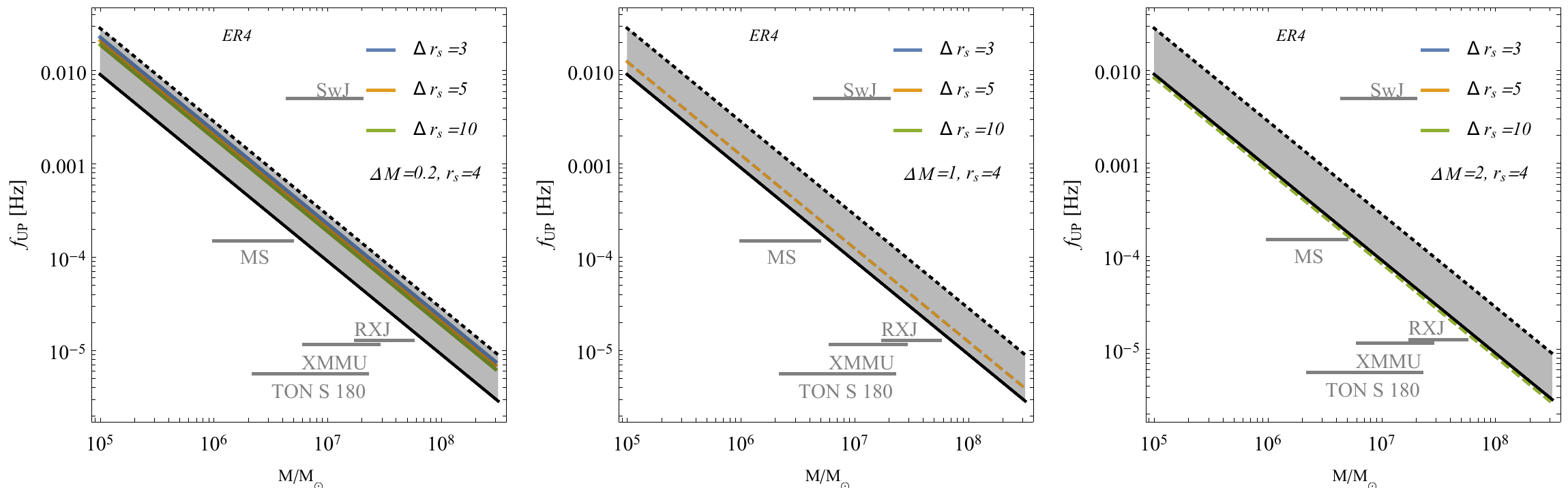}
	\caption{Dependence of the upper frequency on the mass of the central BH in the active galactic nuclei studied in \cite{Smi-Tan-Wag:2021:ApJ:}, given for the ER4 variant of the geodesic model. The solid curves describe $f_\theta:f_r=3:2$ appearing inside the DM shell, and the dashed curves describe $f_\theta:f_r=3:2$  appearing outside the DM shell for various parameters $\dm$, $\drs$ and $\rs=4$.}
	\label{f:f16d}
\end{figure*}

\begin{figure*} 
	\includegraphics[width=\linewidth]{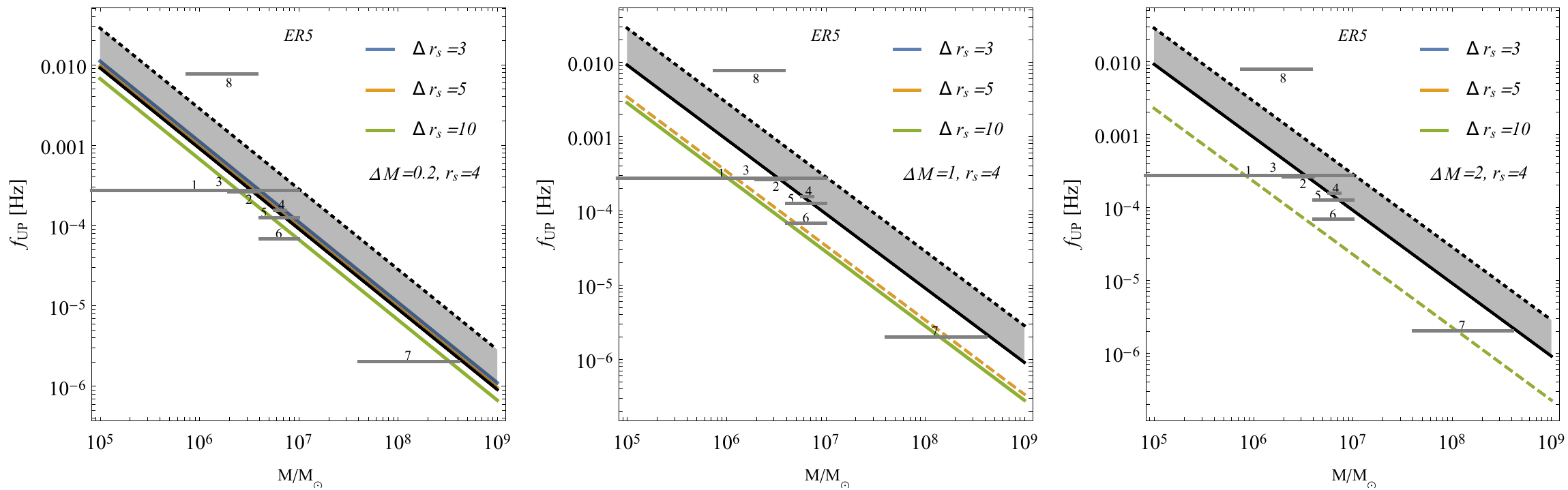}
	\includegraphics[width=\linewidth]{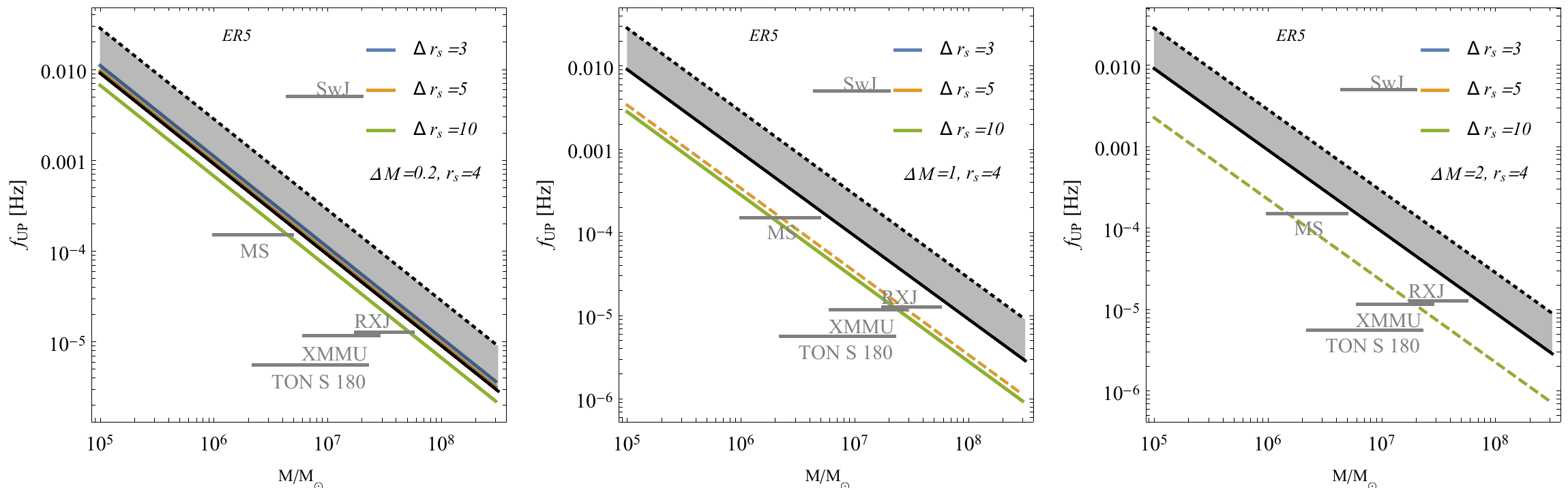}
	\caption{Dependence of the upper frequency on the mass of the central BH in the active galactic nuclei studied in \cite{Smi-Tan-Wag:2021:ApJ:}, given for the ER5 variant of the geodesic model. The solid curves describe $f_\theta:f_r=3:2$ appearing inside the DM shell, and the dashed curves describe $f_\theta:f_r=3:2$  appearing outside the DM shell for various parameters $\dm$, $\drs$ and $\rs=4$.}
	\label{f:f16e}
\end{figure*}

\begin{figure*} 
	\includegraphics[width=\linewidth]{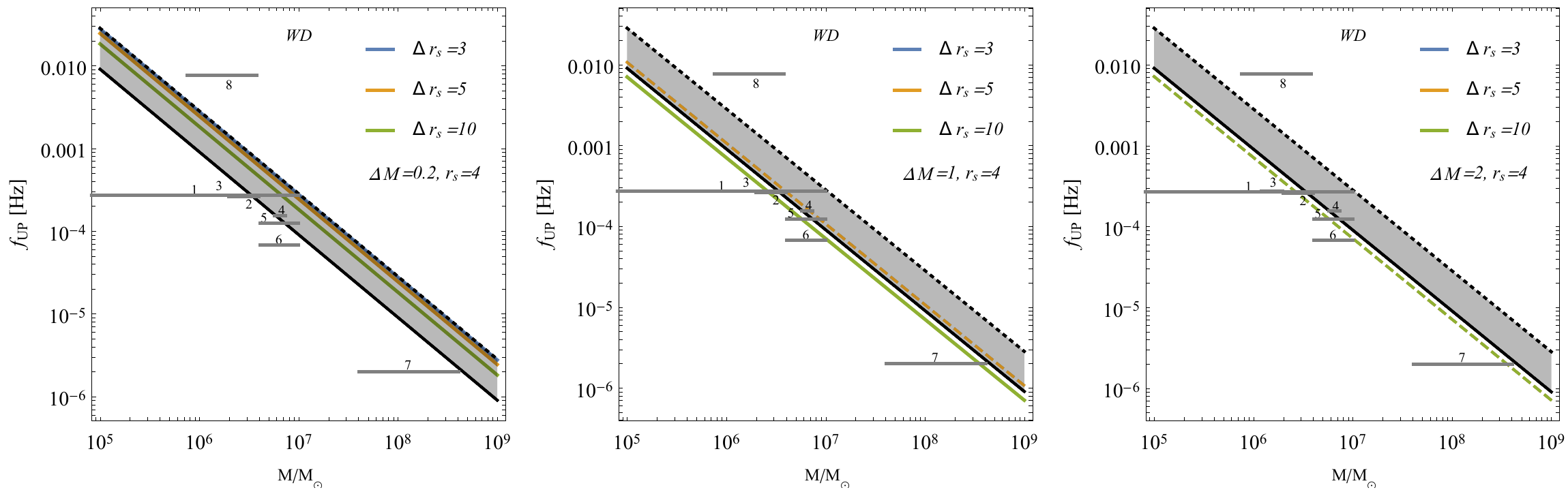}
	\includegraphics[width=\linewidth]{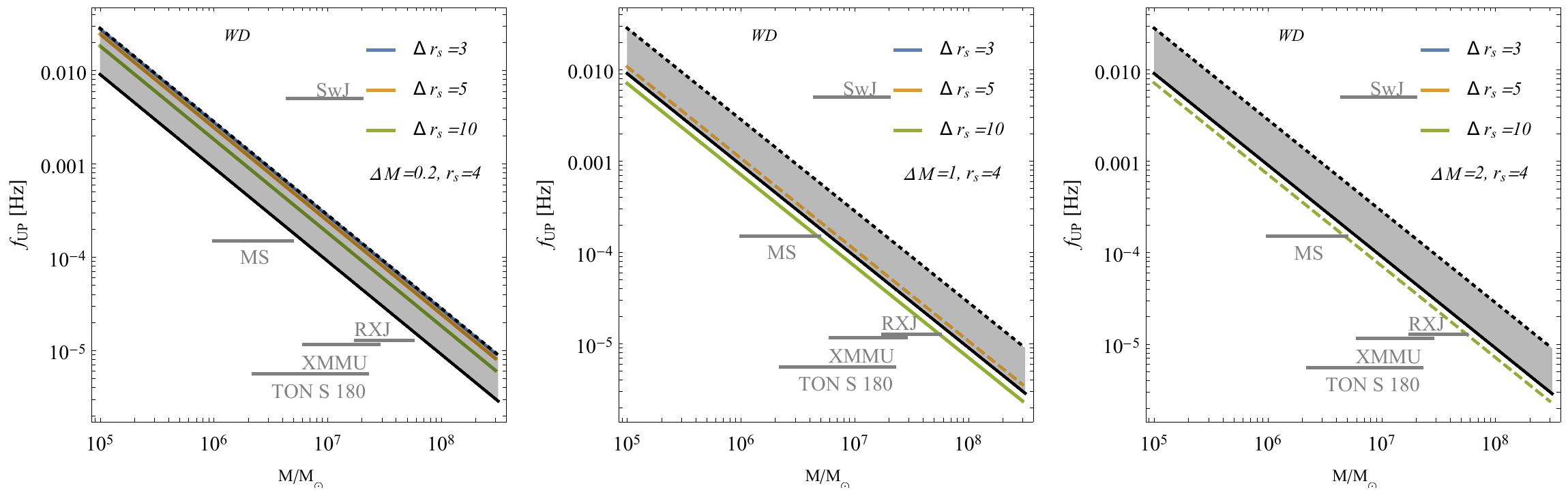}
	\caption{Dependence of the upper frequency on the mass of the central BH in the active galactic nuclei studied in \cite{Smi-Tan-Wag:2021:ApJ:}, given for the WD variant of the geodesic model. The solid curves describe $f_\theta:f_r=3:2$ appearing inside the DM shell, and the dashed curves describe $f_\theta:f_r=3:2$  appearing outside the DM shell for various parameters $\dm$, $\drs$ and $\rs=4$.}
	\label{f:f16f}
\end{figure*}

For the sources containing the supermassive BHs, the effect of the DM shell is always positive from the viewpoint of fitting the data (with exception of the source Sw J164449.31+573451, which seems to be beyond the scope of the geodesic model), and for almost all the considered sources containing an
assumed supermassive BH the considered variants of the geodesic model are able to fit the data if the spacetime parameters are tuned conveniently as follows from situations illustrated in \mbox{Figures \ref{f:f16}-\ref{f:f16f}}. Of course, if the rotational effects on the Schwarzschild geometry modified by the DM shell are considered, we would expect some modification of the shell parameters resulting from the fitting to the rotational version of the spacetime, but we are not expecting fundamental shifts in the positive trends in the fitting of the sources. 

As a final result giving relevant restrictions on the amount of DM that could be contained in the vicinity of supermassive BHs at all the considered sources, we summarize the results of the fitting procedure for each of the sources in Tables \ref{t:tab2}-\ref{t:tab2c}. 

In our fitting procedures and giving our estimates of the spacetime parameters, we assume fixed $\rs=4$, and gave for the considered sources the range of parameters $\drs$ and $\dm$ allowing for fitting the observed mass of the source. Usually, two ranges of allowed values of the parameters were found, one for the lower values of $\dm$, and the other for higher values of $\dm$. Notice that always $\dm \leq 2$ and $\drs \leq 20$ -- details are presented in Tables \ref{t:tab2}-\ref{t:tab2c} where also the classes of the BH+DM shell spacetimes applied in fittings of each of
the considered sources are also denoted in order to obtain an
intuition on the physical situation in the sources. In the case of three sources, namely, TON S 180, ASASSN-14li, and Sw J164449.31+573451, the fitting procedure was not successful; maybe the inclusion of the rotation of the BH+DM spacetime could help in the case of ASASSN-14li and TON S 180, but for Sw J164449.31+573451 the observed frequency seems to be quite out of the possible relation to the geodesic model or its modifications (frequency discrepancy represents more than one order). The same conclusions were obtained in relation to this special source and also in the fitting by the mimicker spacetimes, e.g., in fittings using the wormhole spacetimes \citep{Stu-Vrb:2021:Universe:,Stu-Vrb:2021:JCAP:}. This indicates that the geodesic model and its modifications are not applicable in this very special case. On the other hand, for all the rest of the sources the variants of the geodesic model are successful, or, at least are going in the proper direction and the possible inclusion of the rotational effects in the spacetime geometry are promising for improving of the fitting of the two sources ASASSN-14li and TON S 180. 

We can observe that in positive fittings of the sources in active galactic nuclei almost all of the classes of the BH+DM spacetimes and both disks lying outside and inside the DM shell enter the play. Successful fitting is obtained for the two kinds of the shell as related to its mass -- light with masses $\dm \sim 0.2$ lower than the BH mass, and heavy with mass $1<\dm<2$ slightly overcoming the BH mass. We were not able to find fits for masses strongly exceeding the BH mass. The extension of the shell is estimated to the range $3<\drs<10$ -- its relation to the mass parameter of the shell (its compactness) determines the class of the BH+DM shell spacetime. 

There is a crucial issue of the role of the DM shell in the estimates of the BH mass $M$ that plays a very important role in the geometry model considered in our paper. Of course, such an issue is open for the future more detailed studies of individual sources. Nevertheless, we can point to two fundamentally different approaches. First, if the BH mass estimates are based on the strong-field effects of gravity, as measurements of the BH shadow, we can relate the mass estimate directly to the parameter $M$ and no corrections to our analysis of the HF QPOs are necessary. Second, if the mass estimate is based on the weak-field limit and measurements in distances larger than $\rs + \drs$, we have to relate the mass estimate rather to $M + \dm$ than to $M$, and the results of the data fitting of QPOs have to be reconsidered in any future detailed models of fitting QPOs under the influence of DM concentrated around the central supermassive BHs. Our model can then be directly applied to the fitting procedure in a modified form; having the knowledge of the mass sum $M + \dm$, we can use the assumed ratio $\dm/M$ as a free parameter of the model making the fitting procedure dependent on this free parameter, and deducing the separation of the mass between the BH and the DM shell. 

The results presented in our paper can be considered an indication of a possible explanation for the HF QPO phenomena around supermassive BHs, but inversely, they can also be considered as rough information about the limits on the amount of DM around the supermassive BHs in the considered sources (and the three microquasars), and extension of the region of its highest concentration, thus giving important restrictions on the DM in the central regions of active galactic nuclei and around stellar-mass BH in microquasars. Notice that the obtained limits clearly indicate that around the supermassive BHs’ active galactic nuclei we could find, as expected intuitively, a substantially larger amount of DM in comparison to the BHs in microquasars.



\begin{table*}[h]
\caption{Observations of QPOs around Supermassive BHs \citep{Smi-Tan-Wag:2021:ApJ:}}
\label{t:tab1}
\begin{tabular*}{\textwidth}{@{\extracolsep{\fill}}cllllllll@{}}
\hline
Number  &  Name & BH Spin & log $M_\mathrm{BH}$ ($M_\odot$) & $f_\mathrm{UP}$ (Hz)  \\
\hline
1  &  RE J1034+396	&	0.998	&	$6.0^{+1.0}_{-3.49}$	&	$2.7\times10^{-4}$	\\
2  &  1H0707-495	&	$>0.976$	&	$6.36^{+0.24}_{-0.06}$	&	$2.6\times10^{-4}$	\\ 
3  &  MCG-06-30-15	&	$>0.917$	&	$6.20^{+0.09}_{-0.12}$	&	$2.73\times10^{-4}$	\\
4  &  Mrk 766	&	$>0.92$	&	$6.82^{+0.05}_{-0.06}$	&	$1.55\times10^{-4}$	\\
5  &  ESO 113-G010	&	0.998	&	$6.85^{+0.15}_{-0.24}$	&	$1.24\times10^{-4}$	\\
6  &  ESO 113-G010b	&	0.998	&	$6.85^{+0.15}_{-0.24}$	&	$6.79\times10^{-5}$	\\
7  &  1H0419-577	&	$>0.98$	&	$8.11^{+0.50}_{-0.50}$	&	$2.0\times10^{-6}$	\\
8  &  ASASSN-14li	&	$>0.7$	&	$6.23^{+0.35}_{-0.35}$	&	$7.7\times10^{-3}$	\\
-  &  TON S 180	&	$< 0.4$	&	$6.85^{+0.5}_{-0.5}$	&	$5.56\times10^{-6}$\\
-  &  RXJ 0437.4-4711	&	-	&	$7.77^{+0.5}_{-0.5}$	&	$1.27\times10^{-5}$	\\
-  &  XMMU J134736.6+173403	&	-	&	$6.99^{+0.46}_{-0.20}$	&	$1.16\times10^{-5}$	\\
-  &  MS 2254.9-3712	&	-	&	$6.6^{+0.39}_{-0.60}$	&	$1.5\times10^{-4}$ \\
-  &  Sw J164449.3+573451	&	-	&	$7.0^{+0.30}_{-0.35}$	&	$5.01\times10^{-3}$		\\
\hline
\end{tabular*}
\end{table*}

\begin{table}[h]
\centering
\caption{Best fits of the $\rs=4$, $\drs$, and $\dm$ parameters on observed Supermassive BHs with QPO signature for the ER and ER1 models (Table \ref{t:tab3}). }
\label{t:tab2}
\scalebox{1}{
\begin{tabular}{ccccccc}
\hline
\multicolumn{1}{|c|}{\multirow{2}{*}{Name}} & \multicolumn{2}{c|}{ER}   & \multicolumn{1}{|c|}{\multirow{2}{*}{Class}} & \multicolumn{2}{c|}{ER1} & \multicolumn{1}{|c|}{\multirow{2}{*}{Class}}  \\ \cline{2-3} \cline{5-6} 
\multicolumn{1}{|c|}{}   & \multicolumn{1}{c|}{$\dm$} & \multicolumn{1}{c|}{$\drs$} &\multicolumn{1}{|c|}{}  & \multicolumn{1}{c|}{$\dm$} & \multicolumn{1}{c|}{$\drs$}& \multicolumn{1}{|c|}{}  \\ \hline
(1) RE J1034+396  & $0.2-2$   & $3-10$ & 1,2,4-6 & $0.2-2$   & $3-10$ & $1,2,4-6$ \\
(2) 1H0707-495    & $0.2,1$  &  $10$, $3-10$ & 2,4,5& $1-2$   & $3-10$ & $2,4-6$  \\
(3) MCG-06-30-15   & $1-2$  &  $10$ &2,4 & $1$  &  $10$ & $2$  \\
(4) Mrk 766   & $0.2$  & $10$ & 2 & $1$  & $3-5$ &  $4,5$ \\
(5) ESO 113-G010a   & $0.2,1$   & $10$, $3-10$ & 2,4,5 & $1-2$   & $3-10$ & $2,4-6$ \\
(6) ESO 113-G010b   & $1-2$   & $3-10$ & 2,4-6 & $1,2$  & $10$, $3-10$ & $2,4-6$ \\
(7) 1H0419-577   & $1-2$   & $3-10$ & 2,4-6 &$1-2$   & $3-10$ &  $2,4-6$  \\
(8) ASASSN-14li   & -   & -  & -   & - & - & - \\
TON S 180  & -  & - & -   & - & - & -  \\
RXJ 0437.4-4711   &$1-2$   & $3-10$ & 2,4-6 & $1,2$   &$10$, $3-10$ & $2,4-6$ \\
XMMU J134736.6+173403   & $2$   & $3-10$ & 4-6 &  -   & - & -\\
MS 2254.9-3712    & $1-2$   & $3-10$ & 2,4-6 & $1,2$   &$10$, $3-10$ & $2,4-6$ \\
Sw J164449.3+573451  & -    & -  & -  & - & - & - \\ \hline
\end{tabular}
}
\end{table}

\begin{table}[h]
\centering

\label{t:tab2a}
\scalebox{1}{
\begin{tabular}{ccccccc}
\hline
\multicolumn{1}{|c|}{\multirow{2}{*}{Name}} & \multicolumn{2}{c|}{ER2}   & \multicolumn{1}{|c|}{\multirow{2}{*}{Class}} & \multicolumn{2}{c|}{ER3} & \multicolumn{1}{|c|}{\multirow{2}{*}{Class}}  \\ \cline{2-3} \cline{5-6} 
\multicolumn{1}{|c|}{}   & \multicolumn{1}{c|}{$\dm$} & \multicolumn{1}{c|}{$\drs$} &\multicolumn{1}{|c|}{}  & \multicolumn{1}{c|}{$\dm$} & \multicolumn{1}{c|}{$\drs$}& \multicolumn{1}{|c|}{}  \\ \hline
(1) RE J1034+396  & $0.2-2$   & $3-10$ & 1,2,4-6 & $0.2-2$   & $3-10$ & $1,2,4-6$ \\
(2) 1H0707-495    & $0.2,1$  &  $10$, $3-10$ & 2,4,5 & $1-2$   & $3-10$  & $2,4-6$  \\
(3) MCG-06-30-15   & $1-2$  &  $10$ & 2,4 & $1$  &  $10$  & $2$  \\
(4) Mrk 766   & $0.2$  & $10$ & 2 & $1$  & $3-5$ &  $4,5$ \\
(5) ESO 113-G010a   & $0.2,1$   & $10$, $3-10$ & 2,4-6 & $1-2$   & $3-10$ & $2,4-6$ \\
(6) ESO 113-G010b   & $1-2$   & $3-10$ & 2,4-6  & $1,2$  & $10$, $3-10$ & $2,4-6$ \\2,4-6
(7) 1H0419-577   & $1-2$   & $3-10$ & 2,4-6 &$1-2$   & $3-10$ & $2,4-6$  \\
(8) ASASSN-14li   & -   & -  & -   & - & - & -  \\
TON S 180  & -  & - & -   & - & - & -   \\
RXJ 0437.4-4711   &$1-2$   & $3-10$  &2,4-6  & $1,2$   &$10$, $3-10$ &  $2,4-6$ \\
XMMU J134736.6+173403   & $2$   & $3-10$  & 4-6 &  -   & - &  -\\
MS 2254.9-3712    & $1-2$   & $3-10$ & 2,4-6 & $1,2$   &$10$, $3-10$ & $2,4-6$ \\
Sw J164449.3+573451  & -    & -  & -  & - & - & - \\ \hline
\end{tabular}
}
\caption{Best fits of the $\rs=4$, $\drs$, and $\dm$ parameters on observed Supermassive BHs with QPO signature for the ER2 and ER3 models (Table \ref{t:tab3}).}
\end{table}

\begin{table}[h]
\centering

\label{t:tab2b}
\scalebox{1}{
\begin{tabular}{ccccccc}
\hline
\multicolumn{1}{|c|}{\multirow{2}{*}{Name}} & \multicolumn{2}{c|}{ER4}   & \multicolumn{1}{|c|}{\multirow{2}{*}{Class}} & \multicolumn{2}{c|}{ER5} & \multicolumn{1}{|c|}{\multirow{2}{*}{Class}}  \\ \cline{2-3} \cline{5-6} 
\multicolumn{1}{|c|}{}   & \multicolumn{1}{c|}{$\dm$} & \multicolumn{1}{c|}{$\drs$} &\multicolumn{1}{|c|}{}  & \multicolumn{1}{c|}{$\dm$} & \multicolumn{1}{c|}{$\drs$}& \multicolumn{1}{|c|}{}  \\ \hline
(1) RE J1034+396  &  $0.2-2$ & $3-10$ &1,2,4-6 &$0.2-2$ & $3-10$ & $1,2,4-6$ \\
(2) 1H0707-495    &  $2$ & $3-10$  & 4-6 & $0.2$ & $3-10$&  $1,2,4-6$  \\
(3) MCG-06-30-15   &  - & - & - & $1$ & $3-5$ & $2,4,5$  \\
(4) Mrk 766   & $0.2$ & $3-5$ & 1,2 & $0.2$ & $3-5$ & $1,2,4,5$ \\
(5) ESO 113-G010a   &  $0.2$   & $3-10$  & 1,2 & $0.2$   & $3-10$  & $1,2,4-6$ \\
(6) ESO 113-G010b   & -  & - & - & $0.2,1$   & $10$, $3-10$ & $2,4-6$ \\
(7) 1H0419-577   &  $2$   & $3-10$ & 4-6 & $0.2$, $1-2$   & $10$, $3-10$ & $2,4-6$  \\
(8) ASASSN-14li   & -   & -  & -   & - & - & - \\
TON S 180  & -  & - & -   & - & - & -  \\
RXJ 0437.4-4711   &- & - & - & $0.2$, $1-2$& $10$, $3-10$ & $2,4-6$ \\
XMMU J134736.6+173403   &  - &-  & - & $1-2$ & $3-10$  & $2,4-5$\\
MS 2254.9-3712    & - & - & - &  $0.2$, $1-2$ & $10$, $3-10$ & $2,4-6$ \\
Sw J164449.3+573451  & -    & -  & -  & - & - & - \\ \hline
\end{tabular}
}
\caption{Best fits of the $\rs=4$, $\drs$, and $\dm$ parameters on observed Supermassive BHs with QPO signature for the ER4 and ER5 models (Table \ref{t:tab3}).}
\end{table}

\begin{table}[h]
\centering
\label{t:tab2c}
\scalebox{1}{
\begin{tabular}{ccccccc}
\hline
\multicolumn{1}{|c|}{\multirow{2}{*}{Name}} & \multicolumn{2}{c|}{WD}   & \multicolumn{1}{|c|}{\multirow{2}{*}{Class}} \\ \cline{2-3}
\multicolumn{1}{|c|}{}   & \multicolumn{1}{c|}{$\dm$} & \multicolumn{1}{c|}{$\drs$} &\multicolumn{1}{|c|}{}   \\ \hline
(1) RE J1034+396  &  $0.2-2$ & $3-10$ & 1,2,4-6 \\
(2) 1H0707-495    &  $1-2$ & $3-10$  & 2,4-6  \\
(3) MCG-06-30-15   &  - & - &  - \\
(4) Mrk 766   & $1$ & $3-5$ & 4,5 \\
(5) ESO 113-G010a   &  $1-2$   & $3-10$  &   2,4-6\\
(6) ESO 113-G010b   & $1,2$  & $10$, $3-10$ & 2,4-6  \\
(7) 1H0419-577   &  $1-2$   & $10$ &  2,4 \\
(8) ASASSN-14li   & -   & -  & \\
TON S 180  & -  & - &  \\
RXJ 0437.4-4711   &$1$,$2$ & $10$, $3-10$ & 2,4-6 \\
XMMU J134736.6+173403   &  - &-  & \\
MS 2254.9-3712    & $1$,$2$ & $10$, $3-10$  & 2,4-6 \\
Sw J164449.3+573451  & -    & -  &  \\ \hline
\end{tabular}
}
\caption{Best fits of the $\rs=4$, $\drs$, and $\dm$ parameters on observed Supermassive BHs with QPO signature for the ER4 and ER5 models (Table \ref{t:tab3}).}
\end{table}

\FloatBarrier

\section{Conclusions}

We have discussed the properties of the rough but robust BH+DM shell spacetimes allowing for the inclusion of all kinds of DM in close vicinity to the BH horizon, giving the conditions for their physical reality, discussing the realistic limits on their inner edge, and introducing six classes of their behavior in relation to properties of circular geodesic orbits governing the behavior of Keplerian disks, demonstrating the possibility of the existence of two Keplerian disks, where the inner one has to be located inside the DM shell. 

We have to stress an important finding of our paper, namely,
the existence of spacetime classes containing stable equilibrium
positions where the matter in accretion finishes its evolution in
a stationary state, stopping its orbiting relative to distant
observers. In class III spacetime it is the inner edge and final
state of accretion in the outer (primary) Keplerian disk, and in
classes IV–VI it is the inner edge of the inner (secondary)
Keplerian disk; naturally, in all the cases the equilibrium
position is located inside the DM shell. It would be interesting
to search observations of the regions close to the BH horizon in
active galactic nuclei for the possible existence of the
phenomenon of the existence of nonrotating regions of
Keplerian disks accumulating the accretion matter.

We have given the frequencies of the epicyclic oscillatory
motion around stable circular geodesics in the field of the BH
+DM shell spacetimes, assuming a very rough and simple
model of the spacetime corresponding to a Schwarzschild BH
surrounded by a DM shell with a very simple distribution of
matter. The epicyclic frequencies were applied in all the
relevant variants of the geodesic model of the twin HF QPOs,
e.g., the RP variant related to the hot spot models, and the
epicyclic resonance variant that also reflects oscillation modes
of slender tori, to fitting the observed HF QPO data from
microquasars and active galactic nuclei. The BH+DM spacetime geometry is characterized by three parameters: $\dm/M$ represents the ratio of the shell to BH masses, $\drs$ represents the extension of the shell, and $\rs$ is the position of the inner edge of the shell. In our fittings to the observational data, we have fixed the inner edge of the shell to $\rs=4M$ corresponding to the region of the gravitational binding of the central BH that we consider to be the most natural choice. 

We have demonstrated that in the case of the microquasars the parameters $\rs,\, \drs$, and $\dm$ have a general tendency to decrease the possibility to obtaining satisfactory fits, but for sufficiently low values of $\dm \sim 0,2$ they could mimic the role of the BH spin. In the case of the frequencies observed around supermassive BHs assumed in active galactic nuclei, the parameters act in a positive way, and some combinations of the BH+DM shell spacetime parameters enable satisfactory fits for the supermassive BHs sources with observed QPO data that do not allow for successful fits by the geodesic models assuming vacuum BH spacetimes, as discussed in \cite{Smi-Tan-Wag:2021:ApJ:}. We demonstrate that if the influence of
the DM shell is assumed to be substantial, modifying the spacetime significantly, the fitting can be realized for almost all the considered sources in active galactic nuclei. Both the ER and RP variants of the geodesic model could be very successful. The WD variant demonstrated the lowest tendency to good fitting—a more detailed study taking into account the rotation BH effect is necessary to decide which
variant should be preferred; here, we have tested on the tendency for the good fits. Inversely, our estimates on the mass and extension of the DM shell can be considered as rough upper limits on the presence of the DM located in the vicinity of the supermassive BH in the considered source, or around the stellar-mass BHs in microquasars, confirming a higher tendency of the concentration of DM around supermassive BHs. We summarize the range of allowed values of the spacetime parameters in all the considered sources with supermassive BHs, giving in such a way strong restrictions on the amount of DM allowed around the supermassive BHs in these sources. 

Of course, extensions of our basic study are necessary, which should consider the role of the BH rotation, and the way of estimating the mass of the central BHs that could introduce the necessity of a slight modification of the fitting procedure applied in the present paper. 
\FloatBarrier

\section*{acknowledgments}

The authors acknowledge the institutional support of the Research Centre for Theoretical Physics and Astrophysics, Institute of Physics, Silesian University in Opava. Z.S. acknowledges the Czech Science Foundation grant No. 19-03950S.

\bibliography{references}{}
\bibliographystyle{aasjournal}

\section{Appendix}
Here, we provided an overview of the possibilities of all the variants of the geodesic model of twin HF QPOs with the frequency ratio observed around BHs on the most common value of $3:2$ connected to the most efficient variant of the parametric resonant phenomena \cite{Stu-Kot-Tor:2013:ASTRA:}. We present radial profiles of frequency ratio of $f_U:f_L$ for all eleven (seven after reduction due to spherical symmetry of the spacetime) variants of the geodesic model as given in Table \ref{t:tab3}. 

\begin{figure*} 
	\includegraphics[width=0.85\linewidth]{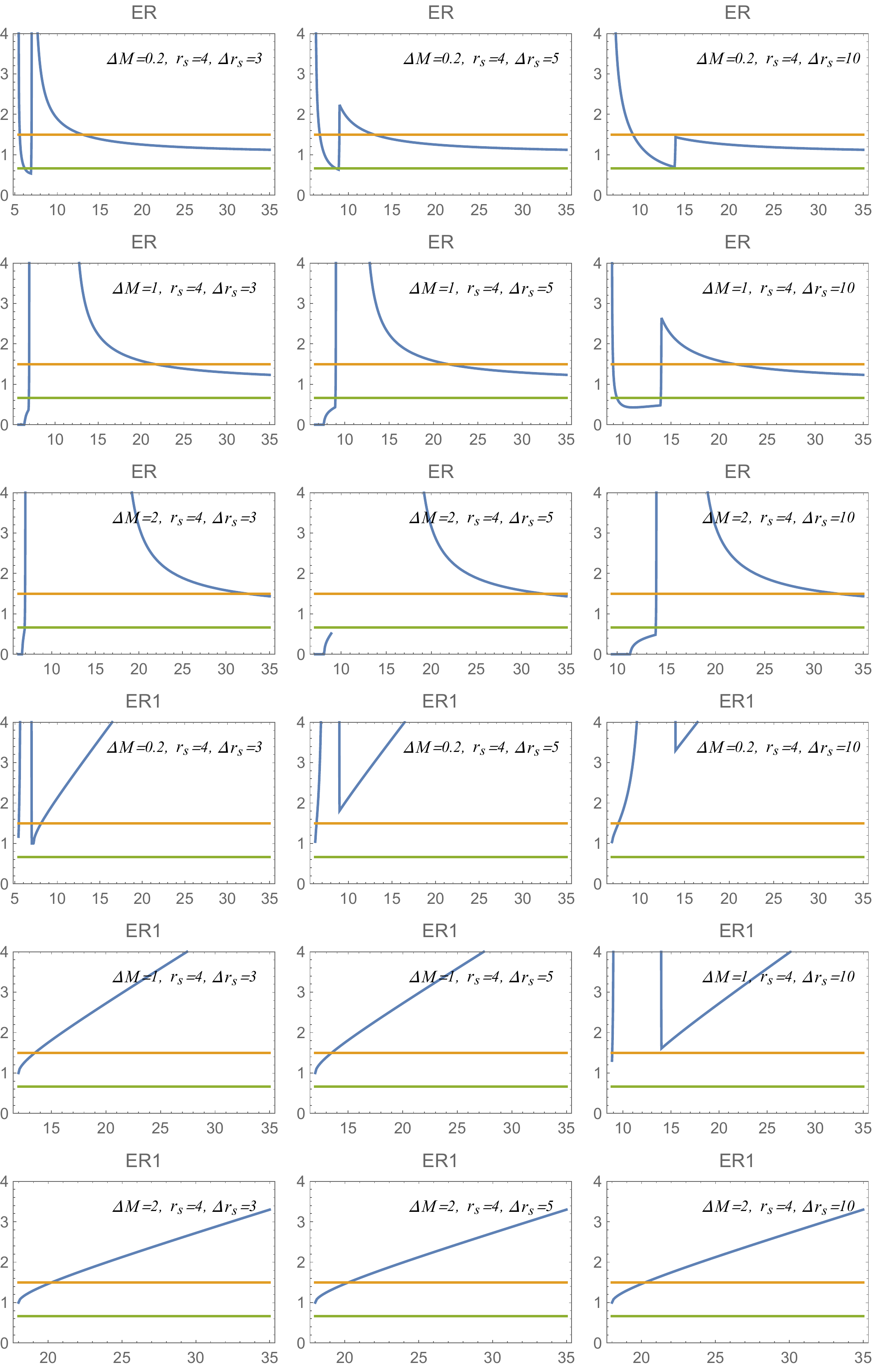}
	\caption{Function of the ratio of $f_U:f_L$ for given selection of the ER and ER1=RP variants of the geodesic model of twin HF QPOs. Orange line denotes the ratio $3:2$, green denotes the ratio 2:3.}
	\label{f:fmod1}
\end{figure*}
\begin{figure*} 
	\includegraphics[width=0.85\linewidth]{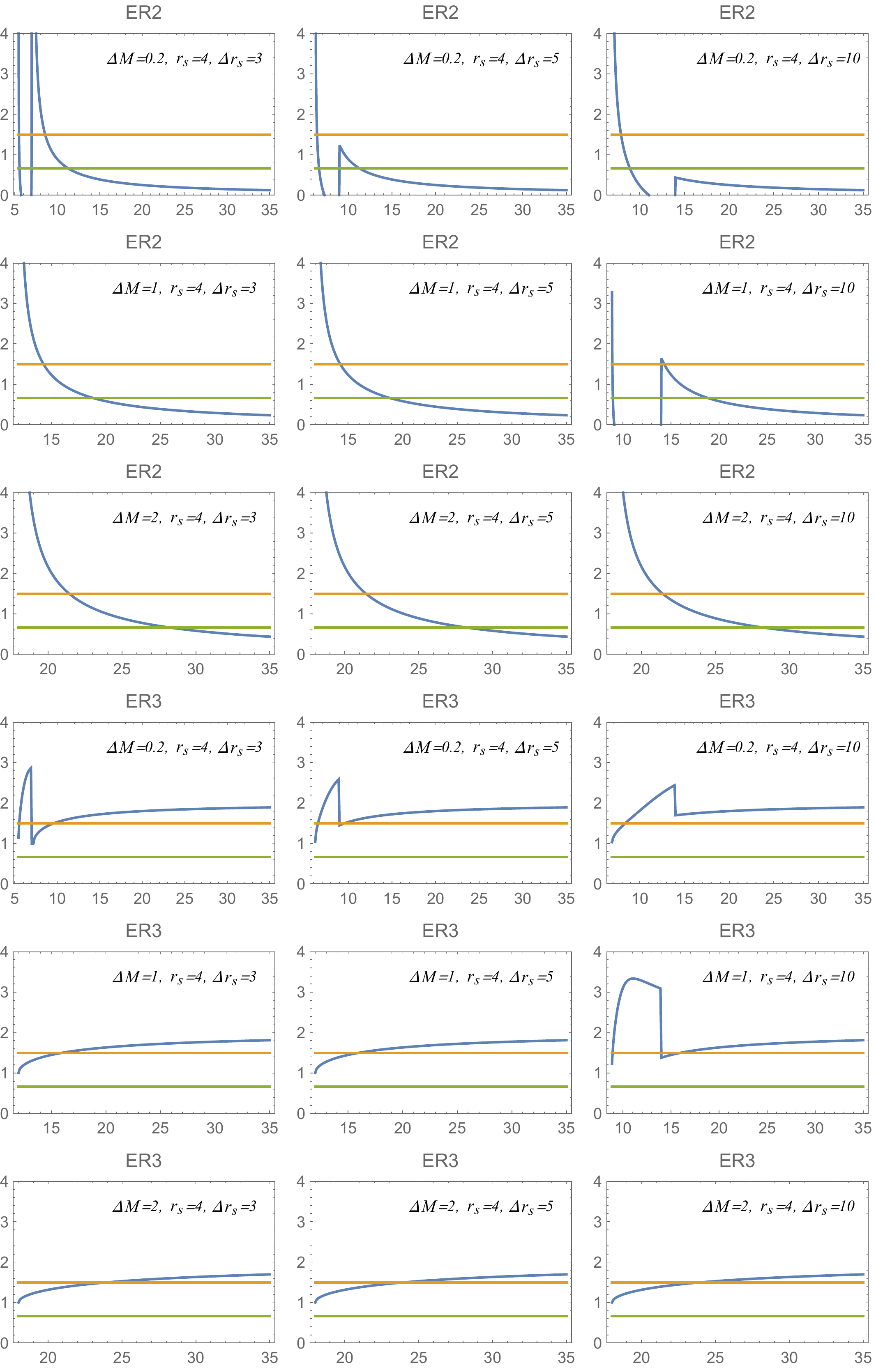}
	\caption{Function of the ratio of $f_U:f_L$ for the ER2 and ER3=TD variants of the geodesic model. Orange line denotes the ratio $3:2$, green denotes the ratio 2:3.}
	\label{f:fmod2}
\end{figure*}
\begin{figure*} 
	\includegraphics[width=0.85\linewidth]{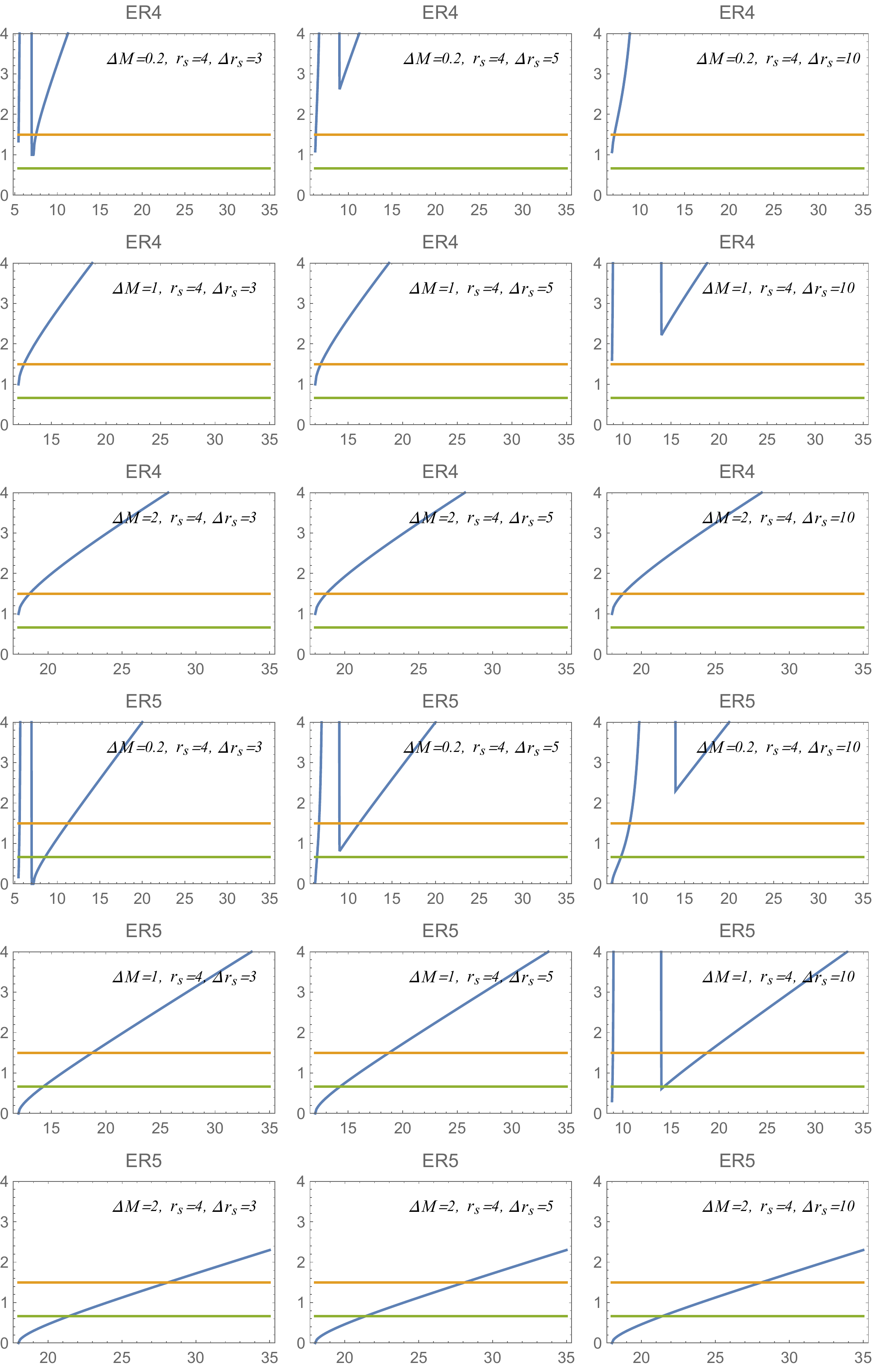}
	\caption{Function of the ratio of $f_U:f_L$ for ER4 and ER5 variants of the geodesic model. Orange line denotes the ratio $3:2$, green denotes the ratio 2:3.}
	\label{f:fmod3}
\end{figure*}
\begin{figure*} 
	\includegraphics[width=0.85\linewidth]{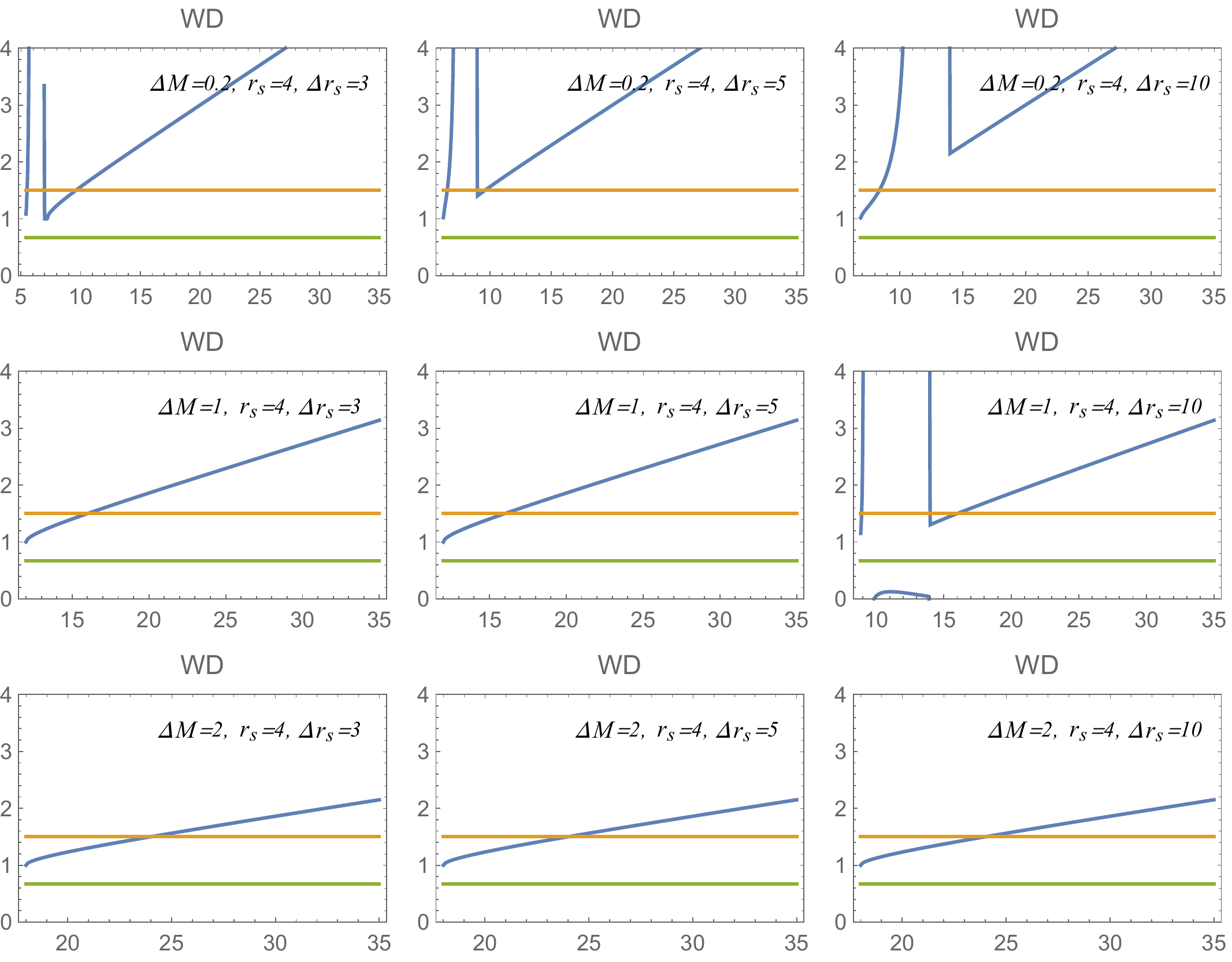}
	\caption{Function of the ratio of $f_U:f_L$ for a WD variants of the geodesic model. Orange line denotes the ratio $3:2$, green denotes the ratio 2:3.}
	\label{f:fmod4}
\end{figure*}
\FloatBarrier

\end{document}